\documentclass[longauth]{aa}
\pdfoutput=1
\usepackage{graphicx, multirow, tablefootnote, longtable}
\usepackage{txfonts, listings}
\usepackage{natbib}
\usepackage{xcolor, tikz}
\usepackage[percent]{overpic}

\bibpunct{(}{)}{;}{a}{}{,}

\begin{document}

  \title{From the bulge to the outer disc: \\
  {\tt StarHorse} stellar parameters, distances, and extinctions for stars in APOGEE DR16 and other spectroscopic surveys}

   \author{A. B. A. Queiroz\inst{1,2}, F. Anders\inst{3,2,1}, C. Chiappini\inst{1,2}, A. Khalatyan\inst{1}, B. X. Santiago\inst{4,2},\\
   M. Steinmetz\inst{1}, M. Valentini\inst{1}, A. Miglio\inst{5}, D. Bossini\inst{6}, B. Barbuy\inst{7}, I. Minchev\inst{1}, D. Minniti\inst{8,9,10}, \\ D. A. Garc\'ia Hern\'andez\inst{11,12}, M. Schultheis\inst{13}, R. L. Beaton\inst{14}, T. C. Beers\inst{15}, D. Bizyaev\inst{16,17}, J. R. Brownstein\inst{18},\\ K. Cunha\inst{19,20}, J. G. Fern\'andez-Trincado\inst{21}, P. M. Frinchaboy\inst{22}, R. R. Lane\inst{23,24}, S. R. Majewski\inst{25}, D. Nataf\inst{26}, \\  C. Nitschelm\inst{27}, K. Pan\inst{16}, A. Roman-Lopes\inst{28}, J. S. Sobeck\inst{29}, G. Stringfellow\inst{30}, O. Zamora\inst{11}}

   \authorrunning{A. Queiroz et al.}
   \titlerunning{New {\tt StarHorse} stellar parameters, distances, and extinctions for spectroscopic surveys}
    \institute{Leibniz-Institut f\"ur Astrophysik Potsdam (AIP), An der Sternwarte 16, 14482 Potsdam, Germany\\
              \email{aqueiroz@aip.de}
	\and{Laborat\'orio Interinstitucional de e-Astronomia - LIneA, Rua Gal. Jos\'e Cristino 77, Rio de Janeiro, RJ - 20921-400, Brazil}
	\and{Institut de Ci\`encies del Cosmos, Universitat de Barcelona (IEEC-UB), Carrer Mart\'i i Franqu\`es 1, 08028 Barcelona, Spain}
     \and{Instituto de F\'\i sica, Universidade Federal do Rio Grande do Sul, Caixa Postal 15051, Porto Alegre, RS - 91501-970, Brazil}
     \and{School of Physics and Astronomy, University of Birmingham, Edgbaston, Birmingham, B 15 2TT, United Kingdom}
     \and{Centro de Astrofísica da Universidade do Porto, Rua das Estrelas, 4150-762 Porto, Portugal}
     \and{Department of Astronomy, Universidade de São Paulo, São Paulo 05508-090, Brazil}
     \and{Departamento de Ciencias Fisicas, Facultad de Ciencias Exactas, Universidad Andres Bello, Av. Fernandez Concha 700, Las Condes, Santiago, Chile}
    \and{Millennium Institute of Astrophysics, Av. Vicuna Mackenna 4860, 782-0436, Santiago, Chile}
    \and{Vatican Observatory, V00120 Vatican City State, Italy}
    \and{Instituto de Astrof\'isica de Canarias (IAC), E-38205 La Laguna, Tenerife, Spain}
    \and{Universidad de La Laguna (ULL), Departamento de Astrof\'isica, E-38206 La Laguna, Tenerife, Spain}
     \and{Observatoire de la C\^ote d'Azur, Laboratoire Lagrange, 06304 Nice Cedex 4, France}
      \and{The Carnegie Observatories, 813 Santa Barbara Street, Pasadena, CA 91101, USA}
     \and{Department of Physics and JINA Center for the Evolution of the Elements, University of Notre Dame, Notre Dame, IN 46556, USA Michigan State University}
     \and{Apache Point Observatory, P.O. Box 59, Sunspot, NM 88349}
      \and{Sternberg Astronomical Institute, Moscow State University, Moscow, Russia}
      \and{Department of Physics and Astronomy, University of Utah, 115 S. 1400 E., Salt Lake City, UT 84112, USA}
     \and{University of Arizona, Tucson, AZ 85719, USA}
     \and{Observat’orio Nacional, Sao Cristóvao, Rio de Janeiro, Brazil}
     \and{Institut Utinam, CNRS UMR 6213, Universit\'e Bourgogne-Franche-Comt\'e, OSU THETA Franche-Comt\'e, Observatoire de Besan\c{c}on, BP 1615, 25010 Besan\c{c}on Cedex, France}
     \and{Department of Physics \& Astronomy, Texas Christian University, Fort Worth, TX 76129, USA}
      \and{Instituto de Astrof\'isica, Pontificia Universidad Cat\'olica de Chile, Av. Vicuna Mackenna 4860, 782-0436 Macul, Santiago, Chile}
      \and{Instituto de Astronomía y Ciencias Planetarias, Universidad de Atacama, Copayapu 485 Copiap\'o
Chile}
      \and{Department of Astronomy, University of Virginia, Charlottesville, VA 22904-4325, USA}
      \and{Center for Astrophysical Sciences, Department of Physics and Astronomy, Johns Hopkins University, 3400 North Charles Street, Baltimore, MD 21218, USA}
      \and{Centro de Astronom\'ia (CITEVA), Universidad de Antofagasta, Avenida Angamos 601, Antofagasta 1270300, Chile}
      \and{Departamento de Física, Facultad de Ciencias, Universidad de La Serena, Cisternas 1200, La Serena, Chile}
      \and{Department of Astronomy, University of Washington, Box 351580, Seattle, WA 98195, USA}
      \and{Center for Astrophysics and Space Astronomy, Department of Astrophysical and Planetary Sciences, University of Colorado, 389 UCB, Boulder, CO 80309-0389, USA}
      }

   \date{Received \today; accepted xx.yy.20zz}

  \abstract
  {We combine high-resolution spectroscopic data from APOGEE-2 survey Data Release 16 (DR16) with broad-band photometric data from several sources, as well as parallaxes from {\it Gaia} Data Release 2 (DR2). Using the Bayesian isochrone-fitting code {\tt StarHorse}, we derive distances, extinctions and astrophysical parameters for around 388,815 APOGEE stars, achieving typical distance uncertainties of $\sim 6\%$ for APOGEE giants, $\sim 2\%$ for APOGEE dwarfs, as well as extinction uncertainties of $\sim 0.07$ mag when all photometric information is available, and $\sim 0.17$ mag if optical photometry is missing. {\tt StarHorse} uncertainties vary with the input spectroscopic catalogue, with the available photometry, and with the parallax uncertainties.
To illustrate the impact of our results, we show that, thanks to {\it Gaia} DR2 and the now larger sky coverage of APOGEE-2 (including APOGEE-South), we obtain an extended map of the Galactic plane, providing an unprecedented coverage of the disk close to the Galactic mid-plane  ($|Z_{Gal}|<1$ kpc) from the Galactic Centre out to $R_{\rm Gal}\sim 20$ kpc. The improvements in statistics as well as distance and extinction uncertainties unveil the presence of the bar in stellar density, as well as the striking chemical duality in the innermost regions of the disk, now clearly extending to the inner bulge.
We complement this paper with distances and extinctions for stars in other public released spectroscopic surveys: 324,999 in GALAH DR2, 4,928,715 in LAMOST DR5,  408,894 in RAVE DR6, and 6,095 in GES DR3.}
\maketitle

\section{Introduction}\label{introd}

The second data release (DR2) of ESA's astrometric flagship mission {\it Gaia} has added an invaluable wealth of astrometric and photometric data for more than a billion stars \citep{GaiaCollaboration2018}. While the DR2 parallax uncertainties are still sufficiently large to hamper detailed tomographic views of the Galaxy beyond $2-3$ kpc around the Sun from {\it Gaia} data alone, the combination of these data with spectroscopic and photometric measurements from various other surveys opens up the possibility of extending the 3D mapping of Galactic stellar populations as far as the Galactic Centre, as well as out to similar heliocentric distances towards the outer disc or directions perpendicular to the disc mid-plane. This enables detailed quantitative comparisons between observed properties in phase and chemical space to chemo-dynamical model predictions (e.g. \citealt{Fragkoudi2018, Frankel2018}). Additionally, for the first time, ages of large numbers of field stars are being determined with sufficient precision, at least within $\simeq 2$ kpc, to impose strong direct constraints on the Galactic star formation history \citep{Bensby2017, Ness2019}.

In \citet[][Q18]{Queiroz2018}, we presented the {\tt StarHorse} code: a {\tt python} tool that uses Bayesian analysis of spectroscopic, photometric, and astrometric data to infer distances, extinction, ages, and masses of field stars. In that paper we illustrated the impact of {\it Gaia} DR1 parallaxes on improving our estimates of distances and extinctions, and generated several value-added catalogues for the spectroscopic datasets APOGEE DR14 \citep{Abolfathi2018}, RAVE DR5 \citep{Kunder2017}, GES DR3 \citep{Gilmore2012a}, and GALAH DR1 \citep{Martell2017}, thus extending the volume for which precise distances are available.

{\tt StarHorse} has been applied in numerous studies concerning different fields of Galactic astrophysics, such as stellar populations in the local neighbourhood \citep[e.g.][]{Anders2018a, Grieves2018, Minchev2018}, the origin of the stellar halo \citep{Fernandez-Alvar2017, Fernandez-Alvar2018}, the physical carriers of diffuse interstellar bands \citep{Elyajouri2019}, Milky Way stellar population kinematics \citep[e.g.][]{Palicio2018, Monari2018, Carrillo2019, Minchev2019}, or recently the chemodynamical study of N-rich stars \citep{Fernandez-Trincado2019}.

In \citet[][A19]{Anders2019}, we used an updated version of {\tt StarHorse}, combining {\it Gaia} DR2 parallaxes and optical photometry with other photometric bands from PanSTARRS-1, 2MASS, and AllWISE, to derive Bayesian distances and extinctions for around 300 million stars brighter than $G=18$. We showed that the addition of complementary information to the {\it Gaia} parallaxes and photometry could lead to a breakthrough where, with the best-quality data, one could start seeing structures like the Galactic bar already in density stellar maps. However, as explained in that paper, the A19 photo-astrometric results were computed with a prior upper limit of 4 mag in $A_V$ extinction, resulting in a limited view of stellar populations towards the innermost regions.

We now have the opportunity to start dissecting the Milky Way, including the central region and the far side of the Galactic disc, by combining Gaia DR2 with the APOGEE DR16 release. The latter includes around 380 000 stars with precise radial velocities, stellar parameters and chemistry from NIR high-resolution spectra taken in both hemispheres \citep{Ahumada2020}. Compared to the earlier releases, the data now include many more targets in general, but especially towards the innermost kiloparsecs of the Galaxy.

In this paper we describe the first value-added catalogues (VACs) generated from {\tt StarHorse} using {\it Gaia} DR2 data in conjunction with APOGEE DR16, as well as public releases of other spectroscopic surveys. We show the first high spatial-resolution chemical maps of our Galaxy covering the entire disc, from 0 to beyond 20 kpc in Galactocentric distance, complementing earlier maps shown in \citet{Anders2014, Hayden2015}, or \citet{Weinberg2019}, who used APOGEE DR10, DR12, and DR14, respectively. We present distances and extinctions, and their associated uncertainties, study the robustness of these quantities to different choices of priors, parameter sets, and systematic corrections, and also compare them to data from other sources.

The paper is organized as follows: in Sect. \ref{method} we provide a brief description of the {\tt StarHorse} code, focusing on the main improvements since Q18 and A19. Section \ref{in2sh} describes the input data (photometry, astrometry and spectroscopy) used in the computations of distances and extinctions for APOGEE DR16. In Sect. \ref{out2sh} we describe the output parameters that resulted from the {\tt StarHorse} calculation, as well as their uncertainties. As an example of science application, Sect. \ref{maps} presents the first density and chemical maps obtained over the entire Galactic plane, and discusses the main science implications  derived from these maps. In Sect. \ref{cats} we discuss the complementary catalogues GALAH, LAMOST, RAVE, and GES,  and the distribution of the resulting parameters. The resulting catalogues are provided in machine-readable form:  the data model can be found in Appendix \ref{datamodel}, a set of validation tests that for our new {\tt StarHorse} APOGEE DR16 results can be found in Appendix \ref{sec:validation}, and summary plots for each survey are shown in Appendix \ref{othervacs}. In Section \ref{conclusion} our main conclusions are summarized.

\section {The {\tt StarHorse} code}
\label{method}
\smallskip

{\tt StarHorse} (in the following sometimes abbreviated as SH) is a Bayesian  isochrone-fitting code that derives distances $d$, extinctions in the V band (at $\lambda=542$ nm) $A_V$, ages $\tau$, masses $m_{\ast}$, effective temperatures $T_{\mathrm{eff}}$,  metallicities [M/H], and surface gravities $\log g$ for field stars. In order to do so, we use as input a set of spectroscopically-measured stellar parameters (typically $T_{\mathrm{eff}}, \log g$, and metallicity [M/H]), photometric magnitudes $m_{\lambda}$, and, when available, parallax $\varpi$, to estimate how close a stellar evolutionary track is to the observed data. We adopt here the latest version of the PARSEC stellar evolutionary model tracks \citep{Bressan2012, Marigo2017}\footnote{Downloaded from http://stev.oapd.inaf.it/cgi-bin/cmd}. For APOGEE DR16 we adopt a fine grid of models to compute the estimated parameters with steps of 0.01 dex in $\log \tau$ and 0.02 dex in metallicity [M/H], covering the ranges: $7.5 < \log \tau < 10.13 $ , $-2.2 <$ [M/H] $< 0.6 $.

To compute the posterior probability distribution function (PDF) for the model parameters given the observed data, we include priors about the geometry, metallicity and age characteristics of the main Galactic components, following previous Bayesian methods to derive distances \citep[e.g.,][]{Burnett2010, Burnett2011, Binney2014a}. The priors adopted here are the same as in Q18 and A19, namely: an IMF from \citet{Chabrier2003} for all Galactic components; exponential spatial density profiles for thin and thick discs (see Section 5 for a discussion on the differences between the geometric and chemical definitions of the thick disc); a spherical halo and a triaxial (ellipsoid+spherical) bulge/bar component, as well as broad Gaussians for the age and MDF priors. The normalisation of each Galactic component, as well as the solar position, were taken from \citet{Bland-Hawthorn2016}.

The code was first described in \cite{Santiago2016}, and later modified to also use {\it Gaia} parallaxes and derive astrophysical stellar parameters in \cite{Queiroz2018}. The latter paper also included extensive validation comparisons with simulations and independent high-quality distance determinations from asteroseismology, eclipsing binaries, and open clusters. Those samples showed precision of $\simeq 10$\% for distances, ages accurate to $\simeq 30$\% and $A_V$ errors of $\simeq 0.1$ mag for stars out to $\simeq 1$kpc, with a continuous worsening of accuracy and precision towards larger distances. Most recently, in A19, we used {\tt StarHorse} to determine photo-astrometric (i.e. not using spectroscopy at all) distances, extinctions, and stellar parameters for {\it Gaia} DR2 stars down to magnitude $G=18$.

For more details about the method, the priors, stellar evolutionary models, code validation and previous released catalogues we refer the reader to Q18. We have since updated the code in some important aspects briefly summarized here, (for more details see A19), namely: we improved the extinction treatment, which now considers the dependence of the extinction coefficient, $A_{\lambda}$, on effective temperature and extinction itself -- as explained in \citet{Holtzman1995, Girardi2008}, for instance;  the latest version also has migrated to {\tt python 3.6}, which made the code faster and compatible with recent libraries. These and other small computational improvements are described in detail in A19, Sect. 3.2.

In this paper, we use the high-quality spectroscopically determined stellar parameters from APOGEE spectra, in conjunction with {\it Gaia} DR2 parallaxes and broad-band photometric measurements.

\section{Input data}\label{in2sh}

We follow a similar configuration to previous {\tt StarHorse} runs \citep{Queiroz2018, Anders2019}, to complement the APOGEE DR16 {\tt StarHorse} catalogue with parameters such as extinction and distances. For that we gather parallaxes from {\it Gaia} DR2 \citep{GaiaCollaboration2018} and photometry from 2MASS \citep{Cutri2003}, WISE \citep{Cutri2013} and PanSTARRS-1 \citep{Chambers2016} and gather this information with spectra from APOGEE DR16 \citep{Ahumada2020}. Here we introduce the input catalogues, the necessary adjustments in these data and the {\tt StarHorse} configuration to produce the final parameters.

\begin{figure*}
\centering
  \includegraphics[width=.9\textwidth]
  {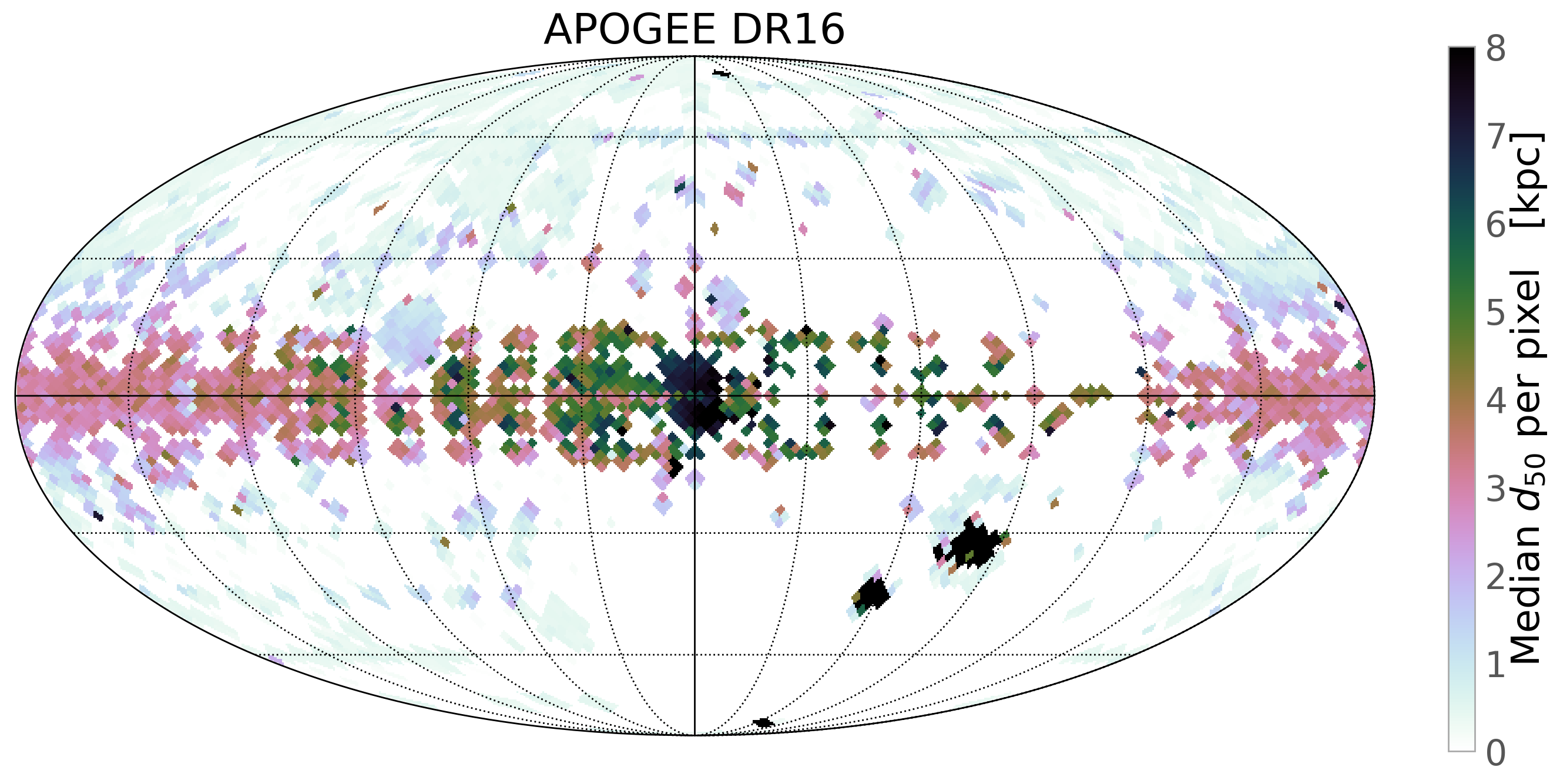} \\
  \includegraphics[width=.9\textwidth]
  {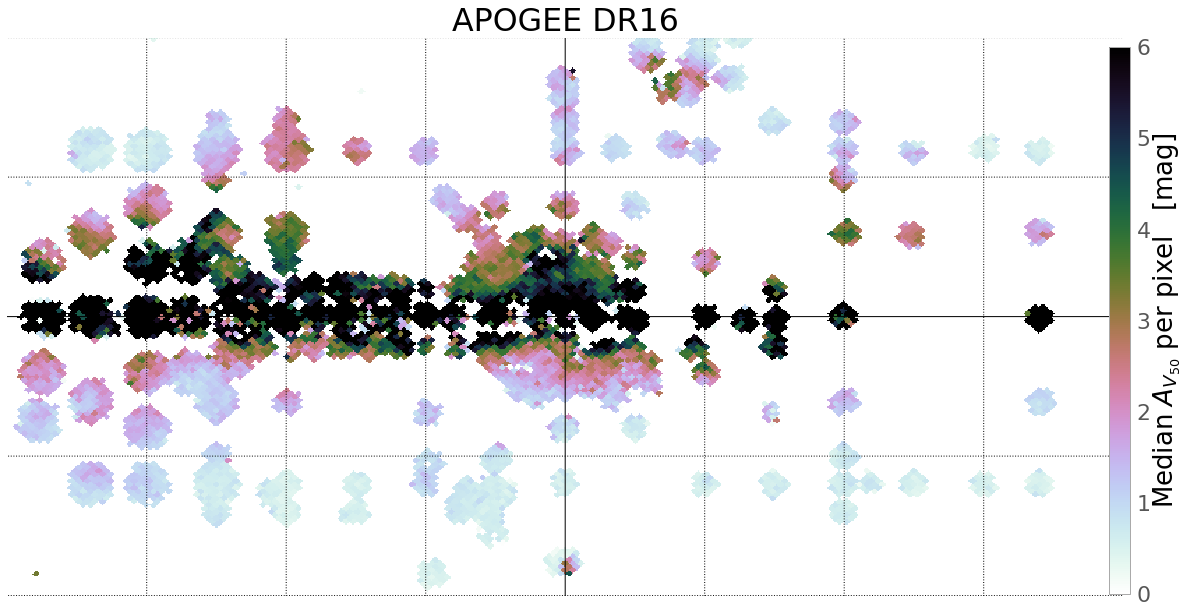} \\
    \caption{Overview of the coverage of the APOGEE DR16 {\tt StarHorse} VAC. Top panel: Median {\tt StarHorse} distance per HealPix cell as a function of Galactic coordinates. Lower panel: Same as in previous panel, but now showing the median $A_V$ as a function of direction in the sky, and zooming in on the  innermost 40 x 20 degrees of the Galactic plane (the line spacing in the lower panel is 10 degrees).} 
  \label{fig:maps}
\end{figure*}

\begin{figure*}
\centering
  \includegraphics[width=.9\textwidth]{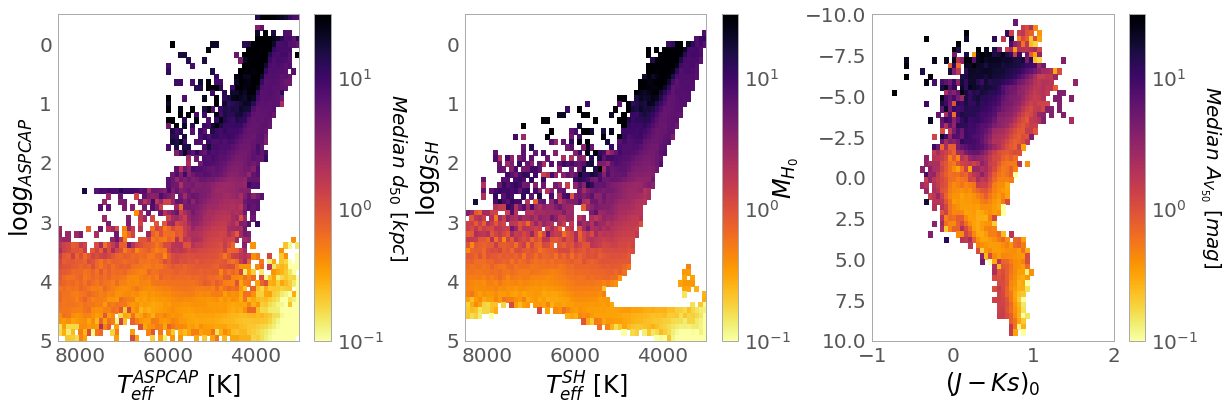}
  \caption{{\it Kiel} diagram colour colored by median {\tt StarHorse} distance as a function of position on the input (left) and output (middle) effective temperatures and surface gravities. The rightmost panel shows the  Colour Magnitude Diagram colored by median {\tt StarHorse} extinction, the color is already corrected by {\tt StarHorse} extinction.}
  \label{fig:kielcmd}
\end{figure*}

\subsection{APOGEE DR16}\label{inapogee}

The Apache Point Observatory Galaxy Evolution Experiment \citep[APOGEE,][]{Majewski2017} started in the third phase of the Sloan Digital Sky Survey \citep[SDSS-III,][]{Eisenstein2011}. APOGEE continues as part of SDSS-IV \citep{Blanton2017}. It is a spectroscopic survey conducted in the near-infrared (NIR), at high resolution ($R\sim 22,500$), and high signal-to-noise ($S/N > 100$) \citep{Wilson2019}. The data reduction pipeline is described  in \citet{Nidever2015}. As a NIR survey, APOGEE is capable to peer into the dusty areas of our Galaxy, such as the Bulge and the central Galactic plane.

The APOGEE survey is collecting data in the Northern Hemisphere since 2011. Since 2015, APOGEE-2 data are also being collected in the Southern Hemisphere. Both hemispheres observations use the twin NIR spectrographs \citep{Wilson2019} on the SDSS 2.5-m telescope at APO \citep{Gunn2006} and the 2.5m du Pont telescope at Las Campanas Observatory (LCO; \citealt{Bowen1973}), respectively. DR16 is the first SDSS-IV data release including data from APOGEE-2 South: it contains a total of 473,307 sources with derived atmospheric parameters and abundances. The pre-processing of the APOGEE DR16 data in preparation for the {\tt StarHorse} run presented here was very similar to the pre-processing of APOGEE DR14 described in Q18.

The APOGEE Stellar Parameter and Chemical Abundance Pipeline \citep[ASPCAP][]{GarciaPerez2016} was optimized for red-giant stars, since this is the main population targeted by the survey. However, we also compute {\tt StarHorse} results for stars in APOGEE DR16 catalogue that fall outside the recommended calibration ranges of ASPCAP. For those stars we use inflated uncertainties of $\sigma_{\log g}\ =\ 0.3$ dex, $\sigma_{T_{\rm eff}} = 200$ K, $\sigma_{[{\rm Fe/H}]}\ =\ 0.15$ dex, and $\sigma_{[\alpha/{\rm Fe}]} \ =\ 0.1$ dex.

As in Q18 (and differently from A19 where no extinction prior was used), we use the APOGEE targeting extinction estimate $A_{K_s}^{\rm Targ}$ as a broad prior for the total line-of-sight extinction: $A_{V_{\rm prior}} = 0.11\cdot A_{K_s}^{\rm Targ}$. {\tt StarHorse} treats this extinction using \citet{Schlafly2016} extinction curve.

Finally, because we use PARSEC stellar models, which at present do not include non-solar $[\alpha/Fe]$ ratio models, we correct for this effect in the input data. For that we use the \citet{Salaris1993} formula, which accounts for  $\alpha$-enhancement by a slight shift of the total metallicity [M/H]:

\begin{equation}\label{salaris}
{\rm [M/H]} = {\rm [Fe/H]} + \log{[C\cdot10^{[\alpha/{\rm Fe}]} + (1-C)]}
\end{equation}
\begin{equation}
\sigma_{[{\rm M/H}]} \simeq \sqrt{\sigma_{{\rm [Fe/H]}}^2 + \sigma_{[\alpha/{\rm Fe}]}^2}
\end{equation}

We choose $C = 0.66101$, in agreement with the scaled solar composition $Y=0.2485+1.78\cdot Z$ used in the PARSEC 1.2S models\footnote{\url{http://stev.oapd.inaf.it/cgi-bin/cmd_3.1}}.

\subsection{{\it Gaia} Data Release 2}\label{gdr2}

The {\it Gaia} astrometric mission was launched in December 2013 and placed  close to the L2 Lagrangian point, about 1.5 million km from the Earth, in July 2014. It is measuring positions, parallaxes, proper motions and photometry for well over $10^9$ sources down to $G \simeq 20.7$, and obtaining physical parameters and radial velocities for millions of brighter stars \citep{GaiaCollaboration2016a}. Its recent Data Release 2 ({\it Gaia} DR2; \citealt{GaiaCollaboration2018}), covers the initial 22 months of data taking (from a predicted total of $>5$ years), with positions and photometry for $1.7\cdot 10^9$ sources, full astrometric solution for $1.3\cdot 10^9$ \citep{Lindegren2018},  $T_{\mathrm{eff}}$, extinction, stellar radii and luminosities for $8\cdot 10^7$ stars \citep{Andrae2018}, and radial velocities for $7\cdot 10^6$ of them \citep{Katz2019}. Particularly important for our purposes are the DR2 parallaxes, since they allow us to better disentangle dwarfs from giants for stars with more uncertain surface gravity measurements.

The {\it Gaia} DR2 parallax precision varies from $< 0.03$ mas for $G \leq 13$ to $\simeq 0.7$ mas for $G=20$, and the parallax zero-point (accuracy) has been shown to be of similar order, and probably dependent on a combination of sky position, magnitude, and colour \citep[e.g][]{Arenou2018, Stassun2018, Zinn2019, Khan2019}. \\

In this work, we adopt the mean zero-point correction of $52.8~\mu$as to the {\it Gaia} DR2 parallaxes determined by \cite{Zinn2019} using red giants co-observed by APOGEE and the {\it Kepler} mission. This is somewhat mid-way between the quasar-derived correction advertised by \citet{Lindegren2018} and the zero-point shift estimated by \cite{Stassun2018}, which is $82~\mu as$. In fact, \cite{Stassun2018} find that their estimate of the offset may in fact be $61~\mu as$, which is much closer to that of \cite{Zinn2019}, if they allow for a possible scale error in the parallaxes. They also note that the larger offset of $82~\mu as$ would be most applicable to the brightest stars, with $G\lesssim 11$, however only $\sim$10\% of the APOGEE sample is so bright. On the other hand, \citet{Khan2019} found that the parallax zero-point shift could actually be smaller: for two $K2$ fields analysed in their paper they found smaller discrepancies between asteroseismic and astrometric parallaxes than in the {\it Kepler} field.

Independent distances measurements, using cepheids and quasars \citep{Riess2018,Lindegren2018} also show that the {\it Gaia} DR2 parallax uncertainties are slightly underestimated, and can suffer from systematics as well as the parallax itself. To account for these effects we have applied the suggested inflation of parallax uncertanties, described in \citet{Lindegren2018}. The inflated error can be written as: 
\begin{equation}
\sigma_{ext} = \sqrt{k\sigma_{in}^2 + \sigma_s^2},
\end{equation}
Where $\sigma_{ext}$ is the inflated uncertainty, $\sigma_{in}$ the uncertainty from {\it Gaia} catalogue, $k=1.08$ is a constant parameter and, $\sigma_s$ is slightly different for different magnitude ranges. For the bright regime ($G < 13.0 mag $) we use $\sigma_s=0.021$  and  for the faint ($G>15.0 mag $) we use  $\sigma_s=0.043$. In between these two regimes we interpolated linearly using $\sigma_s=0.030$.

The {\it Gaia} DR2 catalogue also includes broadband photometry for about $10^9$ sources, although in the case of APOGEE we decided not to include this photometry in the calculations. The reason for this choice is simply because most of APOGEE DR16 is targeting the Galactic plane, and in this region {\it Gaia} DR2 photometry partly suffers from crowding issues. In addition, it should be acknowledged that the {\it Gaia} DR2 photometry for the $G_{BP}$ and $G_{RP}$ bands is essentially aperture photometry, which has been shown to be problematic in regions of high stellar density and/or nebulous emission (e.g. \citealt{Evans2018, Arenou2018}). We therefore followed a conservative approach and did not use this photometry for the APOGEE sample.

\subsection{Photometric Catalogues}\label{photometry}

In all produced catalogues we use infra-red photometry from 2MASS \citep{Cutri2003} and WISE \citep{Cutri2013}. 
Both are all-sky photometric surveys, and 2MASS photometry has almost $100\%$ coverage of the APOGEE catalogue. For that reason we used it as primary photometry when running {\tt StarHorse} (see Q18 details). For both input catalogues, we applied a minimum photometric uncertainty of 0.03 mag. Finally, we assumed the uncertainty of the stars in 2MASS and WISE that have no measured uncertainty to be 0.4 mag.

For the optical regime we use PanSTARRS-1 photometry ($\lambda \sim 3943-10838$ \AA)  \citep{Scolnic2015} with corrected zeropoints according to \citet{Scolnic2015} and minimum photometric uncertainties of 0.04 mag. Furthermore, we only use measurements with reported individual errors for stars fainter than $G=14.5$.

Differently from Q18, we have decided not to use APASS \citep{Henden2014} photometry but only PanSTARRS-1. The motivation for this choice comes from reports that APASS photometry has a high percentage of sources (30 $\%$) with a positional mismatch, especially in the faint regime ($g_{sloan} > 16$) \citep{Marrese2019}.

\begin{figure*}
\sidecaption
\centering
  \includegraphics[width=.66\textwidth]{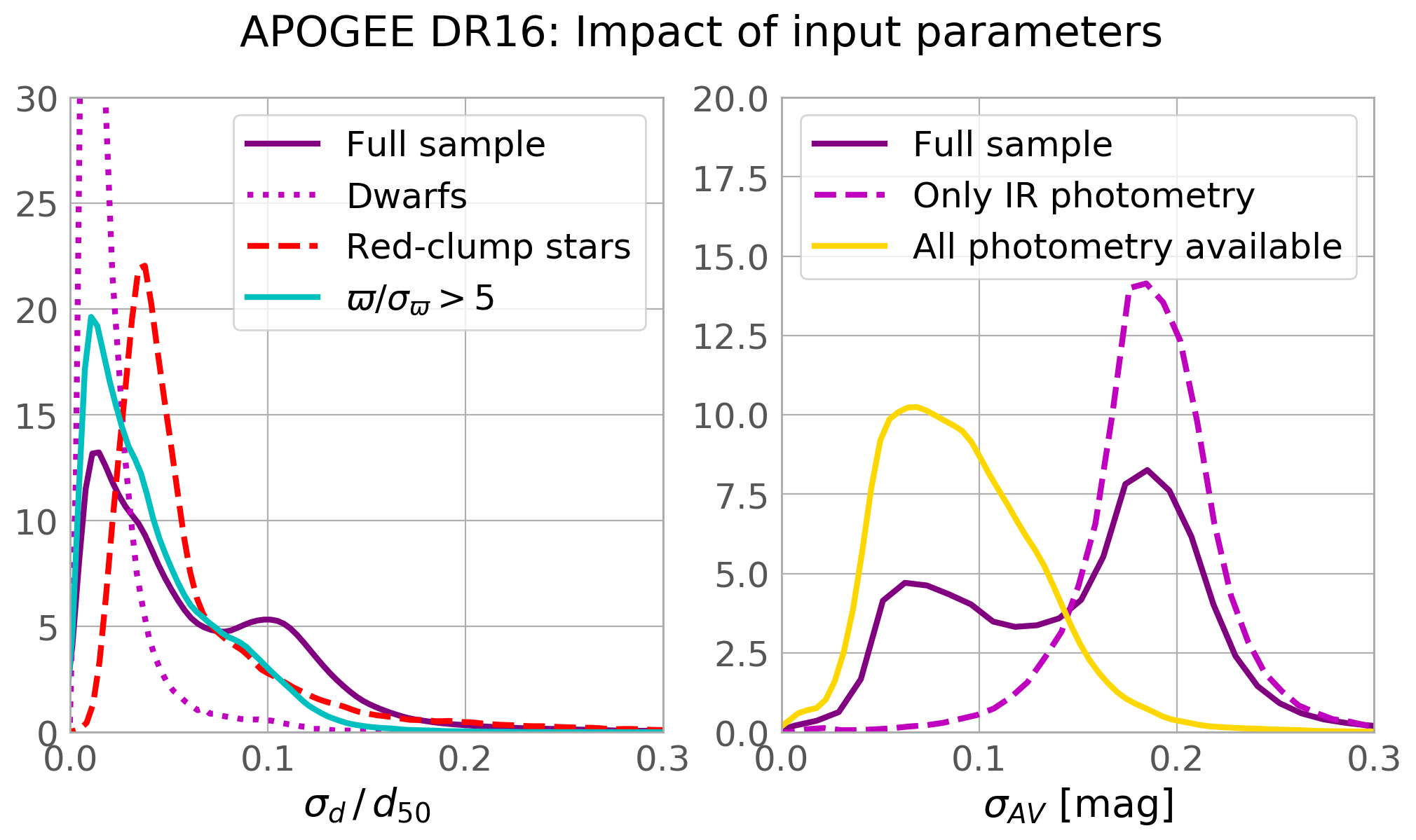}
  \caption{ Kernel-density estimates of the uncertainty distributions for the distances (left panel) and $A_V$ extinctions (right panel) in the APOGEE DR16 {\tt StarHorse} VAC. The different curves show the distributions of distances and extinctions uncertainties for subsets of different data input quality. In the left panel, we highlight the higher distance precision achieved for a) stars with {\it Gaia} DR2 parallaxes more precise than 20\%, b) dwarf stars ($\log g<4$), and c) red-clump stars ($2.3<\log g<2.5$). In the right panel, we show how the (un-)availability of optical photometry drastically improves/worsens the precision of our $A_V$ estimates.}
  \label{fig:uncerts}
\end{figure*}

\section{The APOGEE DR16 {\tt StarHorse} catalogue}
\label{out2sh}

The APOGEE DR16 {\tt StarHorse} catalogue presented here was generated from the processed APOGEE DR16 data, explained in Sect. \ref{inapogee}, cross-matched with {\it Gaia} DR2 ($98\%$), PanSTARRS-1 ($37\%$), 2MASS ($100\%$), and AllWISE ($95\%$). The final produced catalogue contains 388,815 unique stars with derived {\tt StarHorse} parameters, along with their uncertainties.  From the 473,307 APOGEE sources {\tt StarHorse} has converged for 418,715,  and after this we selected unique stars by the highest signal-to-noise. 

Our catalogue will appear as a value-added catalogue of the SDSS DR16 (Ahumada et al. 2020). The catalogue can also be downloaded from Leibniz Institute for Astrophysics (AIP) webpage\footnote{\url{https://data.aip.de/aqueiroz2020}}, similar to what was done in Q18. The description and format of the provided {\tt StarHorse} products are listed in Table \ref{table:outtablefmt}, while the description of the adopted input and output quality flags can be found in Table~\ref{table:flags}.

\subsection{Output parameters - a first look at the catalogue}

\begin{figure*}
\vspace{7cm}
  \begin{overpic}[scale=0.36]{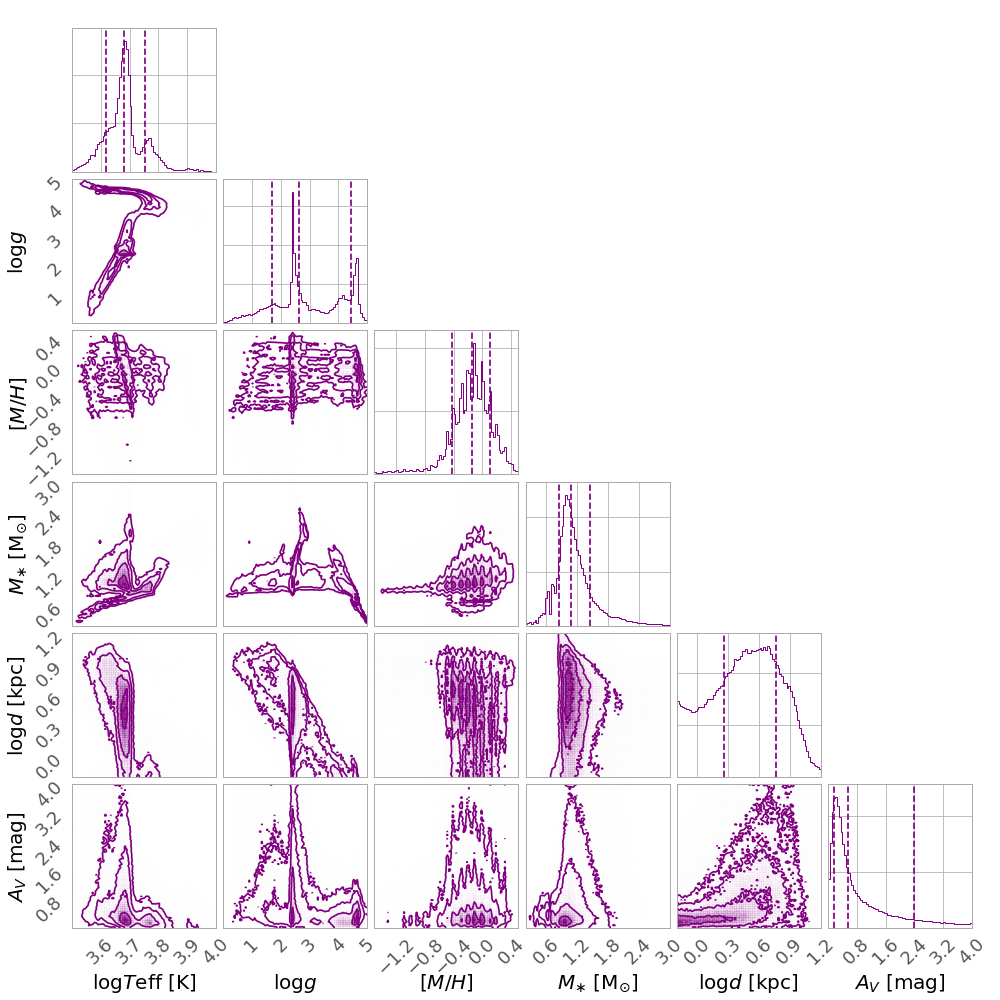}
     \put(52,52){\includegraphics[scale=0.36]{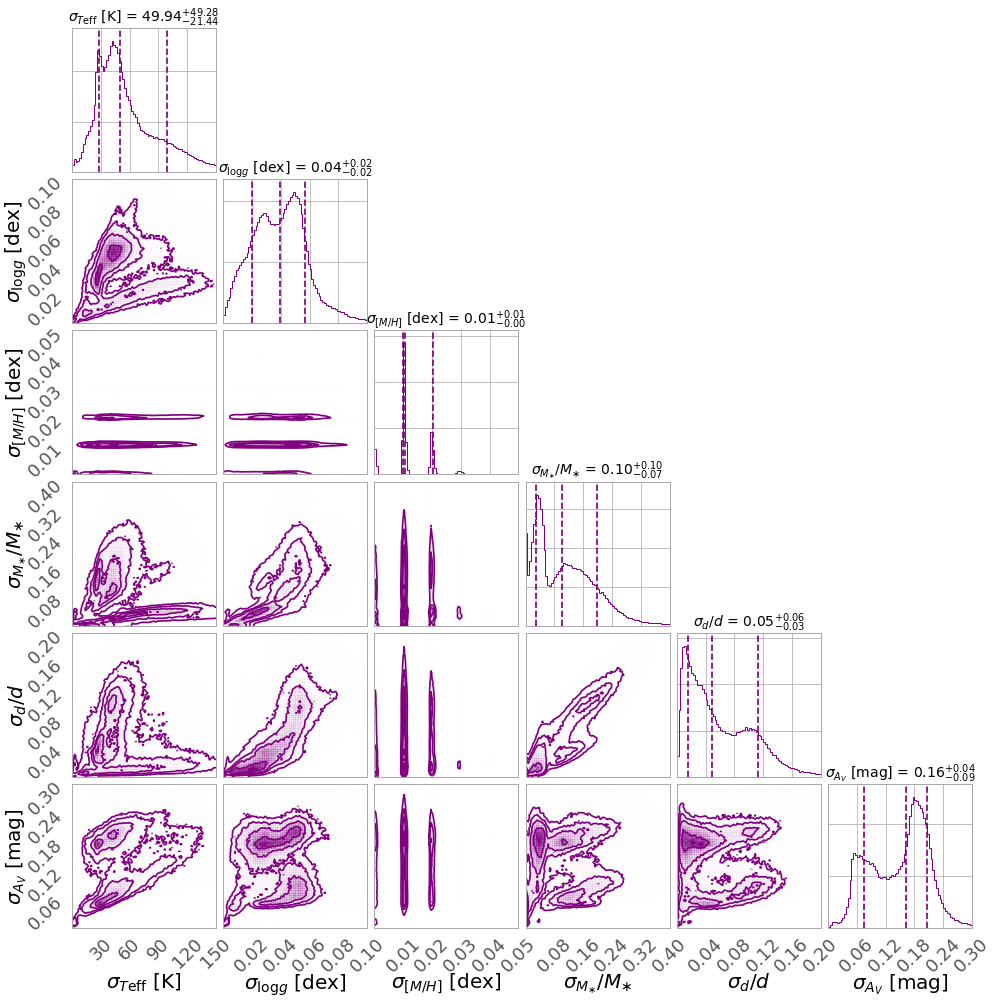}}  
  \end{overpic}
\caption{1D distributions and correlations between {\tt StarHorse} output parameters (bottom left corner plot) and their corresponding uncertainties (top right corner plot) for the APOGEE DR16 sample.}
\label{fig:output_summary}
\end{figure*}

The {\tt StarHorse} output provides the posterior distribution functions of masses, effective temperatures, surface gravity, metallicities, distances, and extinctions -- see Table \ref{table:outtablefmt}. The median value 50th percentile should be taken as the best estimate for that given quantity and the uncertainty can be determined using the 84th and 16th percentiles. The full distribution of the {\tt StarHorse} median output parameters is shown in the left panel of Fig.~\ref{fig:output_summary}.

In addition to the percentile values of the estimated parameters, all released value-added catalogues have columns that describe the {\tt StarHorse} input data, {\tt SH\_INPUTFLAGS}, and the {\tt StarHorse} output data, {\tt SH\_OUTPUTFLAGS} as specified in Table \ref{table:outtablefmt}. The input flags describe which parameters were used in the likelihood calculation for each star. They also indicate if we have used an $A_V$ prior - as the AVprior flag - or if the $A_V$ was determined using the parallax True option (See Q18). The meaning of each string in SH\_INPUTFLAGS can be seen in Table \ref{table:flags}. The output flags inform the number of models which have converged in the likelihood calculation, and also indicate the occurrence of problems in the estimated extinction (see also Table\ref{table:flags}).

In what follows we present some of the basic properties of the APOGEE-DR16 {\tt StarHorse} catalogue (maps involving chemical abundances will be discussed in the next section). In the following figures we have applied a few quality cuts, namely: stars with signal-to-noise ratio greater than 50 ({\tt SNREV} $> 50$), with non-negative posterior extinction ($A_{V_{84}}> 0$), and with a good ASPCAP fit ({\tt ASPCAP\_CHI2} $< 25$). This corresponds to $\simeq 95\%$ of the converged stars.

Figure \ref{fig:maps} shows Galactic maps colour coded by the median of the resulting APOGEE DR16 {\tt StarHorse} distances (upper panel) and extinctions (middle panel). By design, most of the APOGEE pointings are concentrated towards low Galactic latitudes \citep{Zasowski2013, Zasowski2017}, offering a much greater coverage of the thin disc than other surveys. The North-South equatorial asymmetry is also visible, since most of the data so far come from the Northern spectrograph at Apache Point Observatory. Yet, the Magellanic Clouds are clearly visible on the distance map as the distant clumps of sampled stars. Because the density of stars increases towards the Galactic centre, there is also a clear trend of larger median distances in this direction. The $A_V$ map in the bottom panel of Fig. \ref{fig:maps} zooms into the central degrees of the Galactic plane, where the average extinction is higher, and patchy (also visible in this map).

Figure \ref{fig:kielcmd} shows the mean distance per pixel in the spectroscopic {\it Kiel} diagram, using the input parameters from APOGEE (ASPCAP, left panel) and using the {\tt StarHorse} output spectroscopic parameters (middle panel). As expected, stars belonging to the giant branch (comprising most of the APOGEE sample) are found at larger distances than dwarfs since they have brighter absolute magnitude and are therefore detectable in a larger distance range. 
In the giants regime {\tt StarHorse} seems to be detecting asymptotic giant branch stars (AGBs) at very large distances (at $ T_{\rm eff} \sim  4500K $ and $\log g < 1.0$ ), as expected since those are very bright stars. However, those stars should be analysed with care, since the ASPCAP pipeline does not perform well in this range \citep{GarciaPerez2016}. The third panel of Figure \ref{fig:kielcmd} shows also higher extinction for intrinsic brighter and therefore distant stars. The output spectroscopic parameters from {\tt StarHorse} seem to be, as expected, very much in accordance with the input ASPCAP parameters. For the dwarfs stars, which are not ASPCAP calibrated stars and therefore have larger uncertainties, {\tt StarHorse} seems to improve the results finding a smoother solution, as expected because of the use of stellar evolutionary models.

The distribution of distance and extinction uncertainties for the APOGEE DR16 {\tt StarHorse} catalogue are shown in Fig. ~\ref{fig:uncerts}.
Thanks to the availability of {\it Gaia} DR2 parallaxes, the distance uncertainties (left panel of Fig.\ref{fig:uncerts}) are usually smaller than $10\%$. The three peaks at $\simeq 2\%, 4\%$ and $\simeq 10\%$ correspond to nearby dwarf stars within the {\it Gaia} DR2 parallax sphere, red-clump stars, and more distant giant stars, respectively. These distance uncertainties are slightly improved with respect to those from the DR14 APOGEE and Tycho-Gaia astrometric solution (TGAS) sample discussed in Q18, but now are available for a much larger number of stars, covering much larger volumes. Even for distant upper red-giant branch stars with more uncertain parallaxes (e.g. APOGEE targets near the Galactic center), the achieved distance precision is typically within $10\%$.

The extinction uncertainty distribution (right panel of Fig. \ref{fig:uncerts}) is clearly double-peaked, at $A_V \simeq 0.07$ and $A_V \simeq 0.17$, as previously observed by Q18 for APOGEE DR14 combined with TGAS. As shown by the two subsets in the figure, the two peaks correspond to stars with and without available optical magnitudes, respectively. A more detailed discussion of the accuracy of the obtained parameters can be found in  Appendix \ref{sec:validation}.

In Figure \ref{fig:output_summary} we show the correlations between the output parameters and the correlations between the output uncertainties. We see the expected correlations between stellar parameters inherited from the isochrones (e.g. the $\log g$ vs. $T_{\rm eff}$ diagram), as well as stellar population effects, such as the decrease of $\log g$ with increasing distance, or a greater metallicity range for greater distances. Extinction is correlated with increasing mass, metallicity and distance. The doubled peaked uncertainty distribution in extinction is not explained by any other output parameter uncertainty, apart from the completeness of the photometric set as seen in fig. \ref{fig:uncerts}. The uncertainties in the other parameters show approximately linear correlations between $\log g$ and mass, $\log g$ and distance, as well as mass and distance. The distribution of each parameter and its uncertainty can also be seen in the diagonal row of that plot, along with the uncertainty statistics for each of the {\tt StarHorse} output parameters.

\section{Extended Chemical maps in the Galactic plane up to the Bulge}
\label{maps}

\begin{figure*}
  \centering
  \includegraphics[width=1.0\textwidth]{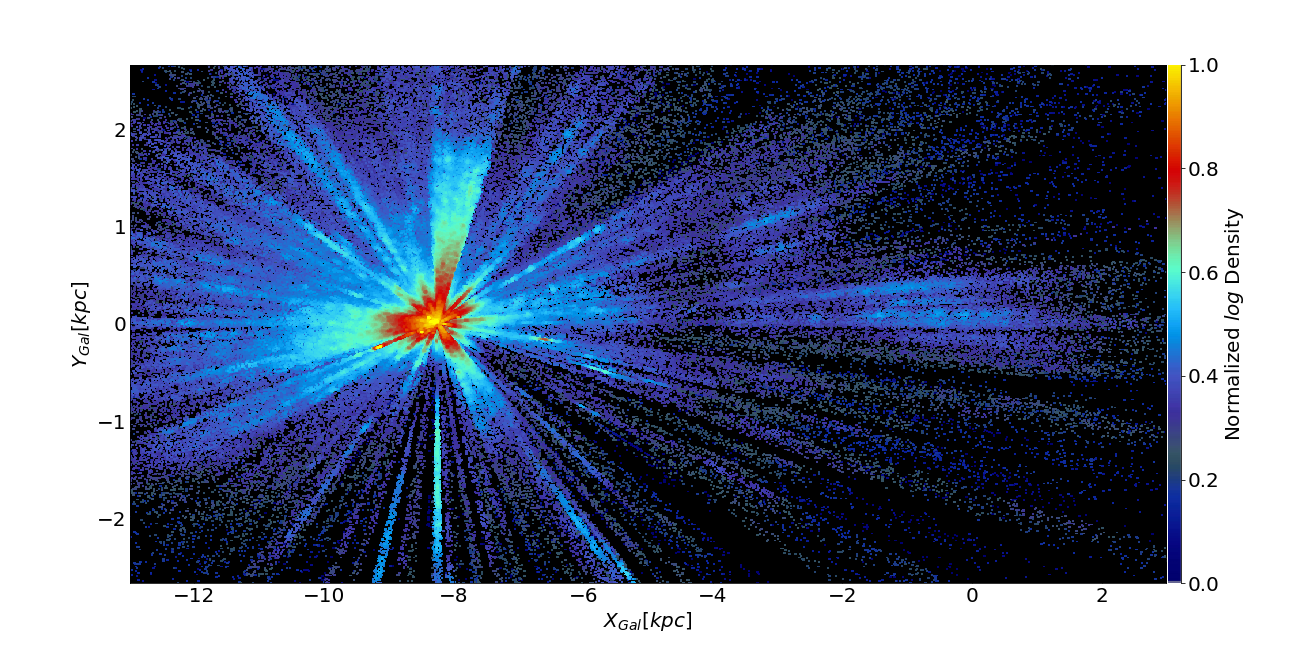}
  \includegraphics[width=1.0\textwidth]{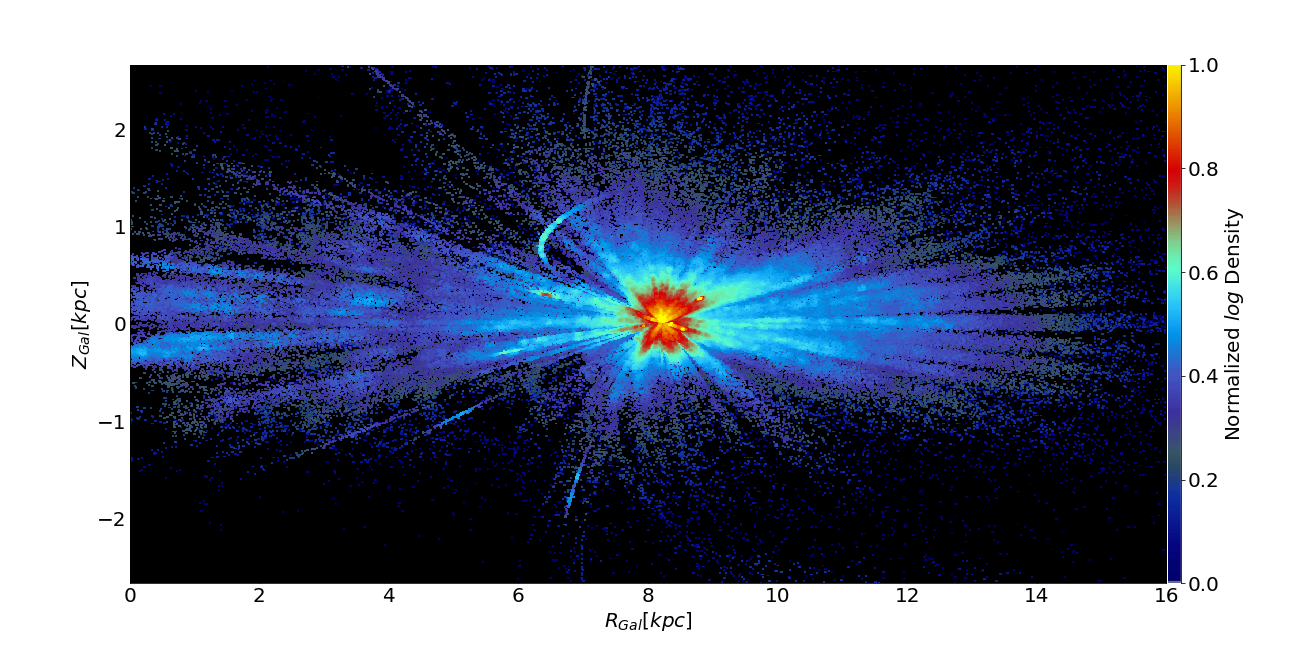}
  \caption{Top panel: Galactocentric Cartesian $XY$ map of the APOGEE DR16 sample, the colour represents the 3D local stellar density estimated by using the  smoothed-particle technique \citep{Monaghan1992} with $N_{ngb}=35$ neigbhours; we use the maximum intensity projection rendering method implemented in {\tt pmviewer} which is described  on \protect\url{http://pmviewer.sourceforge.net}.  Bottom panel: Density distribution in Galactocentric cylindrical $RZ$ coordinates. Some distinct features of APOGEE targeting can be easily discerned: the high target density in the {\it Kepler} field, enhanced density distributions around open clusters (sometimes elongated when the distance precision is low - e.g. ($d\sim 5.2$ kpc) $\omega$ Cen appearing around $R \simeq 6.5$ kpc, $Z \simeq 1$ kpc.)}
  \label{maps2}
\end{figure*}

In this section, we demonstrate the value of the APOGEE DR16 {\tt StarHorse} results by presenting the most extensive and precise chemical-abundance map of the Milky Way disc and bulge to date. The unprecedented coverage of the APOGEE DR16 data can be appreciated in Fig. \ref{maps2}, where we show the density distribution of all DR16 stars with {\tt StarHorse} results in Galactocentric coordinates. 
The figure shows very clearly that the APOGEE DR16 sample covers a large portion of the Galaxy with statistically significant samples, now including also the innermost regions with many more stars close to the Galactic mid-plane ($Z_{\rm Gal}<$0.5 kpc) thanks to the Southern observations taken at Las Campanas. This is an important improvement both in number of targets and in quality of distances and extinction estimates, with respect to previous releases.

To be more quantitative: the stellar density of the APOGEE DR16 sample amounts to over a thousand stars per kpc$^2$ in the complete $R_{\rm Gal}-Z_{\rm Gal}$ plane for $0<R_{\rm Gal}<15$ kpc and $-1 {\rm kpc}<Z_{\rm Gal}<1$ kpc (see Fig. \ref{maps2}), allowing for an unprecedented chemo-kinematic mapping of the inner as well as outer stellar disc. The upper panel of Fig. \ref{maps2} displays a top-down view of the Galactic disc, again demonstrating the exquisite spatial coverage of the APOGEE DR16 sample. The figure also shows a slight but distinct density enhancement in the region of the stellar bar, as observed for the full {\it Gaia} DR2 dataset in \citet{Anders2019}, but with the canonical inclination angle of $\sim 25$ degrees \citep[e.g.][]{Bland-Hawthorn2016}.

Since APOGEE traces around 20 chemical elements at high spectral resolution and provides radial velocities precise to $\sim300$ m/s \citep{Majewski2017}, this dataset will be a legacy for detailed chemodynamical studies of the Milky Way at least for several years.

\begin{figure*}
\centering
  \includegraphics[width=1.\textwidth,origin=c]{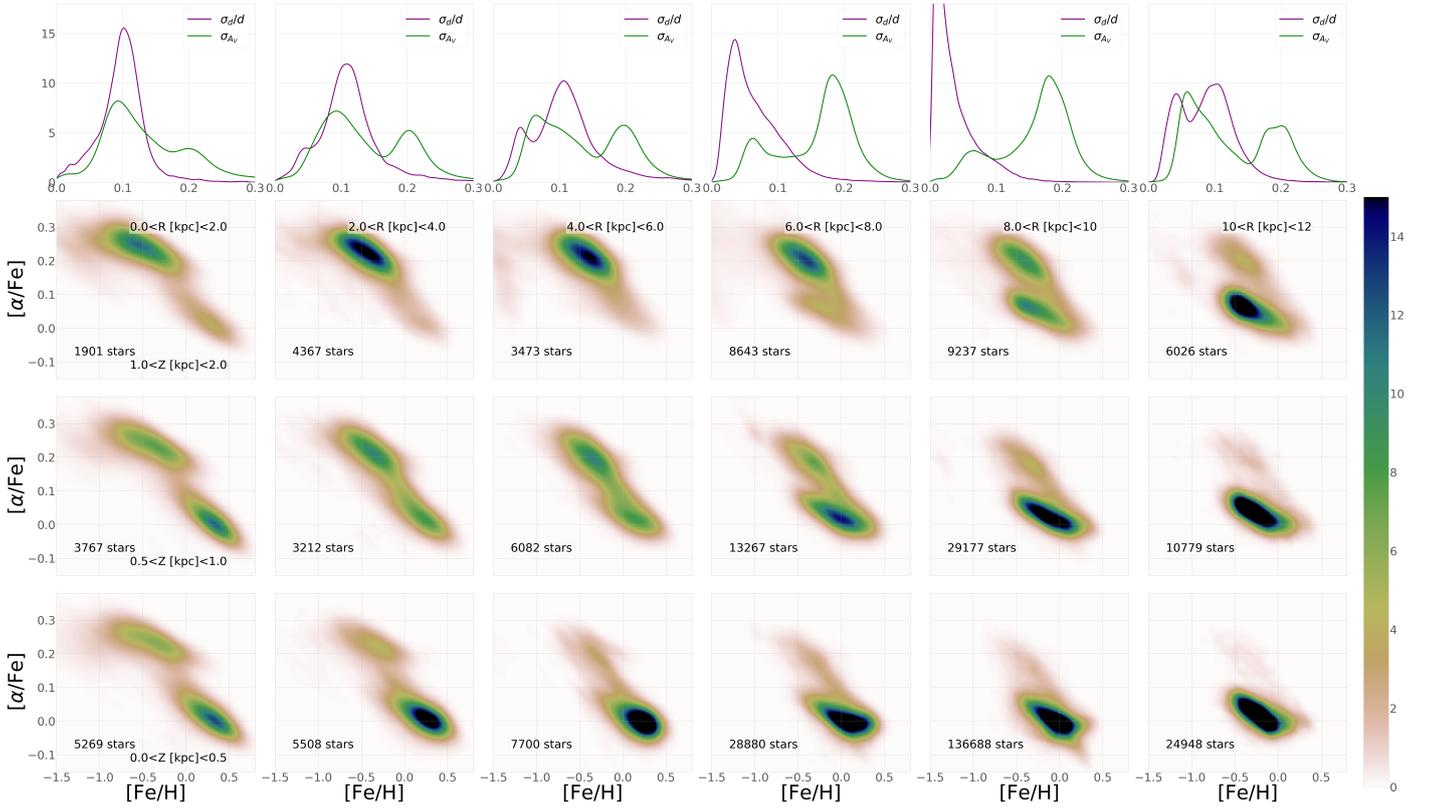}
  \caption{APOGEE DR16 [$\alpha$/Fe] vs. [Fe/H] diagrams in bins of Galactocentric cylindrical coordinates, similar to the chemical maps presented in \citet{Hayden2015}, but extending further into the inner Galaxy. The upper panels show kernel-density estimates of the uncertainties distributions in {\tt StarHorse} extinctions and distances, for each  Galactocentric distance bin (including all $Z_{\rm Gal}$ bins).}
  \label{alphafe_maps_a}
\end{figure*}

\begin{figure*}
\centering
  \includegraphics[width=1.0\textwidth,origin=c]{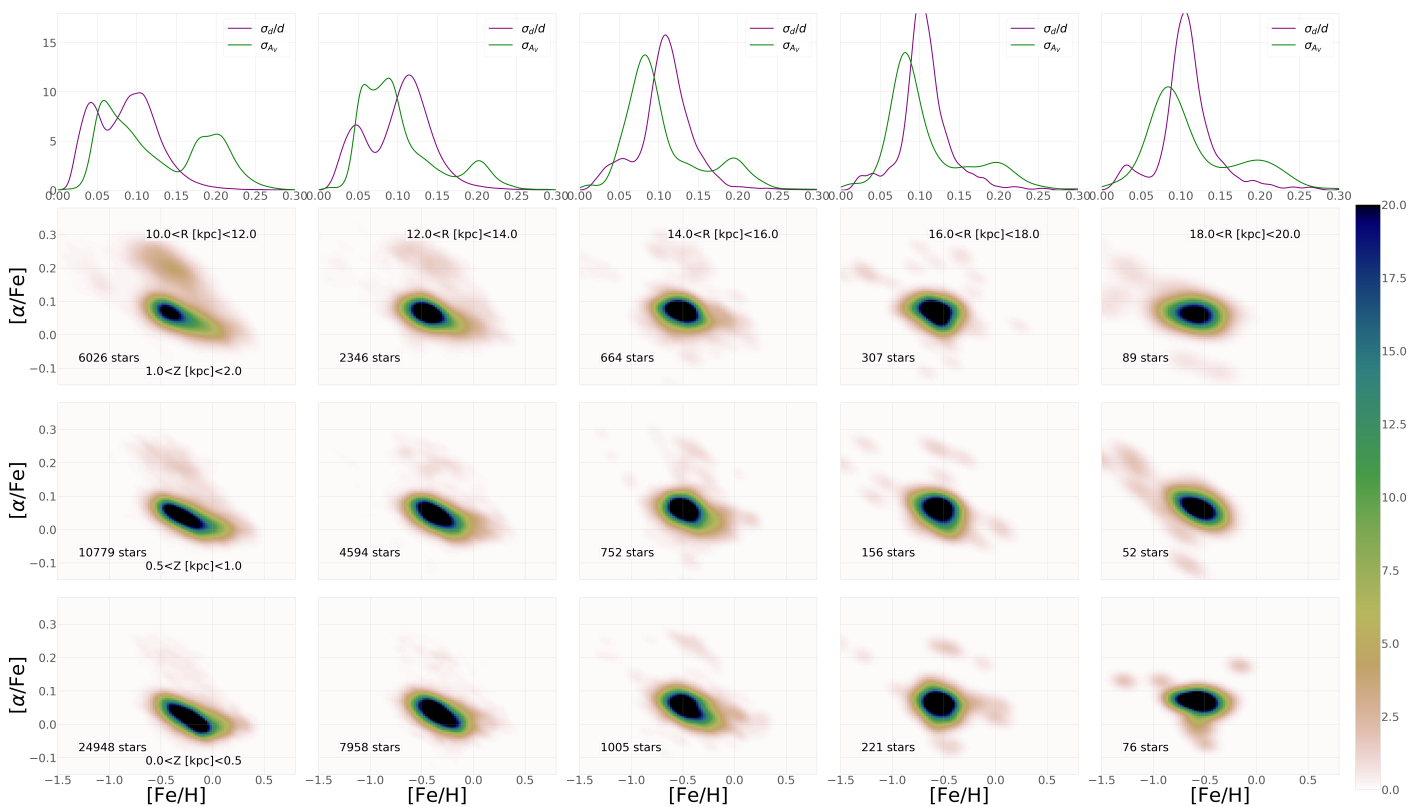}
  \caption{Same as previous Figure, but now extending to the outer disk.}
  \label{alphafe_maps_b}
\end{figure*}

\begin{figure*}
\centering
 \includegraphics[width=1.0\textwidth,origin=c]
{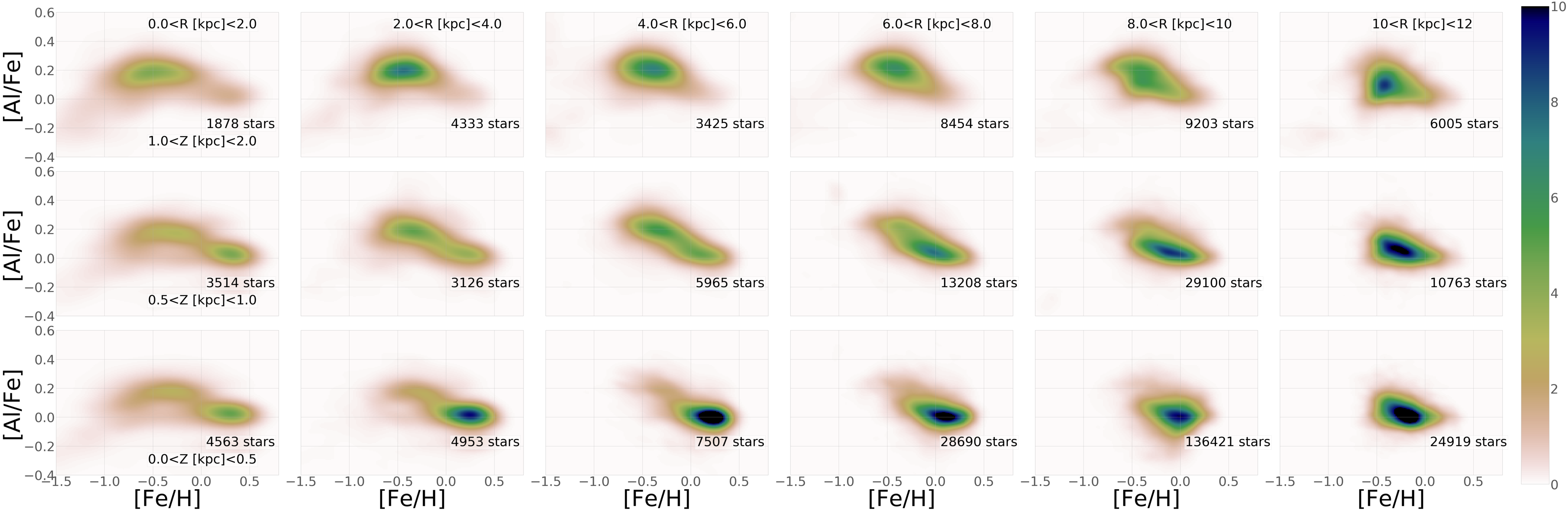}
  \caption{APOGEE DR16 [Al/Fe] vs. [Fe/H] diagrams in bins of Galactocentric cylindrical coordinates out to $R_{\rm Gal}=10$ kpc, similar to Fig. \ref{alphafe_maps_a}.} 
  \label{alfe_maps}
\end{figure*}

\begin{figure*}
\centering
\hspace{1cm}
 \includegraphics[width=1.\textwidth]{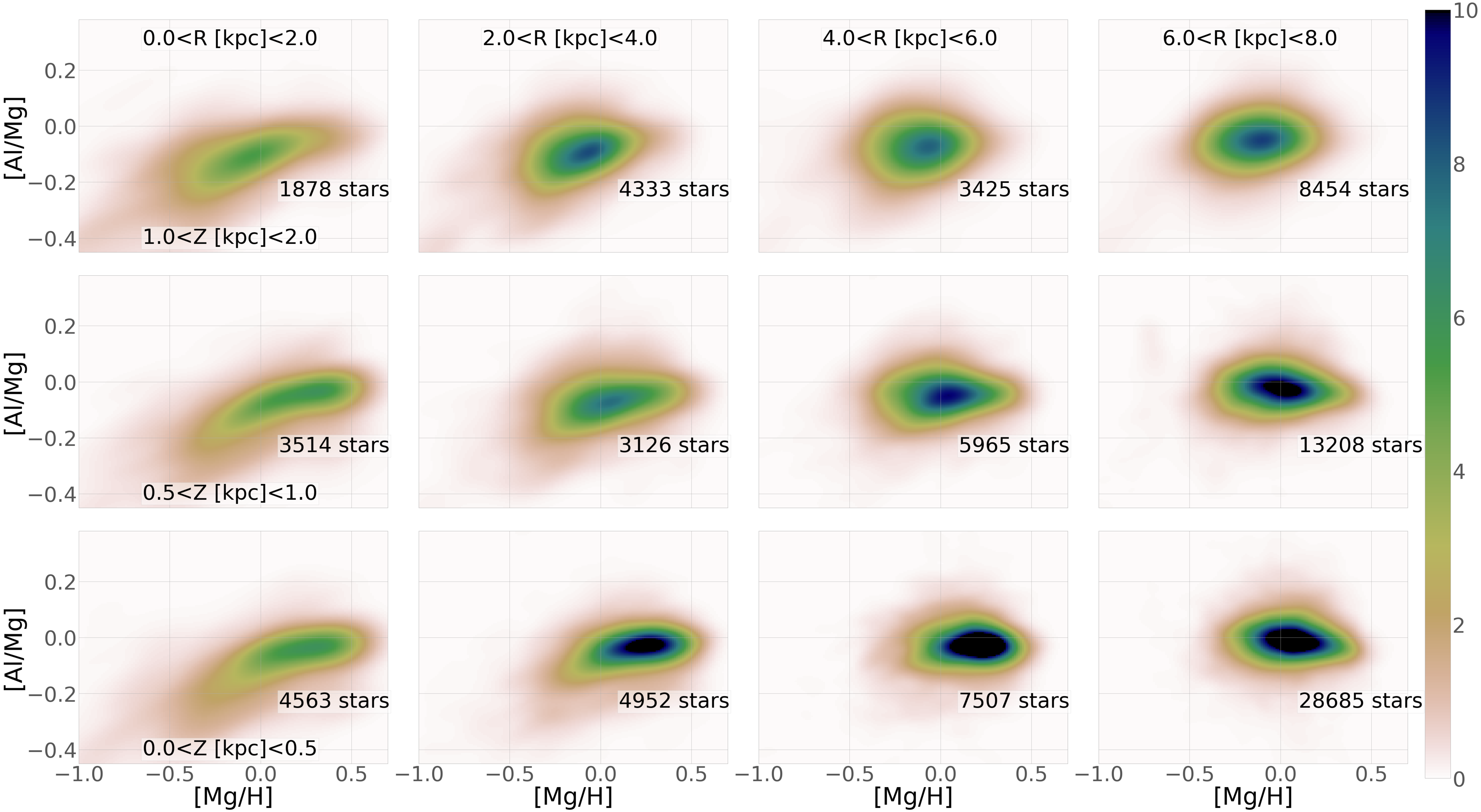} \includegraphics[width=1.\textwidth]{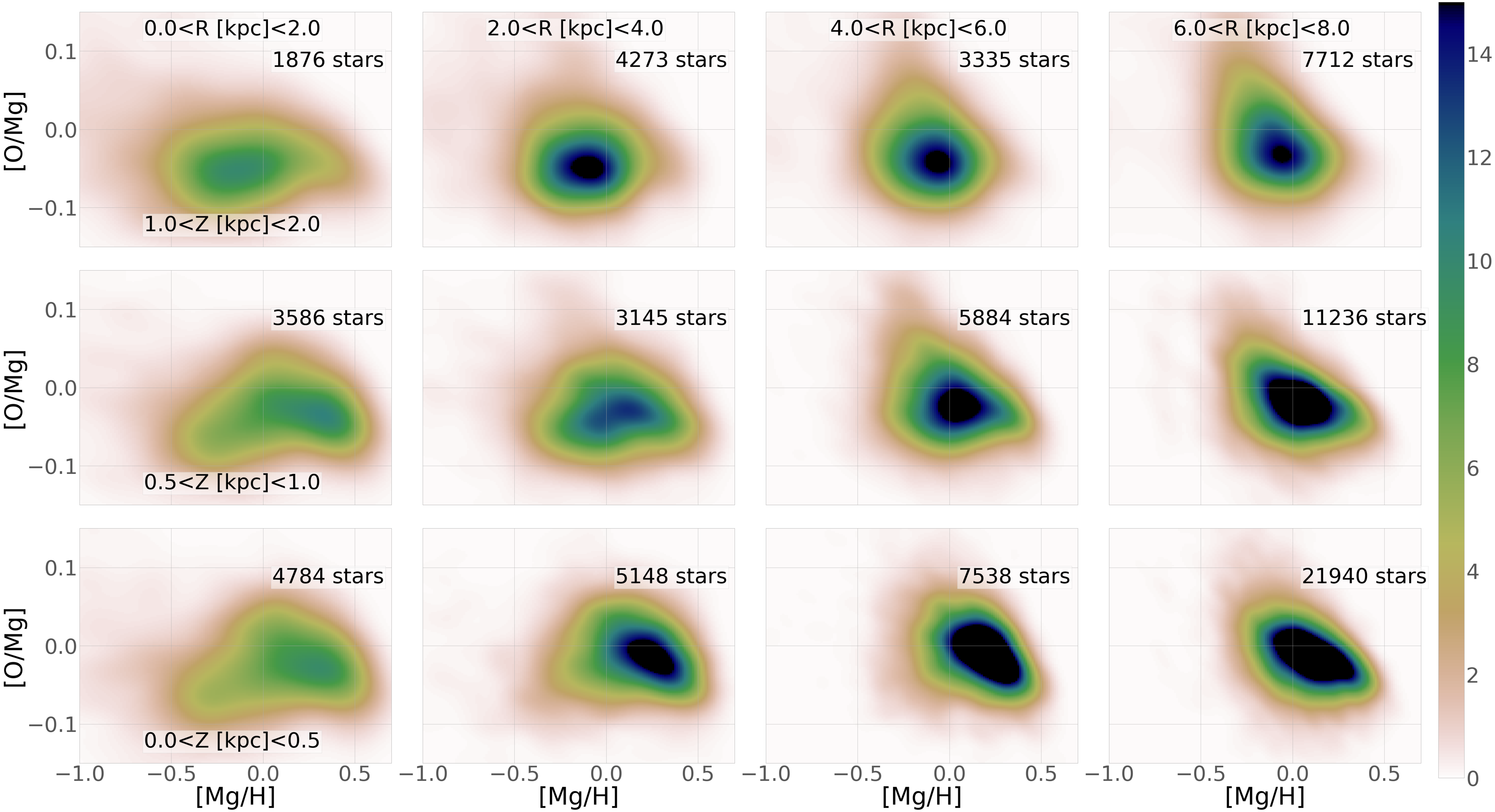}
 \caption{Top panel: APOGEE DR16 [Al/Mg] vs. [Mg/H] diagrams in bins of Galactocentric cylindrical coordinates, similar to Fig. \ref{alphafe_maps_a}, but only out to $R_{\rm Gal}=8$ kpc. Bottom panel: the same for [O/Mg] vs. [Mg/H].} 
  \label{almgo_maps}
\end{figure*}

To illustrate the impact of the APOGEE data released with the Sixteenth SDSS Data Release on the field of Galactic Archaeology, we focus on just a few examples of abundance-ratio maps in bins of Galactocentric cylindrical coordinates $(R_{\rm Gal}, Z_{\rm Gal})$, in a similar manner as the maps presented by \citet{Hayden2015} using DR12 data: a) the standard relative-to-iron abundance diagrams (Figs. \ref{alphafe_maps_a} and \ref{alphafe_maps_b} for [$\alpha$/Fe] vs. [Fe/H] and Fig. \ref{alfe_maps} for [Al/Fe] vs. [Fe/H]); and b) two examples of an abundance ratio as a function of an alpha-element ([Mg/O] vs. [Mg/H] and [Al/Mg] vs. [Mg/H], shown in Fig. \ref{almgo_maps}). These figures show, for different bins of $R_{Gal}$ and $Z_{Gal}$, diagrams of abundances colorcoded by density estimation using a Gaussian kernel. The bandwidths of the kernel density estimates obey Scott's rule \citep{Scott1992}. Figs. \ref{alphafe_maps_a} and \ref{alphafe_maps_b} also show, in the upper plots, the uncertainty distributions in distance and extinction for each $R_{\rm Gal}$ bin.

Due to the pencil-beam nature of the APOGEE survey, and the fact that metal-poor stars are brighter, the relative weight of the sub-populations in each plot may still be slightly affected by the selection function. Therefore, a quantitative interpretation of these spatial chemical maps needs to take into account such biases and will be the subject of future work. 
The so-called $\alpha$ elements are produced by core-collapse supernovae and hence more directly connected with the star formation rate. Recently, \citet{Weinberg2019} have discussed such abundance maps, but based on a much smaller sample of $\sim20,000$ stars from APOGEE DR14, and not including data in the innermost radial bin (0-2kpc), which is now possible.\\

\subsection{The [$\alpha$/Fe] vs. [Fe/H] map}

The [$\alpha$/Fe] vs. [Fe/H] diagram has long served as a tracer of the chemical enrichment timescales of the Milky Way \citep{Matteucci2012}, which are a consequence of the star formation history. A pioneer work to demonstrate the direct connection of the high-[$\alpha$/Fe] "plateau" with old stars was realised by \citet{Fuhrmann1998, Fuhrmann2011} who computed ages for a volume-complete sample of {\it Hipparcos} stars within 25 pc of the Sun. His work clearly showed the stars on the high-[$\alpha$/Fe] plateau to be older than 10 Gyr, whereas stars along the chemical thin-disc sequence were found to be younger. The observed chemical discontinuity in the [$\alpha$/Fe] vs. [Fe/H] diagrams in the solar neighbourhood has important consequences on interpretations related to the assembly history of the Milky Way and similar galaxies (see e.g. \citealt{Chiappini1997, Minchev2013, Mackereth2018a, Buck2019, Nuza2019, Spitoni2019} for discussions).

The mapping of the Milky Way in terms of the [$\alpha$/Fe] vs. [Fe/H] diagram has quickly evolved since then. The first high-resolution spectroscopic samples outside the solar vicinity were small and without age information (e.g. \citealt{Bensby2010, Bensby2011, Alves-Brito2010} - see Figure 14 of \citealt{Anders2014}), but were already able to show the complexity and impact of such maps. For instance, the disappearance of high-[$\alpha$/Fe] stars towards the outer disc could be interpreted as an indication that the (chemical) thick disc had a shorter scale length than the thin disc \citep{Bensby2011, Cheng2012}, contrary to what had been seen for the (geometrically defined) thick discs in other galaxies.

Extended maps, with a much better coverage along the Galactic mid-plane ($|Z_{\rm Gal}| < 0.5$ kpc) only appeared with APOGEE \citep{AllendePrieto2008, Majewski2017}, which already in its first year of data (with around only 20,000 stars of sufficient quality) was able to demonstrate that the chemical discontinuity observed by Fuhrmann was also present far outside the solar neighbourhood \citep{Anders2014, Nidever2014}, confirming also the short scale length of the chemical thick disc. These APOGEE results were complemented by other surveys at larger distances from the galactic mid-plane \citep[e.g.][and references therein]{Bovy2013,Mikolaitis2014}, but without such a good coverage of the inner Galaxy.

Shortly afterwards, \citet{Hayden2015} used a sample of around 70,000  red giants from APOGEE DR12 \citep{Alam2015} to increase the sampled volume with respect to the 2014 maps, covering a Galactocentric distance range between 3 and 15 kpc within 2 kpc of the Galactic plane. By that time it became clear that, towards the outer parts of the disc, one would see flaring, where the low-[$\alpha$/Fe] would dominate even at large heights above the Galactic mid-plane (see \citealt{Minchev2015, Minchev2019} for discussions), implying that the term "thick disc" should be used more carefully: the chemically-defined thick disc (by separating populations in the [$\alpha$/Fe vs. [Fe/H] diagram) is indeed confined to the inner regions, whereas the geometrically-defined thick disc (by a cut in $Z_{\rm Gal}$) is a mixture of flaring mono-age populations, and therefore would show an age gradient (see \citealt{Martig2016, Mackereth2017, Minchev2018}).

The \citet{Hayden2015} chemical-abundance maps were limited by the still poor coverage of the innermost parts of the Milky Way. That paper, along with following APOGEE publications (e.g. \citealt{Zasowski2019} based on DR14) tentatively reported that stars with $R_{\rm Gal}<$ 5kpc seem to lie on a single track, whereas at larger radii two distinct sequences were observed (an observation later interpreted as the fundamental dichotomy between the inner and outer discs by \citealt{Haywood2016, Haywood2018}). \citet{Recio-Blanco2017} using a sample of Gaia-ESO spectra report the existence of low-[$\alpha$/Fe] in the bulge area. With the larger APOGEE sample available from SDSS DR14 \citep{Abolfathi2018}, \citet{Rojas-Arriagada2019} selected stars within 3.5 kpc from the Galactic Center, and reported the detection of a  bimodal sequence in the [Mg/Fe]-versus-[Fe/H], confirming the Gaia-ESO results. The authors also suggested the two sequences to merge above [Fe/H]$\sim0.15$ dex into a single sequence (see \citealt*{Barbuy2018} for a review of other chemical-abundance studies of the Galactic Bulge previous to APOGEE DR14 and {\it Gaia} DR2).

Figures \ref{alphafe_maps_a} and \ref{alphafe_maps_b} present our updated [$\alpha$/Fe] vs. [Fe/H] diagrams in 2kpc-wide bins in $R_{\rm Gal}$ and three narrow bins in $|Z_{\rm Gal}|$, obtained from APOGEE DR16 in combination with {\it Gaia} DR2 and our {\tt StarHorse} distances. These abundance-ratio maps now extend from $R_{\rm Gal}=0$ out to 20 kpc, with excellent statistics (more than $150$ stars per bin) out to $R_{\rm Gal}=18$ kpc, where the target density drops dramatically. To avoid too crowded figures, we divided the chemical-abundance maps into two plots: Fig. \ref{alphafe_maps_a} shows the [$\alpha$/Fe] vs. [Fe/H] diagrams for the inner-disc bins, while Fig. \ref{alphafe_maps_b} shows the outer-disc bins. The distance and extinction uncertainties in each of the radial bins are shown in the top row of the two figures. They show that even in the innermost 2 kpc, {\tt StarHorse} achieves precision of around 10\% in distance and better than 0.1 mag in A$_V$ for most of the targets (the unfortunately less precise extinction estimates in regions closer to the solar position is due to our imposed bright limit for the Pan-STARRS1 photometry).

While the DR16 [$\alpha$/Fe] vs. [Fe/H] diagrams shown in Figs. \ref{alphafe_maps_a} and \ref{alphafe_maps_b} confirm most of the previous analyses, they also show some clear and important differences. 
Figure~\ref{alphafe_maps_a} now shows a much more complete view of the chemical-abundance distribution in the inner disc. Each of the innermost bins ($R_{\rm Gal}<4$ kpc) contains more than 1000 stars now, and especially very close to the Galactic mid-plane these numbers amount to $>5000$ (see the two leftmost bottom panels), potentially allowing also for analyses of azimuthal abundance variations. 

The bimodality reported by \citet{Rojas-Arriagada2019} is clearly confirmed in this improved map: we observe this bimodality in all [$\alpha$/Fe]-[Fe/H] diagrams in the innermost regions ($R_{\rm Gal}<4$ kpc), but especially for stars closest to the Galactic plane ($|Z_{\rm Gal}|<0.5$ kpc). The single sequence reported in \citet{Hayden2015} and \citet{Zasowski2019} for the innermost regions is not confirmed now, as the bins at lower $|Z_{\rm Gal}|$ contain more data. 

In contrast to \citet{Rojas-Arriagada2019}, however, the two blobs define completely detached sequences, without merging, thus showing a true chemical discontinuity. The new maps show that the chemical discontinuity seen in the solar neighborhood bin (mostly studied also by other surveys - middle row, 4th column), extends towards the Bulge, and become completely separated, being very similar to what was found by \citet{Fuhrmann1998, Fuhrmann2011} within 25 pc but now extended to larger metallicities (as expected given the observed abundance gradients in the galactic disc). The more detailed implications of these maps for chemo-dynamical models or the Milky Way will be discussed in future papers. We also caution that this chemical discontinuity is not seen in smaller samples of bulge stars (e.g. \citealt{daSilveira2018}). Biases in small samples, as well as large distance uncertainties, may contribute to the appearance/disappearance of the chemical discontinuity in the bulge. It is difficult, however, to invoke a bias in the APOGEE inner-Galaxy sample (comprised of many thousands of stars) that would artificially increase the chemical discontinuity.

Figure \ref{alphafe_maps_b} shows the [$\alpha$/Fe] vs. [Fe/H] plane for the outermost bins in $R_{\rm Gal}$ (the bin 10-12 kpc is repeated from previous figure because the color scale is slightly different from Fig.\ref{alphafe_maps_a}. Again we observe that for the more distant stars, the addition of the PanSTARRS-1 photometry improves the extinction estimates (compare uncertainty distributions in the top row of the figure). The diagram also clearly confirms the almost total disappearance of the high-[$\alpha$/Fe] sequence around $\sim 14$ kpc. Because the number of stars is small in the very outer disc, the noise in those plots increases, giving more visual weight to outliers.

Finally, we note two other important characteristics of the new maps presented both in Fig.~ \ref{alphafe_maps_a} and Fig.~ \ref{alphafe_maps_b}, when focusing on stars near to the Galactic mid-plane ($|Z_{\rm Gal}|< 0.5$ kpc). 

Firstly, the [$\alpha$/Fe] centroid of the low-[$\alpha$/Fe] distribution gradually shifts to larger values with increasing Galactocentric distance (especially visible in Fig.~\ref{alphafe_maps_b}), corresponding to a positive radial [$\alpha$/Fe] gradient, continuing the trend observed at larger Galactocentric distances \citep{Anders2014, Hayden2014}. 

Secondly, in the innermost bin ($R_{\rm Gal}<$ 2 kpc, and $|Z_{\rm Gal}|<$ 0.5 kpc) the [$\alpha$/Fe] trend for the more metal-rich, low-[$\alpha$/Fe] population ($\sim -0.2 <$ [Fe/H] $<$ 0.5) is linearly decreasing, without any flattening at larger metallicities. This is in agreement with optical studies of the bulge (\citealt{Friaca2017, daSilveira2018} -- see also \citealt*{Barbuy2018}), but remains in stark contrast to what is observed at larger Galactocentric distances (see radial bins from $6<R_{\rm Gal}<12$ kpc, in the same row -- $|Z_{\rm Gal}|<$ 0.5 kpc), where the cloud of data bends, showing a flattening of the abundance-ratio trend beyond solar metallicities. The reason for this bending is the migration of old-metal rich stars from the innermost bins towards the outer regions, populating mostly the $8-12$ kpc bins.

Indeed, the high-metallicity thin disk stars in the outer regions are known to be migrated stars from the inner disc (e.g. \citealt{Grenon1989, Casagrande2011, Anders2017}). For example, according to the chemodynamical model of \citet{Minchev2013,Minchev2014}, the mixture of migrating stars from other Galactocentric distances changes as one moves from the inner to the outer disc, and even in the $8-12$ kpc range there is a large number of migrators from the innermost disc regions. It is around the solar vicinity that a large number of old inner disk stars can be found according to the predictions of \citet{Minchev2014}. This can also be clearly seen in \citet[][their Figure 1]{Anders2017a}.

The larger statistics of the current maps, especially near the Galactic mid-plane, do not support the dichotomy between the inner and outer discs advertised by \citet{Haywood2019}. It suggests instead an inside-out formation of the thin disk, a continuous variation in the chemical properties from the innermost regions towards the outer parts, and significant radial migration (e.g. \citealt{Frankel2018}).

At larger $|Z_{\rm Gal}|$ bins and in the outer disc, the combined effects of radial migration and disc flaring make interpretations more complex, and the multi-element abundance maps available from APOGEE offer a unique opportunity to finally quantify all these processes (see e.g. \citealt{Frankel2018, Frankel2020} for first attempts on constraining radial migration efficiency using APOGEE red-clump giants with statistical age estimates). In the innermost bins, going from low to large $|Z_{\rm Gal}|$ one also sees a smooth transition from a thin disk like component to an old (i.e. [$\alpha$/Fe]-enhanced) thick disk like (or spheroidal) component.

Detailed future investigations should use forward simulations to properly take into account selection effects \citep[see e.g.][for discussions]{Miranda2014, Anders2016, Nandakumar2017, Fragkoudi2018, Frankel2019}. Moreover, the addition of age and kinematical information is also necessary to be able to disentangle these several factors playing a role in these maps namely: radial migration, population mixture, flaring, and details of the nucleosynthetic yields. An illustrative example is provided by the birth-radius estimation technique proposed by \citet{Minchev2018}.

\subsection{The [Al/Fe] vs. [Fe/H] diagram}

Figure \ref{alfe_maps} shows the same type of plot as Fig. \ref{alphafe_maps_a}, but now for the [Al/Fe] abundance ratio instead of [$\alpha$/Fe]. As an additional constraint, we only include stars with well-determined ASPCAP Al abundances ({\tt AL\_FE\_FLAG}=0) in this plot. The maps are similar to the ones in Fig. \ref{alphafe_maps_a}, indicating that overall, Al (being an odd-Z element) behaves like an $\alpha$ element at disc-like metallicities (also previously shown to be the case in the Bulge - for instance see discussion in \citealt{McWilliam2016}). The important difference of Fig. \ref{alfe_maps} with respect to the corresponding Fig. \ref{alphafe_maps_a} is the almost complete absence of the bimodality in the abundance plane for Galactocentric distances $R_{\rm Gal}>2$ kpc.

However, the [$\alpha$/Fe] vs. [Fe/H] discontinuity seen in the very inner regions discussed above is also seen in the [Al/Fe] vs. [Fe/H] diagram: in the $R_{\rm Gal}<$ 2 kpc bin close to the Galactic plane we see essentially two detached [Al/Fe] sequences. This fact provides further evidence for the reality of the chemical discontinuity seen in the heart of the Galactic bulge.

The difference between [$\alpha$/Fe] and [Al/Fe] for the most metal-poor stars is that, whereas the [$\alpha$/Fe] seems to continue raising towards lower metallicities, the [Al/Fe] starts to bend down. This is a consequence of the metallicity-dependent Al yields in massive stars.

\subsection{The [Al/Mg] vs. [Mg/H] and [Mg/O] vs. [Mg/H] diagrams}

As an example illustrating the wealth of new chemical-abundance information contained in DR16, we now discuss the behavior of the ratios between two $\alpha$-like elements, not using iron but magnesium as a reference element. Because Mg is mainly a product of core collapse supernovae, its increase with time follows the star formation rate more closely than iron, which can keep increasing even if the star formation stops due to the contribution of type Ia supernovae released on longer timescales. From the observational side, magnesium is also a convenient element because the calibrated ASPCAP [Mg/H] abundances show small dispersions, very small trends with effective temperature, and they follow the expected trends in the abundance diagrams.

Fig.\ref{almgo_maps} shows both an [Al/Mg] vs. [Mg/H] and an [O/Mg] vs. [Mg/H] map of the Galaxy, focusing on the inner disc and bulge region ($R_{\rm Gal}<8$ kpc). In both plots, we again only plot stars with high-quality DR16 ASPCAP abundances, by requiring the corresponding abundance flag entries ({\tt MG\_FE\_FLAG} and, respectively, {\tt AL\_FE\_FLAG} and {\tt O\_FE\_FLAG}) to equal zero.

The main point of Fig. \ref{almgo_maps} is to showcase the vast amount of new high-quality APOGEE data, especially for the inner disc. To appreciate the increase of the sample with respect to DR14, Fig. \ref{almgo_maps} should be compared to Figs. 4 and 5 of \citet{Weinberg2019}, which was based on a small sample of 20,000 stars with only slightly stricter quality requirements (3700 K $<T_{\rm eff}<4600$ K, ${\tt SNREV}>80$, no “ASPCAP bad” flags, {\tt EXTRATARG}=0). The new data clearly allow us to study the very heart of our Galaxy in much more detail, even when the same quality cuts are applied.

The main isotopes of both O and Mg are produced during the hydrostatic phases of high-mass stars. This ratio in then mostly sensitive to details of related to the stellar yields (mass loss and rotation in the case of oxygen, and convection treatment in the case of Mg), but is expected to remain close to solar \citep{Woosley2002, Sukhbold2016, Groh2019}.

Two things can be noted in the [O/Mg] vs. [Mg/H] diagrams in the inner Galaxy (Fig. \ref{almgo_maps}), namely: a systematic slight increase of the [O/Mg] median value from the innermost regions towards the solar neighborhood for stars in the upper row (1 kpc $<|Z_{\rm Gal}|<$ 2 kpc); and b) a less pronounced presence of the low-[Mg/H] low-[O/Mg] population towards the mid-plane that remains visible only in the innermost bin.

In order to understand if this is due to O or Mg, we next check the [Al/Mg] diagrams (bottom panel of Fig.\ref{almgo_maps}). Similarly, the median [Al/Mg] ratio in the top row (1 kpc $<|Z_{\rm Gal}|<$ 2 kpc) increases with Galactocentric distance, reaching the solar value at the solar ring. Moreover, [Al/Mg] also increases with metallicity in the smallest Galactocentric distance bins. 

Taking both results at face value (without considering further biases that could be affecting proportions of stars in the different loci of these diagrams), the results suggest that there is an increase of Mg towards larger metallicities (or a relative decrease of both O and Al - e.g. \citealt{Groh2019}). \\

\begin{figure*}
\centering
 \includegraphics[width=1.\textwidth]{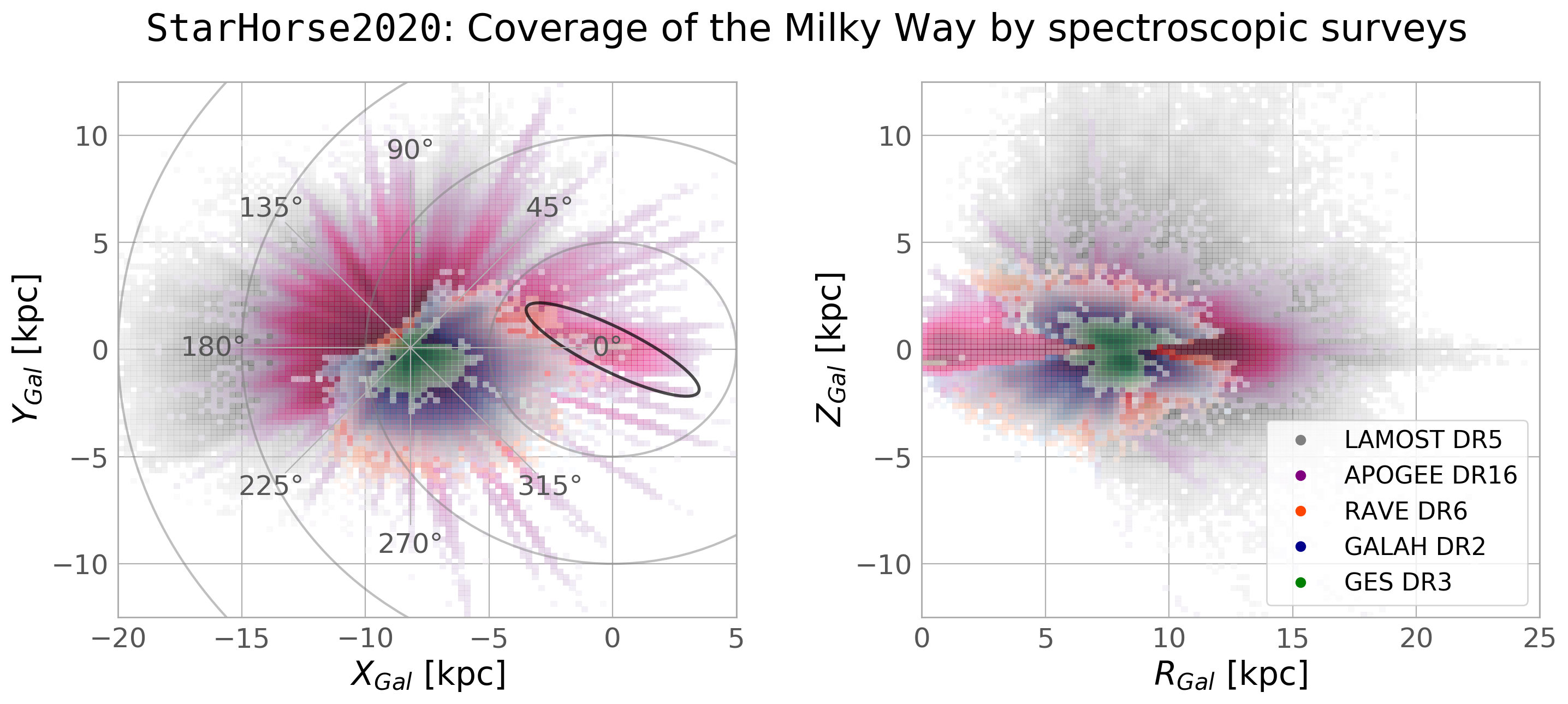}  \caption{Survey coverage of the catalogues presented in this paper in Galactocentric coordinates. In both panels, the colours encode the different surveys (grey: LAMOST DR5, magenta: APOGEE DR16, red: RAVE DR6, blue: GALAH DR2, and green: GES DR3) as well as the relative density of observed stars (bins with less than five stars are left blank). To guide the eye, grey circles are placed in multiples of 5 kpc around the Galactic Centre, the expected location of the Galactic bar (e.g. \citealt{Bland-Hawthorn2016}) is marked by the black ellipse, and a heliocentric Galactic longitude frame is overplotted. Left panel: Cartesian $XY$ coordinates. Right panel: Cylindrical $RZ$ coordinates.}
  \label{survey_maps}
\end{figure*}

\section{{\tt StarHorse} results for other publicly released spectroscopic surveys}
\label{cats}

In this paper we also provide distances and extinctions for different spectroscopic surveys, namely for: GALAH DR2 \citep{Buder2018}, LAMOST DR5 \citep{Xiang2019}, RAVE DR6 \citep{Steinmetz2020}, and GES DR3 \citep{Gilmore2012}. Here we again used {\it Gaia} DR2 parallaxes \citep{GaiaCollaboration2018}. Moreover, we have also included photometry from APASS \citep{Henden2014}, not included in the APOGEE run. Also, since none of these surveys extend to the very extincted regions, we have  used Gaia DR2 photometry in this case.

{\it Gaia} contains three passbands $G$, $G_{BP}$ and $G_{RP}$ in the respective wavelengths: 320-1050 nm, 320-680 nm, 610-1070 nm \citep{GaiaCollaboration2016a, Weiler2018}. Even though this photometry is very precise, there are some discrepancies between observations and the sensitivity curves published. To correct for this effect, we followed the recommendations of \citet{MaizApellaniz2018}; these are the same corrections as applied in \citet[][see their Table 1]{Anders2019}.

We have computed distances and extinctions, in the same way as for APOGEE DR16, for which we present catalogues in the same format as before (Table \ref{table:outtablefmt}). Fig. \ref{survey_maps} shows the resulting spatial coverage of the surveys analysed here, and Fig. \ref{survey_dists} shows the corresponding distance and distance uncertainty distributions. In addition, in Appendix \ref{othervacs} we provide summary plots similar to Figs. \ref{fig:maps}, \ref{fig:kielcmd}, and \ref{fig:output_summary} demonstrating the sky coverage and the quality of the results for each of the surveys. In the following subsections we describe the assumptions made in each of these catalogues.

\begin{figure*}
\sidecaption
\centering
 \includegraphics[width=.66\textwidth]{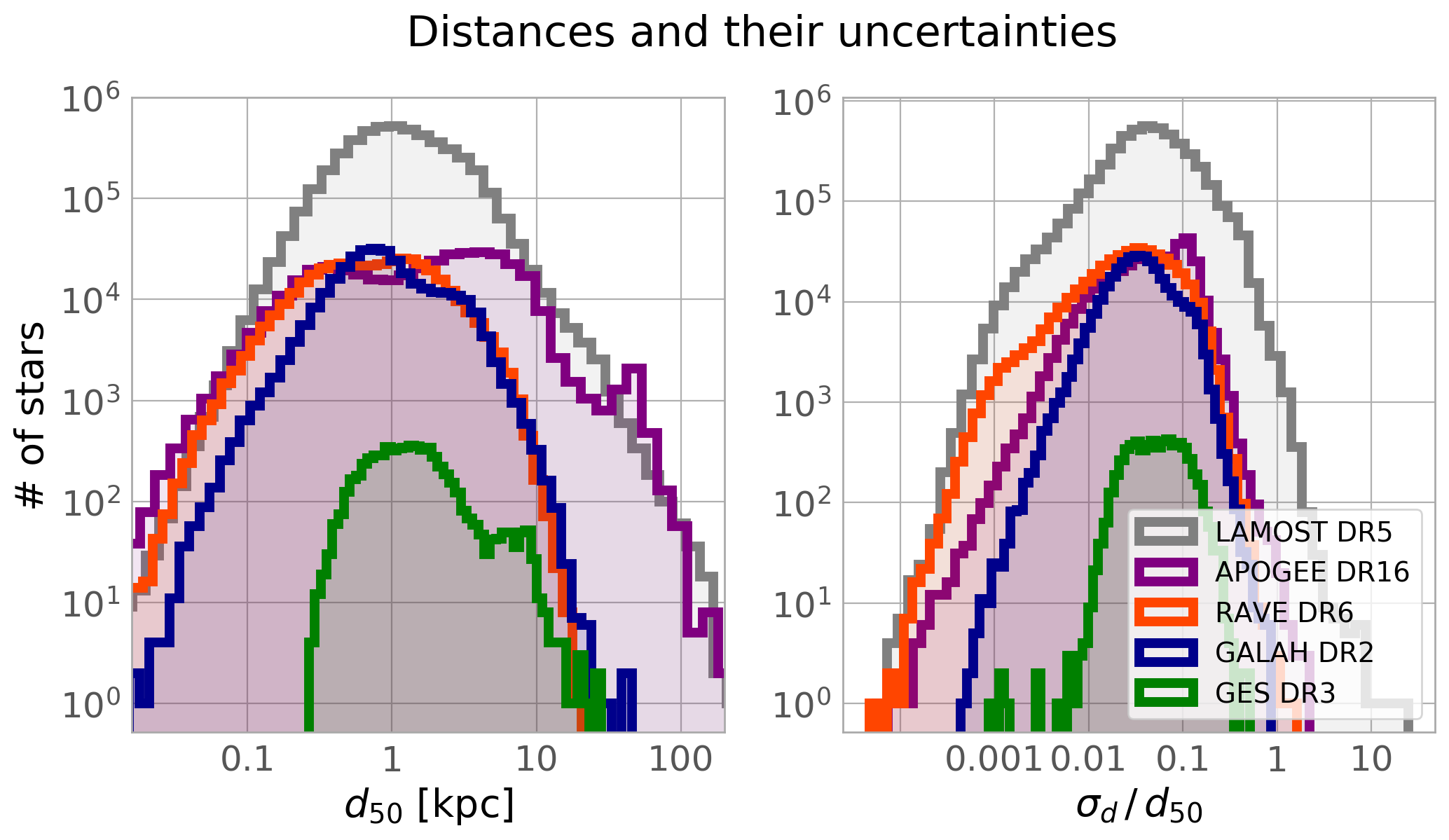}  \caption{ Distribution of posterior distances (left) and their corresponding relative uncertainties (right), for the catalogues presented in this paper. In both panels, the axes are logarithmic and the colours are the same as in Fig. \ref{survey_maps} (grey: LAMOST DR5, magenta: APOGEE DR16, red: RAVE DR6, blue: GALAH DR2, and green: GES DR3).}
  \label{survey_dists}
\end{figure*}

\subsection{GALAH DR2}\label{ingalah}

The Galactic Archaeology with HERMES \citep[GALAH,][]{deSilva2015, Martell2017} is a large spectroscopic survey that aims to identify stellar groups that were born together, by searching for similarity on the chemical patterns of the stars. Therefore GALAH spectra were obtained with the high resolution and multi-band spectrograph HERMES \citep{Barden2010}, which is capable to deliver abundances for up to 23 chemical elements. Its latest data release, GALAH DR2 release in April 2018, contains radial velocities, atmospheric parameters and abundances for a total of 342,682 unique stars \citep{Buder2018}.

GALAH maps all stellar populations between magnitudes $(12<V<14)$ and avoiding the Galactic plane $|b| > 10$. In Q18 we computed distances and extinctions using the GALAH DR1 parameters combined with {\it Gaia} DR1. Now we have available much more data both in GALAH DR2 and {\it Gaia} DR2. We here follow the same procedure as in Q18 to run this latest public GALAH data. The atmospheric parameters were treated as they come in the catalogue. We again use the \citep{Salaris1993} correction for stars that have [$\alpha$/Fe] (see Section \ref{inapogee}). For those without a reported [$\alpha$/Fe] ratio, we assumed [M/H]$=$[Fe/H].

We chose to run GALAH with APASS photometry since its faint limits are still too bright to be able to use PanSTARRS-1 (due to saturation). We also run StarHorse with parallax True mode (see Q18 section 3.2.1), since more than 90\% of the catalogue contains parallaxes uncertainties better then 20\%. From the input catalogue a total of 324,999 stars converged (94\%) with solutions of distances, extinctions and astrophysical parameters that can be downloaded via the CDS.

\subsection{LAMOST DR5 DD-Payne VAC}\label{inlamost}

The Large Sky Area Multi-Object Fiber Spectroscopic Telescope \citep[LAMOST,][]{Cui2012,Zhao2012} is one of the largest scale spectroscopic surveys, and the first large astronomical device in China. It has been collecting data since 2012, and now after about 8 years the survey has released 9 million spectra in the  wavelength range of 3690-9100 \AA  and spectral resolution of $R \sim 1800$. These 9 million spectra contain stars, galaxies, quasars, as well as non-classified sources.

We have adopted the recently published DR5 DD-Payne VAC\footnote{\url{http://dr5.lamost.org/doc/vac}} \citep{Xiang2019} catalogue. This catalague contains stellar parameters and individual elemental abundances for 6 million LAMOST DR5 stars, obtained with a data-driven approach incorporating constraints from theoretical spectra and trained on GALAH DR2 and APOGEE DR14 results.

From this catalogue we only select stars with stellar parameters with uncertainties in gravity, surface temperature, metallicity and [$\alpha$/Fe] ratios smaller than $\sigma_{\log g}\ <\ 1$ dex, $\sigma_{T_{\rm eff}} < 800$ K, $\sigma_{[{\rm Fe/H}]}\ <\ 1.0$ dex, and $\sigma_{[\alpha/{\rm Fe}]} \ <\ 1.0$, respectively. The goal was to avoid stars with too large uncertainties, and save computing time.

For LAMOST DR5 we have combined the spectra again with {\it Gaia} parallaxes and photometry. We complemented the input data with  photometry from PanSTARRS1, 2MASS and WISE. We also run LAMOST with parallax true mode since most parallaxes in LAMOST also have uncertainties better then 20\%. From 5,651,710 sources with available parallaxes {\tt StarHorse} converged for 4,928,715 stars (87\%). One of the reasons for a smaller convergence in the case of LAMOST is the fact that we use a thicker spaced PARSEC model grid (0.05 Gyr in age and 0.05 dex in [M/H]). The solutions of distances, extinctions and astrophysical parameters can be downloaded via the CDS.

\subsection{RAVE DR6}\label{inrave}

RAVE spectra were obtained with the multi-object spectrograph deployed on $1.2-m$ UK Schmidt Telescope of the Australian Astronomical Observatory (AAO). The spectra have a medium resolution of $(R \sim 7.500)$ and cover the CaII-triplet region (8410-8795\AA). We use the final RAVE data release,  DR6 \citep{Steinmetz2020}, and in particular, the purely spectroscopically derived stellar atmospheric parameters subscripted {\tt cal\_madera} \citep{Steinmetz2020a}. The uncertainties that we use are, in general, the maximum between the calibrated and not calibrated parameters given in the catalogue or a fiducial maximum. These corrections are very similar to the ones applied to run RAVE DR5 combined with {\it Gaia} DR1 in Q18. We then combined RAVE DR6 with {\it Gaia} DR2 parallaxes and the photometric data used in this case is the same as for LAMOST. We configured {\tt StarHorse} to use the {\tt parallax=true} option and the same coarser isochrone grid we used for LAMOST, since the uncertainties of these surveys are larger. From the input catalogue of DR6 (488,233 unique objects), 408,894 stars converged, and we make their derived astrophysical stellar parameters avaible here. Due to the significantly smaller formal uncertainties of the DR6 MADERA stellar parameters compared to DR5, the number of stars for which {\tt StarHorse} converged is slightly smaller than for DR5.

\subsection{Gaia-ESO survey DR3}\label{inges}

The Gaia-ESO survey \citep[GES][]{Gilmore2012} is a large public spectroscopic survey with high resolution that covers all Milky Way components and open star clusters of all ages and masses. The final GES release is expected to include about $10^5$ stars. We downloaded the Gaia-ESO data release 3 (DR3) from the ESO catalogue facility. This catalogue contains a total of 25533 stars, including the Milky-Way field, open clusters, and calibration stars. We select only the stars in the Milky Way field to produce our {\tt StarHorse}, which is about 7870 stars. In this case we also made a quality criteria cut, namely: $\sigma_{\log g}\ <\ 0.4$ dex, $\sigma_{T_{\rm eff}}/T_{\rm eff} < 0.05$ K, $\sigma_{[{\rm Fe/H}]}\ <\ 0.2$ dex. The final catalogue used as {\tt StarHorse} input contains then 6316 stars, The complementary photometric data used in this case is the same as for LAMOST. We then run the code again with parallax True mode, and {\tt StarHorse} converged for 6,095 stars. The {\tt StarHorse} astrophysical parameters for the GES DR3 stars are also  at the CDS.

\section {Conclusions}
\label{conclusion}

With this paper we present a set of value-added catalogues derived from the stellar spectroscopic surveys APOGEE, GALAH, LAMOST, RAVE, and GES. In particular our APOGEE DR16 VAC, released as part of SDSS DR16 \citep{Ahumada2020}, was produced by running the {\tt StarHorse} code, described in detail by Q18, on the DR16 ASPCAP catalog matched to {\it Gaia} DR2, and with the addition of photometry from PanSTARRS-1, 2MASS, and AllWISE. This VAC contains distance and extinction estimates for 388,815 unique stars out of a total of 437,485 unique objects contained in the DR16 catalogue.
Our code was validated extensively in \citet{Santiago2016, Queiroz2018}, and \citet{Anders2019}. In Appendix \ref{sec:validation} we provide some additional tests showing that the newly derived parameters for APOGEE DR16 generally compare well to results obtained from asteroseismology, open clusters, and other spectroscopic surveys. There is evidence for slightly overestimated extinctions for our APOGEE DR16 VAC, which we attribute in part to the missing reliable optical photometry for most of this sample, and in part to an offset in the ASPCAP temperature scale, especially outside the recommended calibration regime.

 In  Appendix \ref{sec:validation} we also show that our distances are less biased towards the inner Galactic disc than the neural-network based distances of \citet{Leung2019} (see Fig.~\ref{fig:astroNN}).
The typical uncertainties for the APOGEE DR16 sample are of the order of $\simeq 10\%$ in distance and of $0.16$ mag in $A_V$. A clearly bimodal distribution of extinction uncertainties is observed, with the peak at $\sigma_{A_V} \simeq 0.06$ found for stars with available optical magnitudes from PanSTARRS-1, while the peak at larger $\sigma_{A_V}$ is made by stars with no such measurements. The typical distance uncertainties are also different for dwarfs ($\simeq 2\%$) and giants ($\simeq 5\%$).
The scientifc results from the first analysis of the {\tt StarHorse} APOGEE DR16 catalogue can be summarized as:
\begin{itemize}
  \item Using the {\tt StarHorse} VAC we have demonstrated that the APOGEE DR16 sample represents a major leap in terms of coverage of the Galactic disc with high-resolution spectra. The density of APOGEE targets exceeds a dozen stars per kpc$^2$ everywhere in the $R_{\rm Gal}-Z_{\rm Gal}$ plane for $0<R_{\rm Gal}<18$ kpc and $-3\ {\rm kpc}<Z_{\rm Gal}<3$ kpc, allowing for an unprecedented chemo-kinematic mapping of the inner as well as outer stellar disc, with significant azimuthal coverage.
  \item From the improved APOGEE coverage and {\tt StarHorse} distances we can see, in the density maps projected in XY Galactocentric coordinates, a bar signature as well as in A19 but with a smaller angle with respect to the Galactic plane, more consistent with previous studies about the Galactic bar structure \citep{Bland-Hawthorn2016}.
  \item the extended chemical-abundance maps in Fig. \ref{alphafe_maps_a} confirms, for the first time with good statistics of thousands of stars, a chemical bimodality in the very inner Galaxy $0<R_{\rm Gal}<2$ kpc and $0<|Z_{\rm Gal}|<1$ kpc. Which is different from previous analyses that reported a single sequence \citep{Hayden2015, Zasowski2019}, but with much less populated samples. 
  \item The two groups visible in the [$\alpha$/Fe]-[Fe/H] plane in the innermost bin define completely detached sequences, implying a true chemical discontinuity. The larger statistics of the current maps, especially near the  Galactic mid-plane, do not support the dichotomy between the inner and outer discs. On the other hand, it suggests the chemical discontinuity to be a clear property of the global chemical-enrichment history of the Milky Way.
  \item The chemical maps of [$\alpha$/Fe]-[Fe/H] extend to the very outer disk, $R_{\rm Gal}>20$ kpc, and also show the complete disappearance of a high-alpha population further than $R_{\rm Gal}>14$ kpc. This confirms the shorter scale length of the Galactic thick disk concerning the Galactic thin disk, following previous studies \citep{Cheng2012,Anders2014}.
  \item There is an indication for a positive radial $[\alpha/Fe]$ gradient, observed from the fact that the $[\alpha/Fe]$ centroid of the $\alpha$-poor sequence in the inner Galaxy gradually shifts to larger values with increasing Galactocentric radius observed in Figures  \ref{alphafe_maps_a} and \ref{alphafe_maps_b} continuing the trend reported by \citet{Anders2014, Hayden2014}.
 \item The maps of $[\alpha/Fe]$ show some evidence for radial migration of old-metal rich stars from the inner Galaxy to the Outer Galaxy, this is seen by the flattening of the abundance-ratio trend beyond the solar metallicities.
\item The chemical duality in the inner bins is also confirmed in maps using aluminum and iron, $[Al/Fe]-[Fe/H]$. This is not seen for larger Galactocentric distances, where the disk chemical bimodality disappears in this abundance regime. This indicates a strong chemical duality in the inner Galaxy. Those diagrams also show metallicity-dependent Al yields in massive stars, with $[Al/Fe]$ starting to bend down towards lower metallicities.
\item The resulting maps using $\alpha$-elements and magnesium as a reference instead of iron, show an increase of Mg with respect to Galactocentric distance. Since Mg follows the star formation more closely than iron, this suggests an inside out formation.
\end{itemize}
The data produced here and made publicly available allow for much more sophisticated chemical-abundance studies over much larger disc volumes than previous data releases. New studies gathering also kinematic information will enable unprecedented constraints for chemodynamical models of the Milky Way, espetially in the inner-most and outer-most Galaxy.

All the newly produced {\tt StarHorse} catalogues are available for download from \url{https://data.aip.de/aqueiroz2020} please use the DOI to quote the data: $doi:10.17876/data/2020\_2$.

\begin{acknowledgements}

The {\tt StarHorse} code is written in python 3.6 and makes use of several community-developed python packages, among them {\tt astropy} \citep{AstropyCollaboration2013}, {\tt ezpadova}\footnote{\url{https://github.com/mfouesneau/ezpadova}}, {\tt numpy} and {\tt scipy} \citep{Virtanen2019}, and {\tt matplotlib} \citep{Hunter2007}. The code also makes use of the photometric filter database of VOSA \citep{Bayo2008}, developed under the Spanish Virtual Observatory project supported from the Spanish MICINN through grant AyA2011-24052.\\

Funding for the SDSS Brazilian Participation Group has been provided by the Minist\'erio de Ci\^encia e Tecnologia (MCT), Funda\c{c}\~ao Carlos Chagas Filho de Amparo \`a Pesquisa do Estado do Rio de Janeiro (FAPERJ), Conselho Nacional de Desenvolvimento Cient\'{\i}fico e Tecnol\'ogico (CNPq), and Financiadora de Estudos e Projetos (FINEP).\\

Funding for the Sloan Digital Sky Survey IV has been provided by the Alfred P. Sloan Foundation, the U.S. Department of Energy Office of Science, and the Participating Institutions. SDSS-IV acknowledges support and resources from the Center for High-Performance Computing at the University of Utah. The SDSS web site is \url{www.sdss.org}.\\

SDSS-IV is managed by the Astrophysical Research Consortium for the Participating Institutions of the SDSS Collaboration including the Brazilian Participation Group, the Carnegie Institution for Science, Carnegie Mellon University, the Chilean Participation Group, the French Participation Group, Harvard-Smithsonian Center for Astrophysics, Instituto de Astrof\'isica de Canarias, The Johns Hopkins University, 
Kavli Institute for the Physics and Mathematics of the Universe (IPMU) / University of Tokyo, Lawrence Berkeley National Laboratory, 
Leibniz-Institut f\"ur Astrophysik Potsdam (AIP),  
Max-Planck-Institut f\"ur Astronomie (MPIA Heidelberg), 
Max-Planck-Institut f\"ur Astrophysik (MPA Garching), 
Max-Planck-Institut f\"ur Extraterrestrische Physik (MPE), 
National Astronomical Observatory of China, New Mexico State University, 
New York University, University of Notre Dame, 
Observat\'ario Nacional / MCTI, The Ohio State University, 
Pennsylvania State University, Shanghai Astronomical Observatory, 
United Kingdom Participation Group,
Universidad Nacional Aut\'onoma de M\'exico, University of Arizona, 
University of Colorado Boulder, University of Oxford, University of Portsmouth, 
University of Utah, University of Virginia, University of Washington, University of Wisconsin, 
Vanderbilt University, and Yale University.\\

Guoshoujing Telescope (the Large Sky Area Multi-Object Fiber Spectroscopic Telescope LAMOST) is a National Major Scientific Project built by the Chinese Academy of Sciences. Funding for the project has been provided by the National Development and Reform Commission. LAMOST is operated and managed by the National Astronomical Observatories, Chinese Academy of Sciences.\\

Funding for RAVE has been provided by: the Australian Astronomical Observatory; the Leibniz-Institut f\"ur Astrophysik Potsdam (AIP); the Australian National University; the Australian Research Council; the French National Research Agency; the German Research Foundation (SPP 1177 and SFB 881); the European Research Council (ERC-StG 240271 Galactica); the Istituto Nazionale di Astrofisica at Padova; The Johns Hopkins University; the National Science Foundation of the USA (AST-0908326); the W. M. Keck foundation; the Macquarie University; the Netherlands Research School for Astronomy; the Natural Sciences and Engineering Research Council of Canada; the Slovenian Research Agency; the Swiss National Science Foundation; the Science \& Technology Facilities Council of the UK; Opticon; Strasbourg Observatory; and the Universities of Groningen, Heidelberg and Sydney. The RAVE web site is at \url{https://www.rave-survey.org}.

This work has made use of data from the European Space Agency (ESA)
mission {\it Gaia} (\url{http://www.cosmos.esa.int/gaia}), processed by the {\it Gaia} Data Processing and Analysis Consortium (DPAC,
\url{http://www.cosmos.esa.int/web/gaia/dpac/consortium}). Funding
for the DPAC has been provided by national institutions, in particular the institutions participating in the {\it Gaia} Multilateral Agreement. \\

This work has also made use of data from {\it Gaia}-ESO based on data products from observations made with ESO Telescopes at the La Silla Paranal Observatory under programme ID 188.B-3002.\\

FA is grateful for funding from the European Union's Horizon 2020 research and innovation programme under the Marie Sk\l{}odowska-Curie grant agreement No. 800502 H2020-MSCA-IF-EF-2017. CC acknowledges support from DFG Grant CH1188/2-1 and from the ChETEC COST Action (CA16117), supported by COST (European Cooperation in Science and Technology).\\

DAGH acknowledges support from the State Research Agency (AEI) of the Spanish Ministry of Science, Innovation and Universities (MCIU) and the European Regional Development Fund (FEDER) under grant AYA2017-88254-P.
J.G.F-T is supported by FONDECYT No. 3180210.\\

AB thanks Cosmos.

\end{acknowledgements}


\bibliographystyle{aa}
\bibliography{StarHorse2020-print}

\newpage

\appendix

\newpage

\section{{\tt StarHorse} Data model}
\label{datamodel}

The tables in this appendix describe our data model for the APOGEE DR16 {\tt StarHorse} VAC (Table \ref{table:outtablefmt}), as well as the meaning of the human-readable flags {\tt SH\_INPUTFLAGS} and {\tt SH\_OUTPUTFLAGS} (Table \ref{table:flags}).

\begin{table*}[h!]
\caption{ Data model for the {\tt StarHorse} catalogues described in this paper}.
\centering
\begin{tabular}{c|c|c }
 Column & Description & Unit \\ 
 \hline
 \hline
 ID & Unique object identifier & string \\ 
 glon & Galactic longitude & deg \\
 glat & Galactic latitude & deg \\
 mass16 & 16th percentile of {\tt StarHorse} stellar mass PDF  & $M_\odot$ \\
 mass50 & 50th percentile of {\tt StarHorse} stellar mass PDF &  $M_\odot$ \\
 mass84 & 84th percentile of {\tt StarHorse} stellar mass PDF & $M_\odot$ \\
 teff16 & 16th percentile of {\tt StarHorse} effective temperature PDF & K \\
 teff50 & 50th percentile of {\tt StarHorse} effective temperature PDF & K \\
 teff84 & 84th percentile of {\tt StarHorse} effective temperature PDF & K \\
 logg16 & 16th percentile of {\tt StarHorse} surface gravity PDF & dex \\
 logg50 & 50th percentile of {\tt StarHorse} surface gravity PDF & dex \\
 logg84 & 84th percentile of {\tt StarHorse} surface gravity PDF & dex \\
 met16 & 16th percentile of {\tt StarHorse} metallicity PDF & dex \\
 met50 & 50th percentile of {\tt StarHorse}  metallicity PDF & dex \\
 met84 & 84th percentile of {\tt StarHorse}  metallicity PDF & dex \\
 dist16 & 16th percentile of {\tt StarHorse} distance PDF & kpc \\
 dist50 & 50th percentile of {\tt StarHorse} distance PDF & kpc \\
 dist84 & 84th percentile of {\tt StarHorse} distance PDF & kpc \\
 AV16 & 16th percentile of {\tt StarHorse} extinction in the V band PDF & mag \\
 AV50 & 50th percentile of {\tt StarHorse} extinction in the V band PDF & mag \\
 AV84 & 84th percentile of {\tt StarHorse} extinction in the V band PDF & mag \\
 SH$\_$INPUTFLAGS & {\tt StarHorse} flags specifying catalogue input completeness and quality & string \\
 SH$\_$OUTPUTFLAGS & {\tt StarHorse} flags specifying output quality & string \\
\end{tabular}
\label{table:outtablefmt}
\end{table*}

\begin{table*}[h!]
\caption{ Description of the contents of the {\tt StarHorse} flags.}
\centering
\begin{tabular}{c|c}
 SH\_INPUTFLAGS & Description \\ 
 \hline
 "TEFF.." & calibrated spectroscopic parameters (e.g. TEFF) were used \\ 
 "uncalTEFF.." & uncalibrated spectroscopic parameters (e.g. TEFF) + inflated uncertainties were used \\ 
 "PARALLAX" & {\it Gaia} DR2 parallaxes + recalibrated zeropoint and uncertainties were used \\
 "JHKs" &  2MASS photometry was used\\
 "W1W2" &  WISE photometry was used \\
 "BVgri" & APASS photometry was used \\
 "gps1\_rps1.." & PanSTARRS-1 photometry was used \\
 "AV\_prior" & extinction prior (e.g. from APOGEE targeting) was used \\
 \hline
 \hline
 SH\_OUTPUTFLAGS &  \\ 
 \hline
 "NEGATIVE\_EXTINCTION" & bad extinction estimates \\ 
 "NUMMODELS\_HIGH" & high number of stellar models compatible with observations within $3\sigma$ \\ 
 "NUMMODELS\_LOW" & low number of stellar models compatible with observations within $3\sigma$ \\
\end{tabular}
\label{table:flags}
\end{table*}

\section{Validation}
\label{sec:validation}

At the level of spectroscopic stellar surveys, it is difficult to perform truly independent benchmark tests for the resulting distance, extinction, and stellar parameter scales \citep{Jofre2019}. Most comparison samples are themselves affected by significant systematic uncertainties. Especially for the APOGEE survey, meaningful comparisons with fundamental physical parameters such as interferometric temperatures or masses of detached eclipsing binaries are unavailable. In \citet{Santiago2016} and Q18 we performed fundamental accuracy and precision tests using simulated stars, nearby eclipsing binaries, astrometric distances, among others. In this section we therefore limit our validation to new, but slightly less fundamental tests: consistency with input parallaxes, asteroseismology (using the CoRoT-APOGEE sample), open clusters (using {\it Gaia} DR2 results), an inter-survey comparison, and a comparison with results obtained by \citet{Leung2019}.

\subsection{Comparison to input parallaxes}

\begin{figure}
\centering
  \includegraphics[width=.49\textwidth]
  {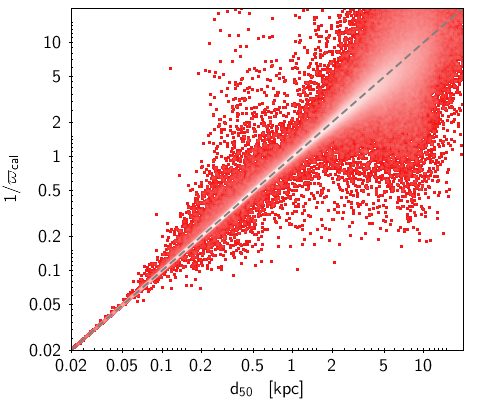} 
  \includegraphics[width=.49\textwidth]
  {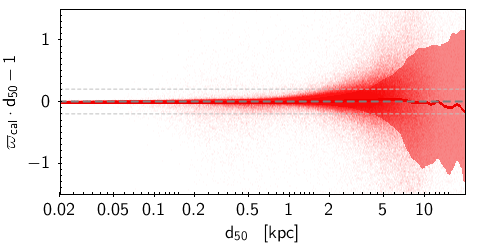} \\
    \caption{Comparison of {\tt StarHorse} DR16 distances to naive distances obtained by inverting the recalibrated {\it Gaia DR2} parallaxes. Top panel: One-to-one comparison of posterior with naive $1/\varpi$ distances. Bottom panel: Residuals between pure astrometric and spectro-photo-astrometric ($1/d_{50}$) parallaxes. The red line shows the smoothed running median, while the shaded region shows the corresponding 1$\sigma$ variations.}
  \label{fig:1overpi}
\end{figure}

As a first simple consistency check, we show in Fig. \ref{fig:1overpi} a comparison between our spectro-photo-astrometric distances with the recalibrated {\it Gaia} DR2 input parallaxes. 
We canonically allow {\tt StarHorse} to converge to a solution that deviates from the input measurements by maximum 4$\sigma$, using trimmed Gaussians in the likelihood computation. We therefore expect an almost perfect agreement with the input parallaxes within the corresponding uncertainties. Fig. \ref{fig:1overpi} shows that this expectation is fulfilled: The upper panel compares our posterior estimates with the naive $1/\varpi$ distances (which is biased estimator of the true distance; see \citealt{Luri2018}), while the lower panel demonstrates that there are minimal residuals between the input and the posterior parallaxes within the {\it Gaia} DR2 parallax sphere (the region where parallax uncertainties are within $10-15\%; d\lesssim2.5$ kpc). We only see slight systematic trends appearing for distances $d\gtrsim10$ kpc. In the regime in-between, the parallax information is successfully complemented by APOGEE, delivering less biased and more precise {\tt StarHorse} distances.

\subsection{Asteroseismology: The CoRoT-APOGEE sample}

\begin{figure*}
\centering
  \includegraphics[width=.49\textwidth]
  {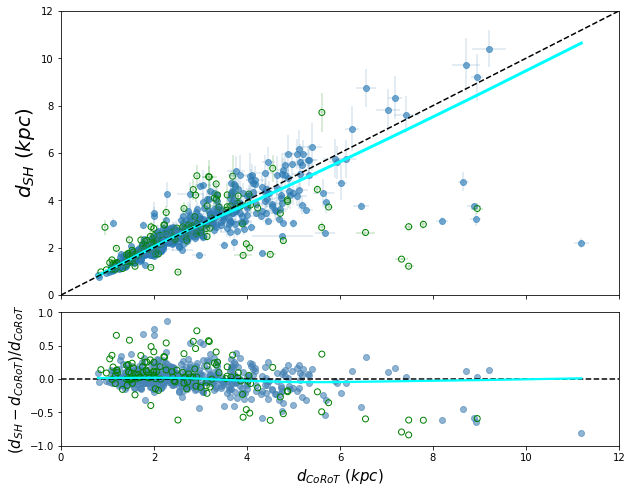} 
  \includegraphics[width=.49\textwidth]
  {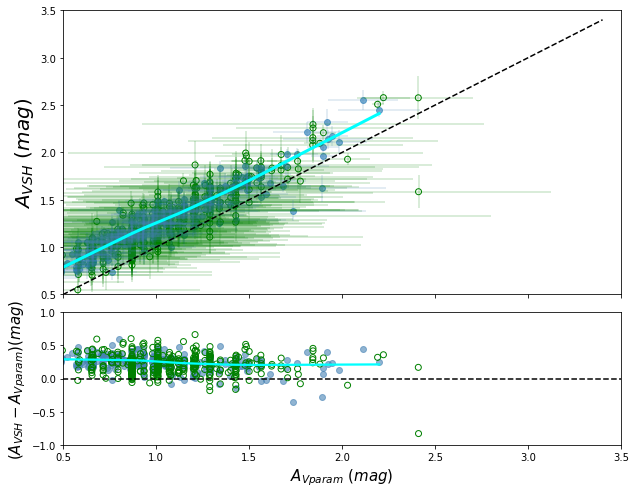} \\
  \includegraphics[width=.49\textwidth]{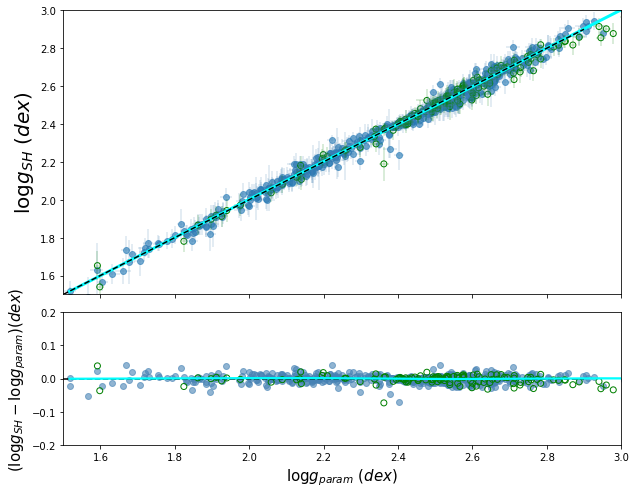} \\
    \caption{Comparison between distances and extinctions obtained here and the ones obtained from asteroseismology for CoRoT stars with APOGEE spectra (CoRoGEE sample) using an updated version of the PARAM code \citep{Rodrigues2017}. Blue filled dots are all stars with PARAM tension flags equal zero, for which the PDF of the estimated quantities does not contain multiple peaks. The cyan line is the locally linear adjust of the blue filled dots. In the case of extinction, right upper panel, the blue dots represents the subset of PARAM tension flags equal zero and stars for which all photometric filters were available. Green open symbols are all stars that do not satisfy the conditions of the blue dots.}
  \label{fig:corot}
\end{figure*}

\begin{figure*}
\centering
\includegraphics[width=.7\textwidth]{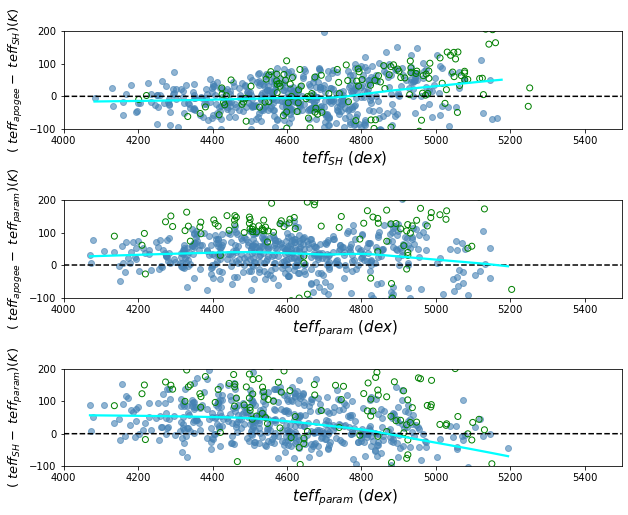} \\
\caption{Comparison between temperatures from APOGEE DR16, output PARAM code, and output {\tt StarHorse}. Blue filled dots are all stars with PARAM tension flags equal zero, for which the PDF of the estimated quantities does not contain multiple peaks. The cyan line is the locally linear adjust of the blue filled dots. Green open symbols are all stars that do not satisfy the conditions of the blue dots.}
\label{fig:corotteff}
\end{figure*}

In Fig. \ref{fig:corot} we show a direct comparison of the distances, $A_V$ and surface gravity for stars in common between the APOGEE DR16 {\tt StarHorse} results and the CoRoGEE sample \citep{Anders2017a}, which contains stars observed by both APOGEE and the CoRoT space mission \citep{Baglin2006}. The CoRoT data allow us to determine stellar masses and radii from asteroseismology, thus also providing more precise distances outside the {\it Gaia} parallax sphere.

A similar comparison was shown in Sect. 5.2 of Q18, but the present one is significantly different in two ways: 1. The CoRoGEE distances were obtained with an updated version of the PARAM code \citep{Rodrigues2017}, with the configuration where the input parameters were the two seismic parameters ($\Delta\nu$ and $\nu_{\rm max}$) and the APOGEE DR16  temperatures, metallicities, and [$\alpha$/Fe] values. No {\it Gaia} parallaxes were used. 2. In contrast to the {\tt StarHorse} run shown in Fig. 9 of Q18 (which used the PARAM distances as an input), we now compare to the {\tt StarHorse} results obtained without any input from neither asteroseismology nor PARAM. In summary, we compare the result of two independent distance codes, one of which uses spectroscopy and asteroseismology (PARAM), and the other uses spectroscopy and astrometry ({\tt StarHorse}).

In Figure \ref{fig:corotteff} we show the comparisons between the input temperatures from APOGEE DR16 and the output temperatures from PARAM and {\tt StarHorse} codes. We see a systematic shift between PARAM and APOGEE DR16 temperatures even for PARAM tension flags equal zero. Differently from PARAM, {\tt StarHorse} output temperatures are very similar from the APOGEE input, since the spectroscopic errors are small and {\tt} StarHorse does not rely on the seismic information. The systematic difference in Temperatures between PARAM and APOGEE maybe due to the different calibration scales and model choices, which in the case of PARAM is MESA \citep{Paxton2011}. 

Considering this systematic shift between PARAM output temperatures and APOGEE DR16, and looking at simulation tests with StarHorse (See Fig. 6 of Q18, bottom left panel) we expect a shift in the extinction itself -- which is seen in the left upper panel of Fig. \ref{fig:corot}. The magnitude of this shift in the extinction scale, however, exceeds our expectation: for a systematic +50 K shift in $T_{\rm eff}$ we would expect not more than 0.1 mag difference in extinction. We therefore tentatively attribute this difference to the missing reliable optical photometry for the APOGEE DR16 sample. Distances and superficial gravity are in very good agreement with the ones derived by PARAM using asteroseismic measurements.

\begin{figure*}
\centering
  \includegraphics[width=\textwidth]{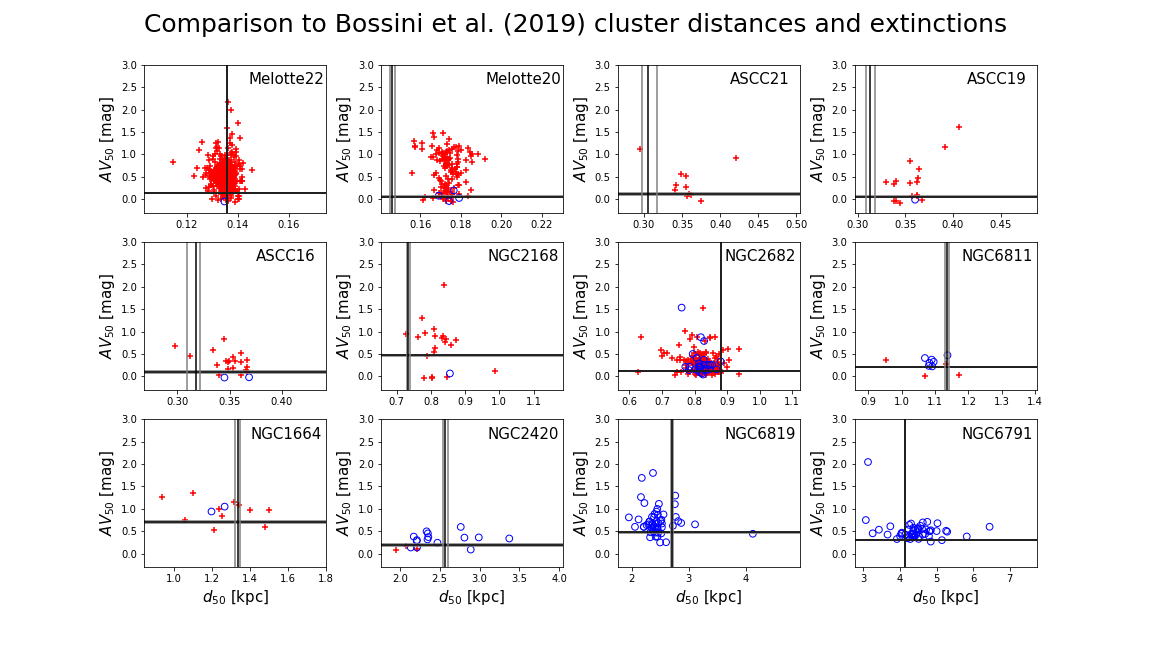} 
    \caption{Comparison between distances and extinctions obtained in this paper with those obtained by \citet{Bossini2019}, but in this case for the same input photometry as {\tt StarHorse} and with PARSEC models, for open clusters. Each panel corresponds to an open cluster with more than 10 member candidates observed by APOGEE. The median {\tt StarHorse} results for individual stars in each cluster are shown as red crosses (dwarfs) and blue open circles (giants)}. The horizontal and vertical lines correspond to the median values and 1$\sigma$ limits inferred by \citet{Bossini2019} through isochrone fitting.
  \label{fig:bossini}
\end{figure*}

\begin{figure*}
\sidecaption
\centering
  \includegraphics[width=.66\textwidth]{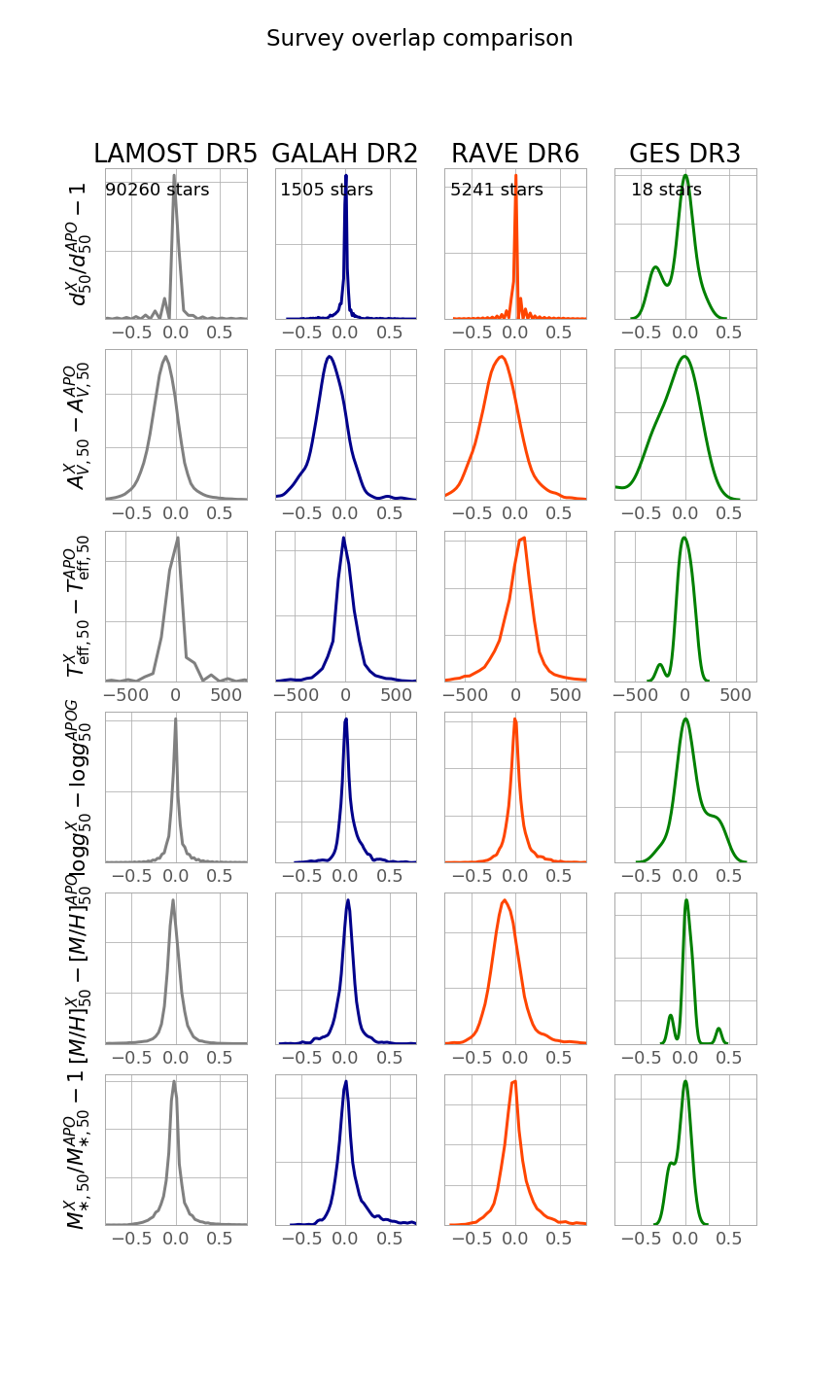} 
    \caption{Inter-survey comparison of the derived {\tt StarHorse} results, using stars co-observed by APOGEE and LAMOST (left column), GALAH (second column), RAVE (third column), and GES (fourth column). Each panel shows a generalised histogram of differences of the posterior parameters obtained by {\tt StarHorse} indicated in the y-axis of each row. For each survey, the number of stars in common with APOGEE DR16 is given in the top panel.}
  \label{fig:surveycomparison}
\end{figure*}

\begin{figure*}
\centering
  \includegraphics[width=.49\textwidth]{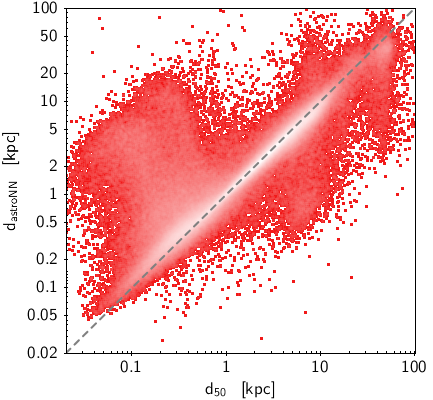} 
  \includegraphics[width=.49\textwidth]{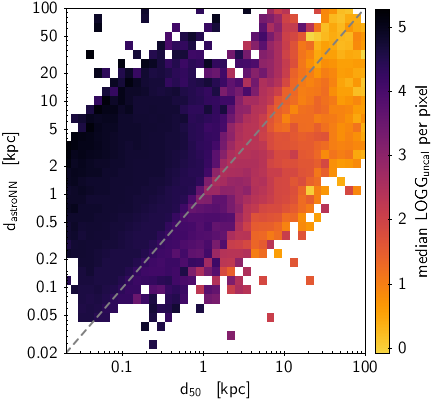} \\
  \includegraphics[width=.49\textwidth]{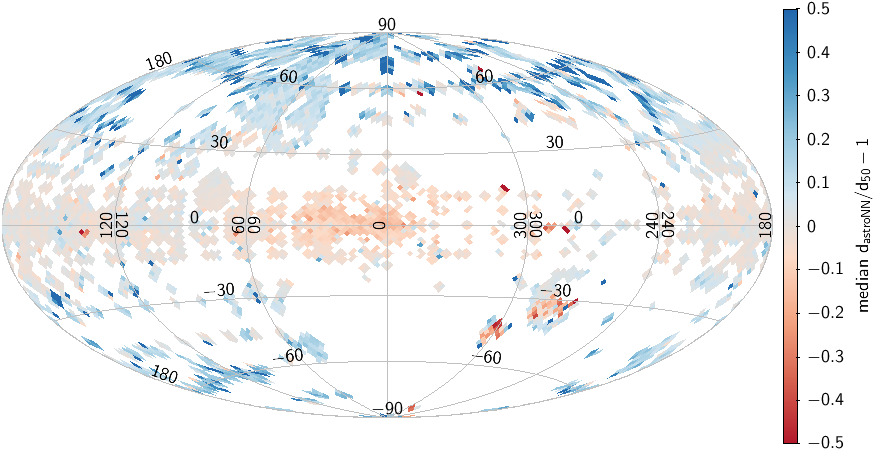} 
  \includegraphics[width=.49\textwidth]{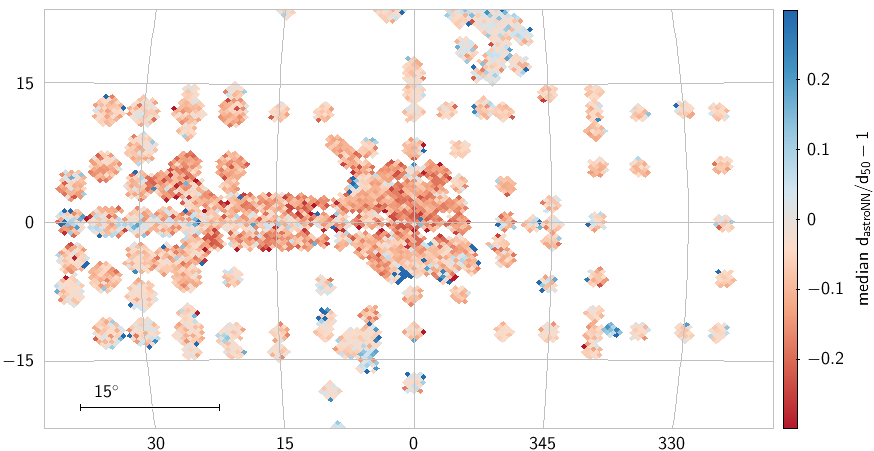} \\
    \caption{Comparison with {\tt astroNN} distances presented by \citet{Leung2019}. Top panel 1-to-1 comparison (left: density distribution, right: colour-coded by median uncalibrated surface gravity determined by ASPCAP, showing that {\tt astroNN} is overstimating distances to dwarf stars.). Bottom panels: relative distance differences as a function of sky position (left: whole sky, right: zoom into the inner Galaxy).} 
  \label{fig:astroNN}
\end{figure*}

\subsection{Open clusters}

In A19, we presented a detailed comparison of {\tt StarHorse} results (without using spectroscopic data) with open-cluster parameters derived from {\it Gaia} DR2 data (specifically, \citealt{Cantat-Gaudin2018} and \citealt{Bossini2019}). \citet{Cantat-Gaudin2018} determined membership probabilities for 1229 Galactic open clusters, while \citet{Bossini2019} published revised Bayesian cluster parameters for 269 of those clusters, based on the same membership list. Here we again compare with the results obtained by \citet{Bossini2019}, keeping in mind now that the APOGEE DR16 {\tt StarHorse} results were obtained from higher-quality data.

In Fig. \ref{fig:bossini}, we compare the APOGEE {\tt StarHorse} results obtained for the most certain cluster members of \citet{Cantat-Gaudin2018} to the distances and extinctions determined similar to  \citet{Bossini2019} with same input photometry as we use in {\tt StarHorse} and using PARSEC models in the PARAM code. The figure shows a cluster-by-cluster comparison for the 12 most populated clusters observed by APOGEE, ordered by distance. In general, and in accordance with A19, we observe good agreement of the distance scales (within 20\%). Some discrepancies are noticeable both in extinction and distance, which could be related to differential reddening, impure membership, and bad photometry, though this is mostly within the accuracy limits of the open cluster distance scale of \citet{Bossini2019}. For the closest clusters, we see a very strong systematic difference in extinction estimates (up to $>1$ mag). Its origin, however, is different from the shift seen in the comparison with the CoRoGEE sample: the ASPCAP $T_{\rm eff}$ scale of the M dwarf stars is offset from the PARSEC scale by over 200 K, thus forcing {\tt StarHorse} to converge to a solution with higher extinction (see Q18, Fig. 6).
\\

\subsection{Inter-survey comparison}

Some of the stars observed by APOGEE have also been observed by other spectroscopic surveys, be it as a part of a dedicated cross-calibration effort or by chance. These stars are also useful to test the consistency of the {\tt StarHorse} results. Therefore, in Fig. \ref{fig:surveycomparison} we show the distribution of differences in {\tt StarHorse} output parameters for stars co-observed by APOGEE DR16 and LAMOST DR5, GALAH DR2, RAVE DR6, and GES DR3, respectively (using simple cross-matches based on the {\it Gaia} DR2 {\tt source\_id}), colour-coded in the same way as Figs. \ref{survey_dists} and \ref{survey_maps}. The distances obtained from the different input spectroscopic parameters show very satisfactory consistency (first row of Fig. \ref{fig:surveycomparison}), with systematics at the 1-2\%-level, and standard deviations typically below the quoted uncertainties.

In accordance with the previous tests above, the extinction comparison for the survey overlap stars (second row of Fig. \ref{fig:surveycomparison}) shows that the APOGEE DR16 extinctions are on a slightly offset scale with respect to the ones obtained from LAMOST DR5, GALAH DR2, and RAVE DR6. As explained above, we suggest this to be due to a combination of a slight systematic offset of the ASPCAP $T_{\rm eff}$ scale with respect to the one of the PARSEC models, and the missing reliable optical photometry for most of the DR16 sample.

The comparison of the other {\tt StarHorse} output parameters ($T_{\rm eff}, \log g$, [M/H], and mass), is shown in the bottom rows of fig. \ref{fig:surveycomparison}, showing a very satisfactoy agreement in the parameter scales of the different surveys.

\subsection{{\tt astroNN} distances}

Finally, in Fig. \ref{fig:astroNN} we compare our APOGEE DR16 distances with the ones obtained with the neural-network spectral analysis code {\tt astroNN} \citep{Leung2019}. These authors claimed that "there is no doubt that our distances have higher precision and accuracy than those determined using stellar models and density priors, such as the BPG distances", based on a comparison with the pre-{\it Gaia} distances published in \citet{Santiago2016} prior to Gaia. Here we repeat their comparison with our new results, now including Gaia DR2, revealing a more complex picture.

The top left panel of Fig. \ref{fig:astroNN}  shows that there is a generally very good agreement between the distances derived by the two codes for the bulk of the sample up to $\sim 10$ kpc (density colour coding in this plot is logarithmic). There are, however, groups of stars which deviate considerably from the one-to-one relation: 1. dwarf stars located mostly at high latitudes (see $\log g$-coloured plot in the top right panel and sky distribution of distance residuals in the lower left panel) for which {\tt astroNN} determines too high distances (compare to Fig. \ref{fig:1overpi}), and 2. giant stars in the Inner Galaxy, for which systematic differences on the order of $10-20\%$ are visible (in the sense that the \citealt{Leung2019} distances are significantly smaller; see lower right panel).

The first group of stars can be explained by the limited training set used by \citet{Leung2019}, which were comprised almost exclusively of red-giant stars. 
The second effect was indeed also noticed by \citet{Bovy2019} who corrected the systematic offset of the {\tt astroNN} distances heuristically (see their Fig. 1).

\section{Summary plots for GALAH, RAVE, GES, and LAMOST}
\label{othervacs}

\begin{figure*}
\centering
  \includegraphics[width=.7\textwidth]
  {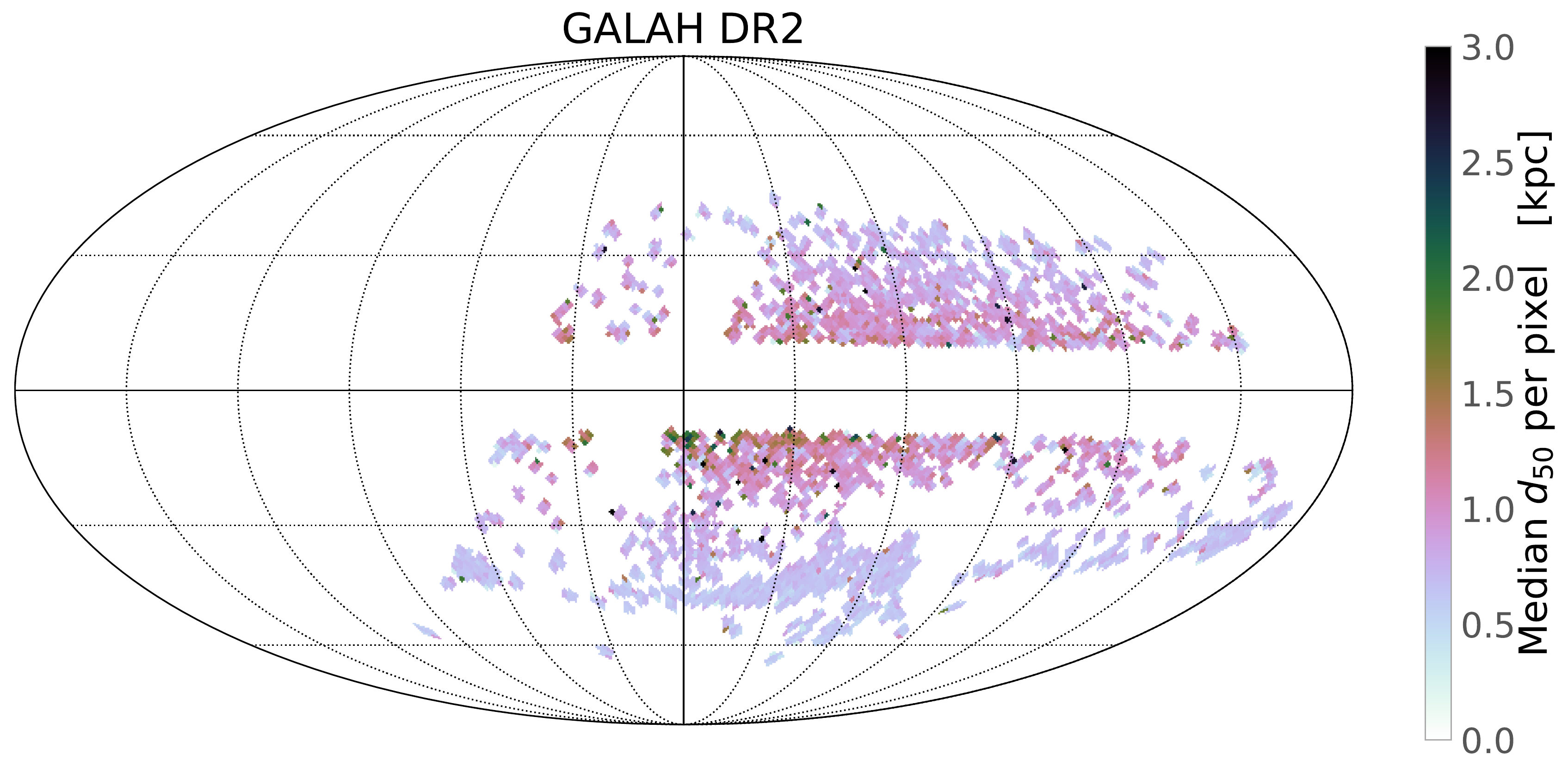} \\
  \includegraphics[width=.7\textwidth]
  {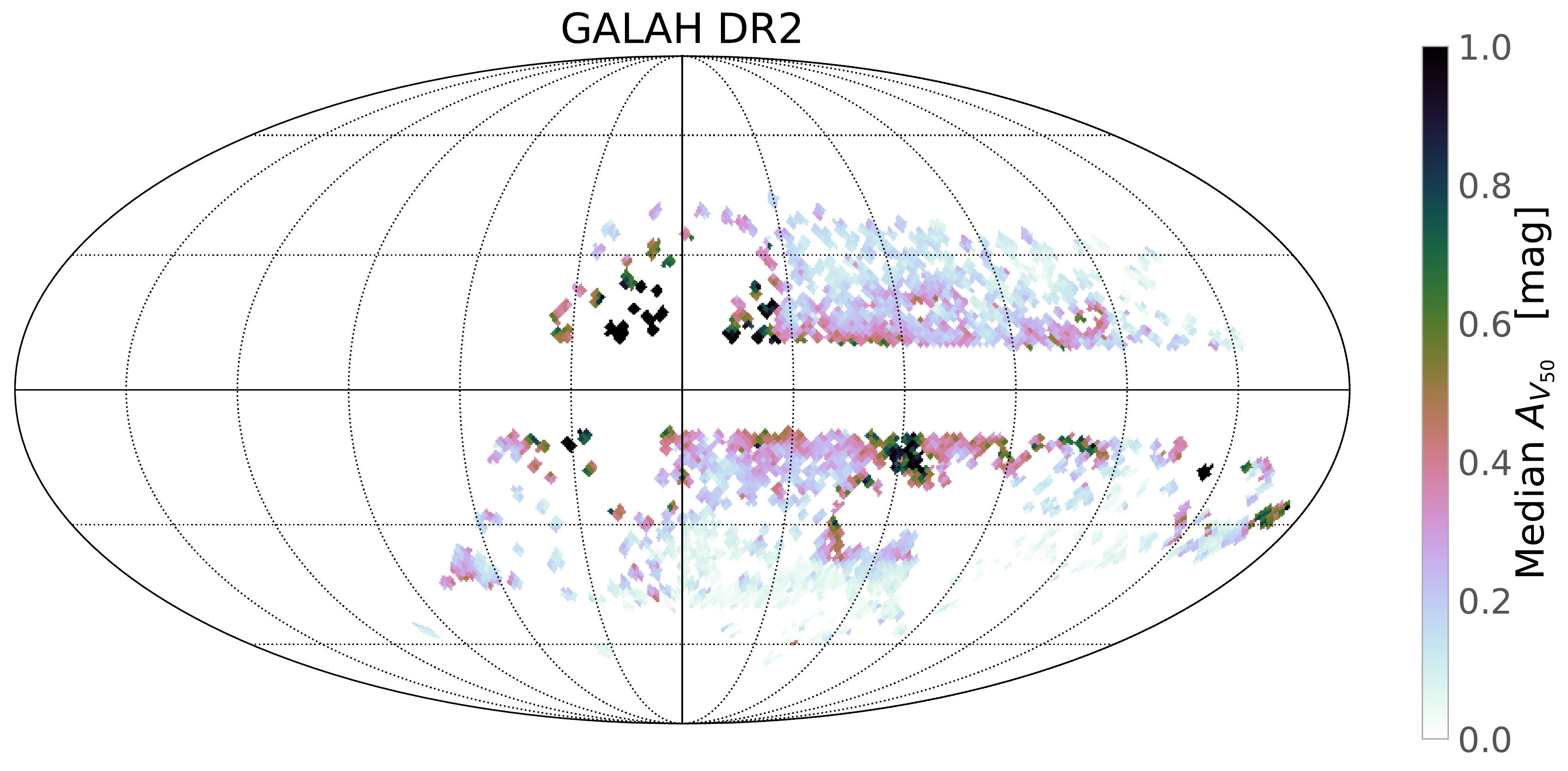} \\
  \includegraphics[width=.9\textwidth]
  {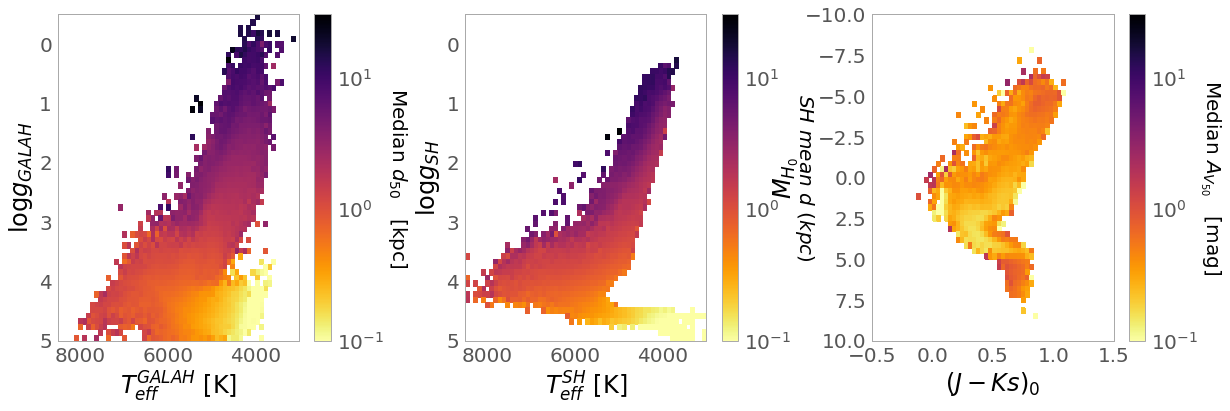} \\
  \caption{Similar to Figs. \ref{fig:maps} and \ref{fig:kielcmd}, but now for GALAH DR2.}
  \label{galahmaps}
\end{figure*}

\begin{figure*}
\centering
  \includegraphics[width=.7\textwidth]
  {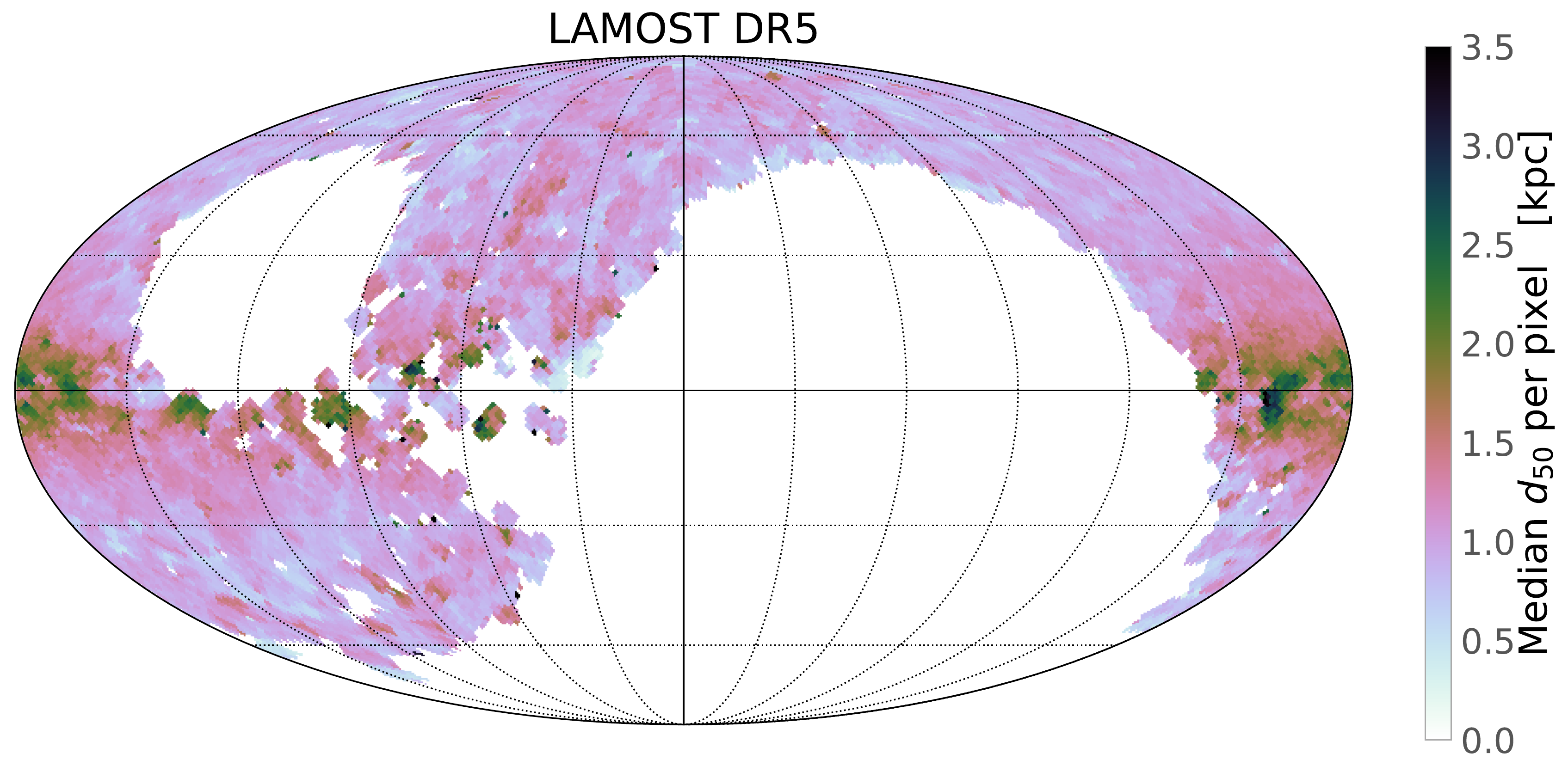} \\
  \includegraphics[width=.7\textwidth]
  {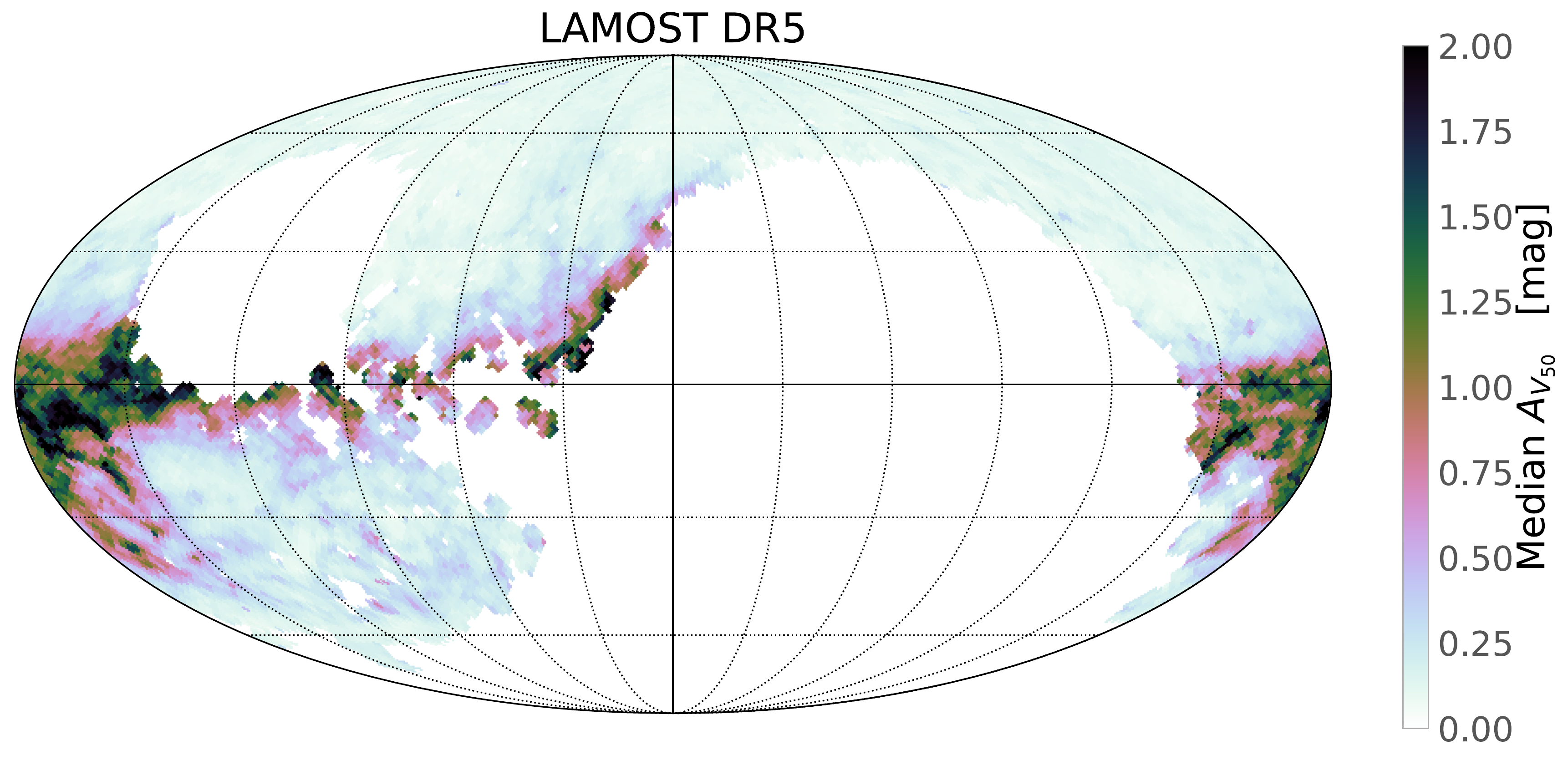} \\
  \includegraphics[width=.9\textwidth]
  {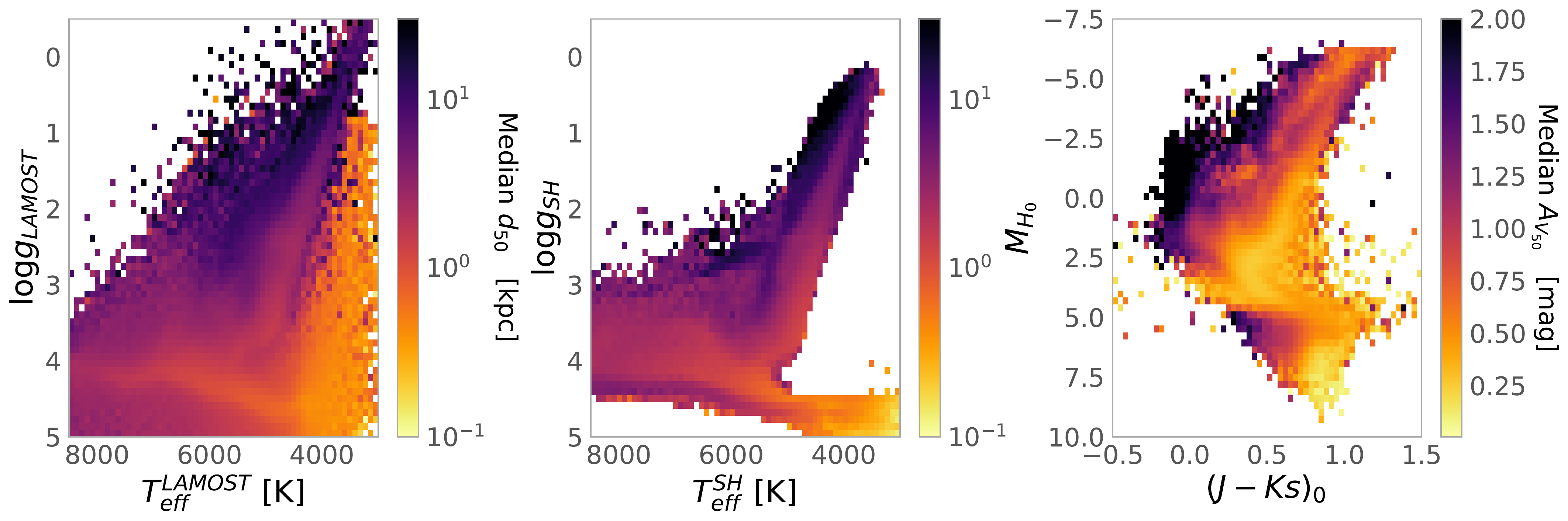} \\
  \caption{ Similar to Figs. \ref{fig:maps} and \ref{fig:kielcmd}, but for LAMOST data. The CMD shown in the right panel does not include sources fainter than $K_s = 14.5$}.
  \label{LAMOSTmaps}
\end{figure*}

\begin{figure*}
\centering
  \includegraphics[width=.7\textwidth]
  {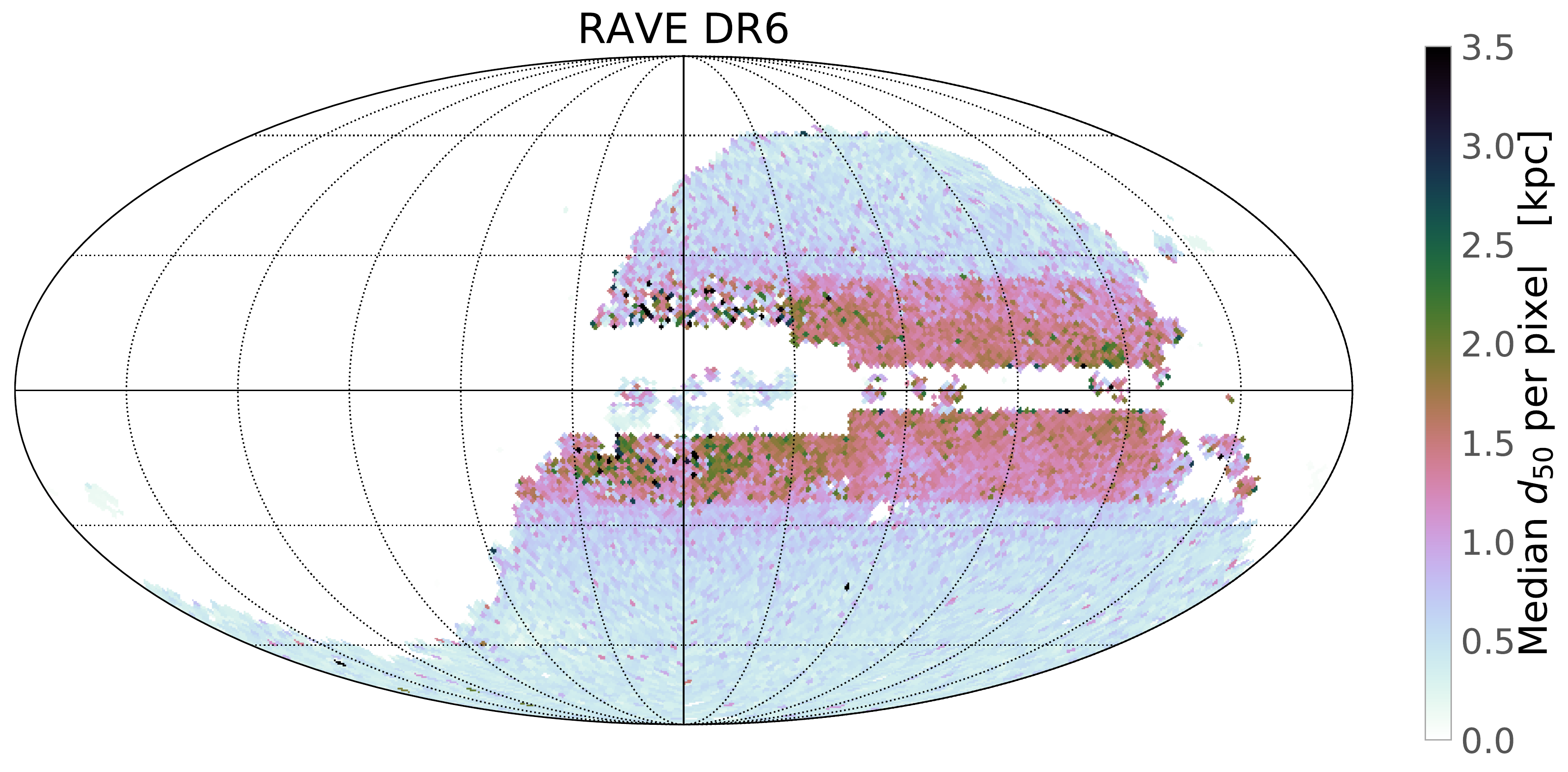} \\
  \includegraphics[width=.7\textwidth]
  {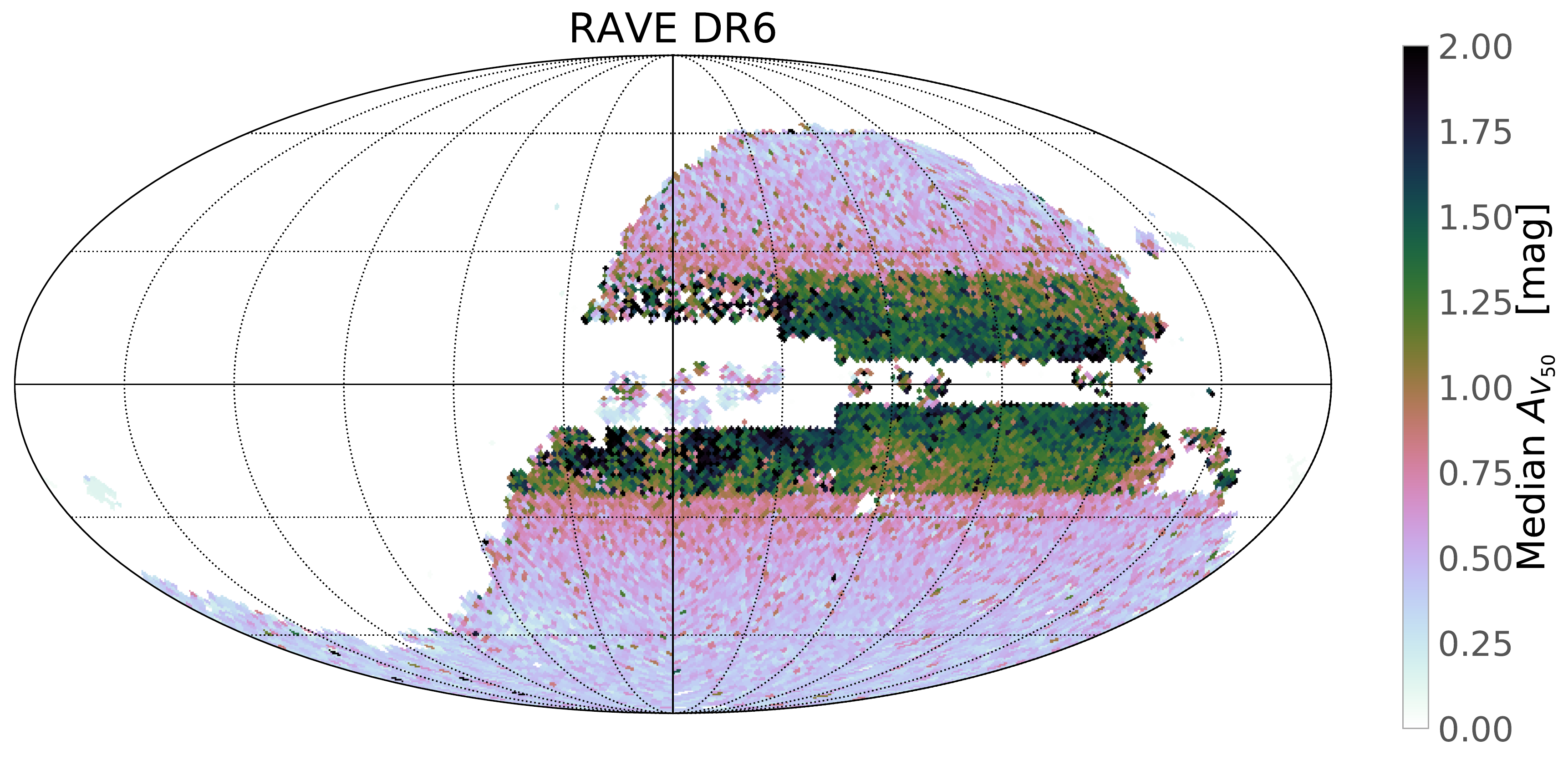} \\
  \includegraphics[width=.9\textwidth]
  {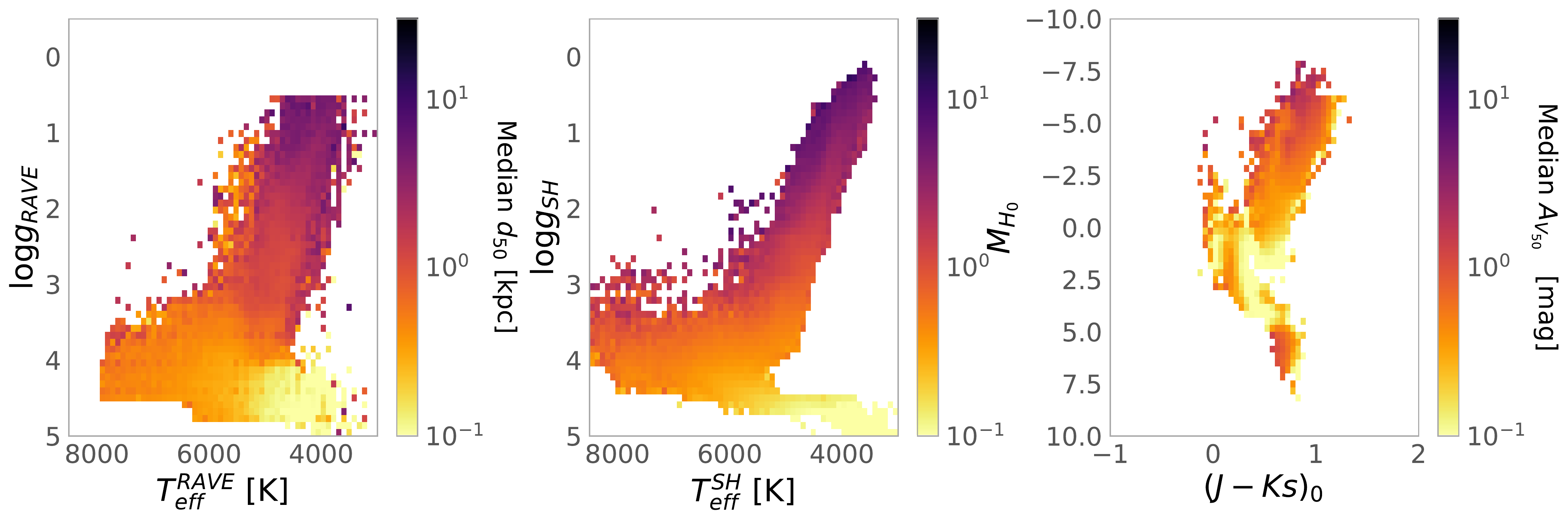} \\
  \caption{ Similar to Figs. \ref{fig:maps} and \ref{fig:kielcmd}, but for RAVE DR6 data.}
  \label{ravemaps}
\end{figure*}

\begin{figure*}
\centering
  \includegraphics[width=.7\textwidth]
  {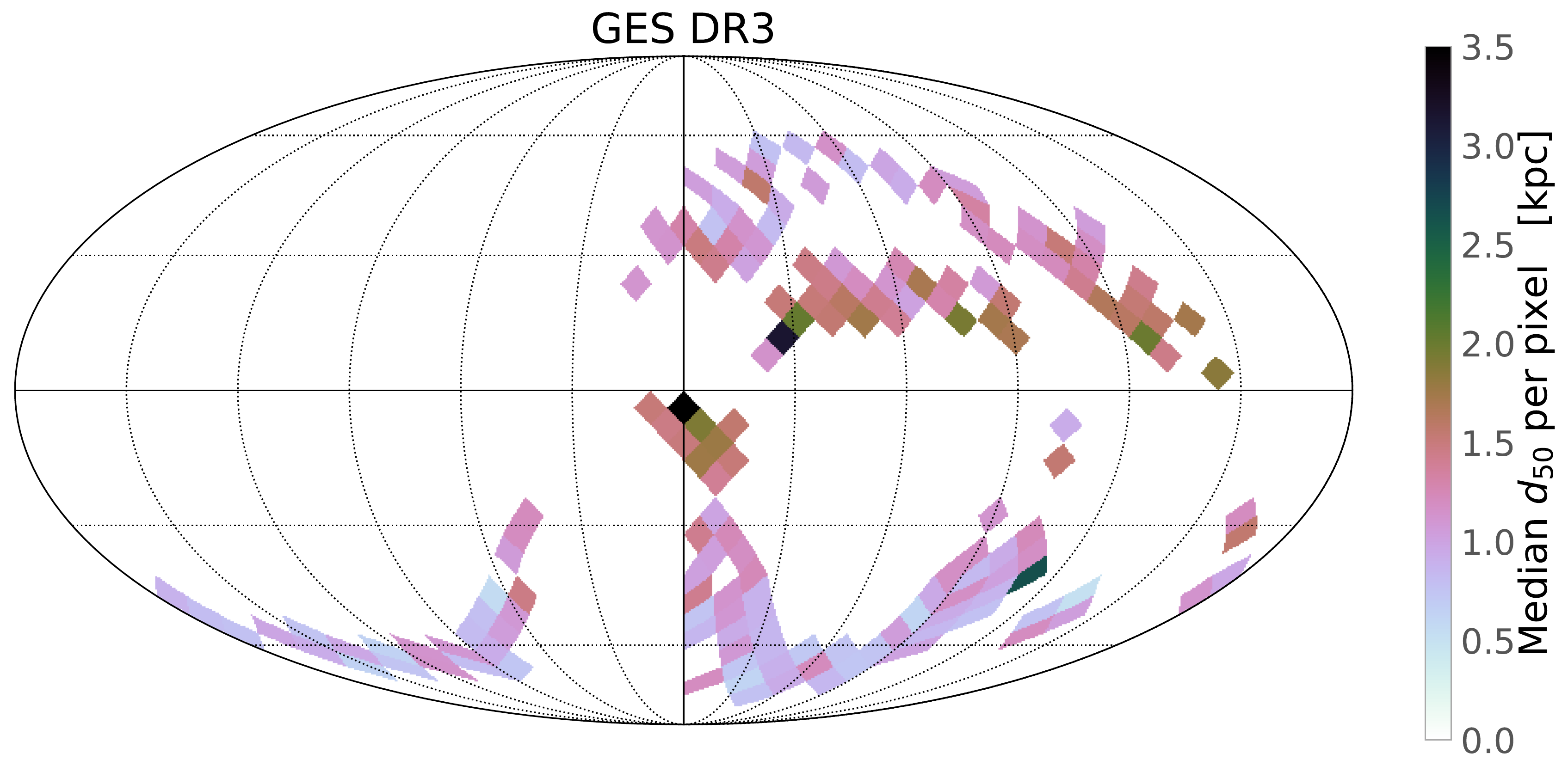} \\
  \includegraphics[width=.7\textwidth]
  {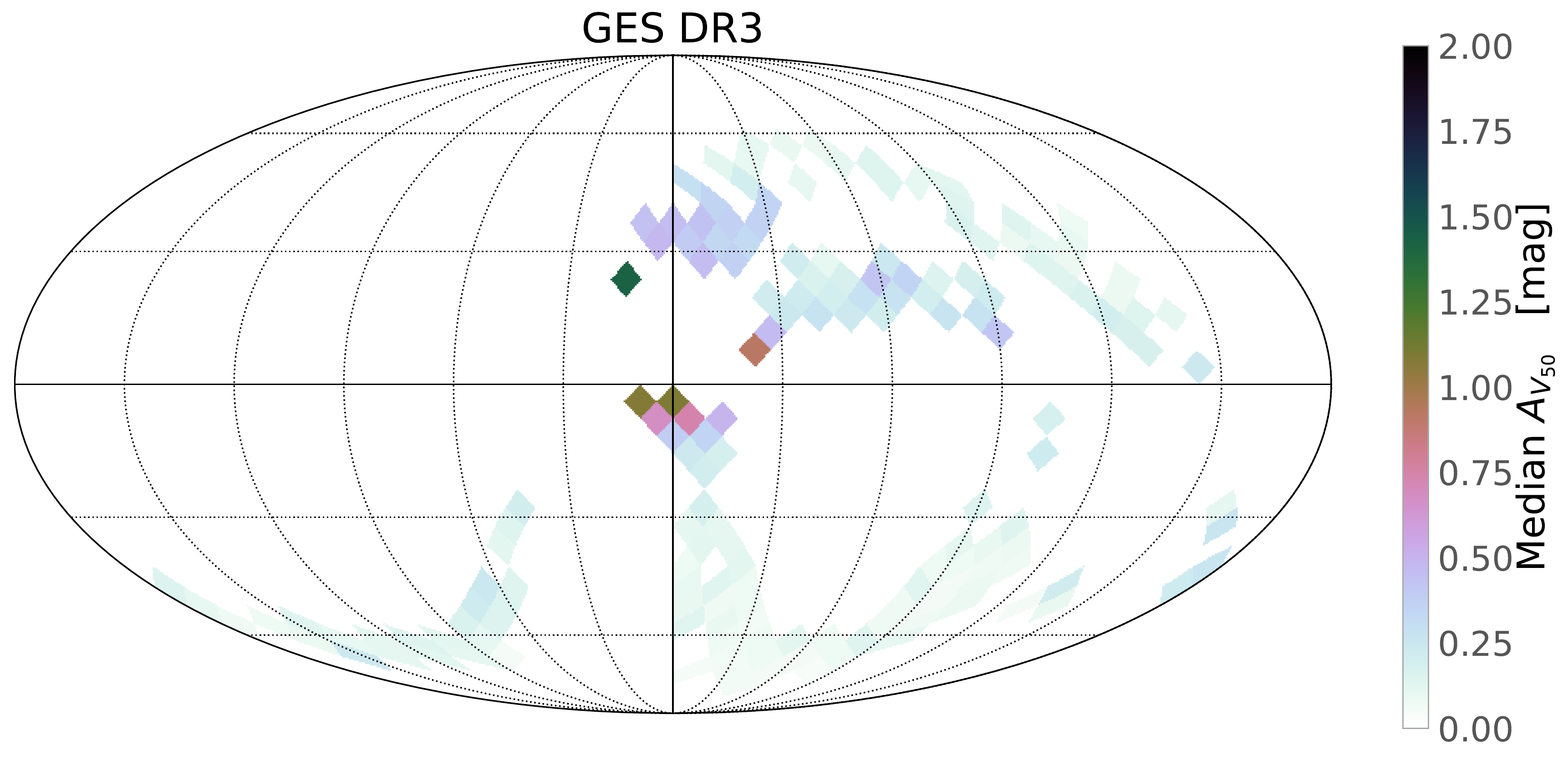} \\
  \includegraphics[width=.9\textwidth]
  {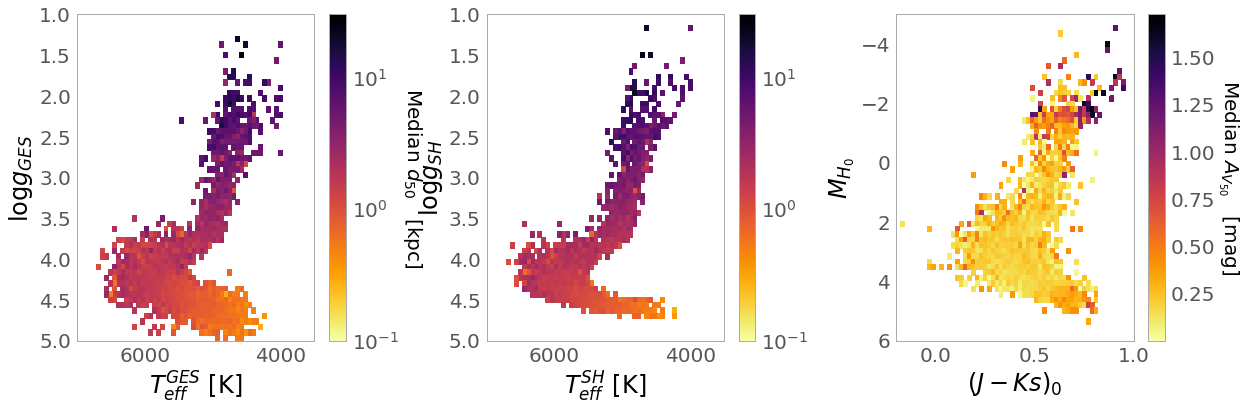} \\
  \caption{Similar to Figs. \ref{fig:maps} and \ref{fig:kielcmd}, but for GES DR3 data.} 
  \label{GESmaps}
\end{figure*}

\begin{figure*}
\vspace{7cm}
  \begin{overpic}[scale=0.36]{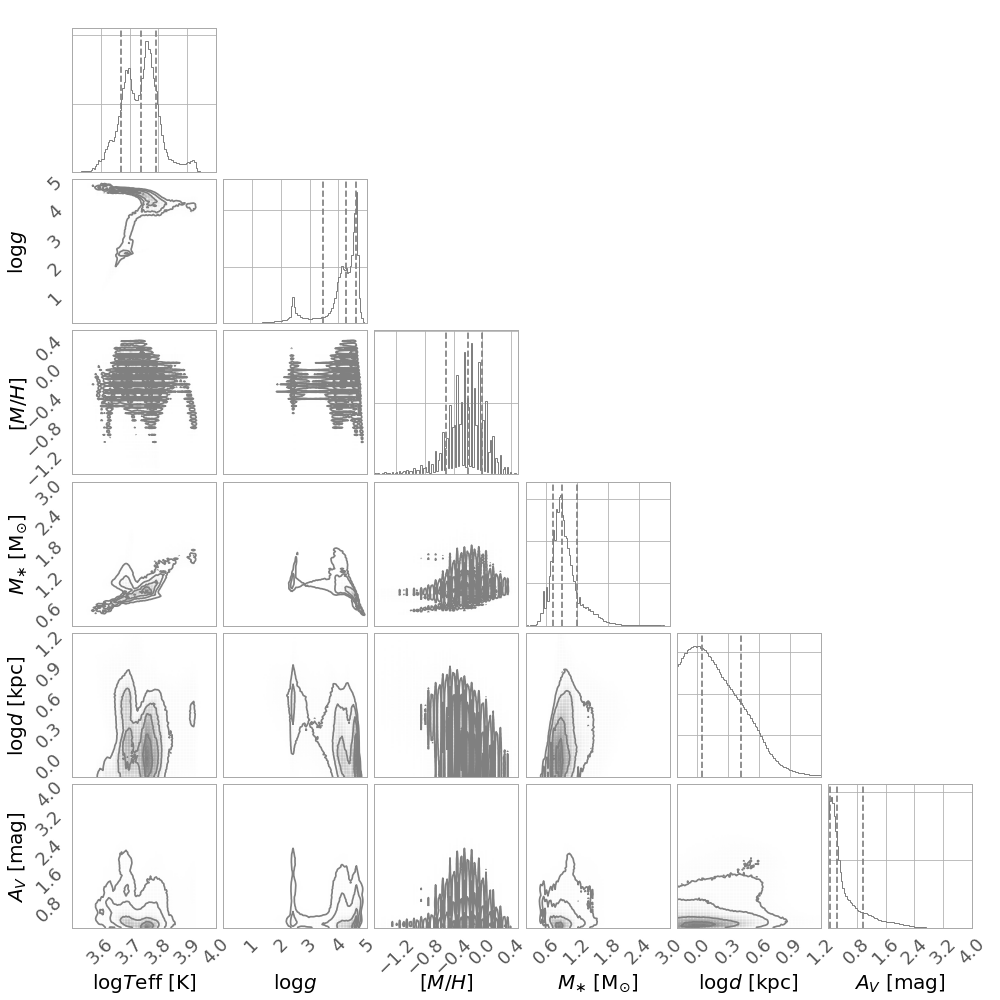}
     \put(52,52){\includegraphics[scale=0.36]{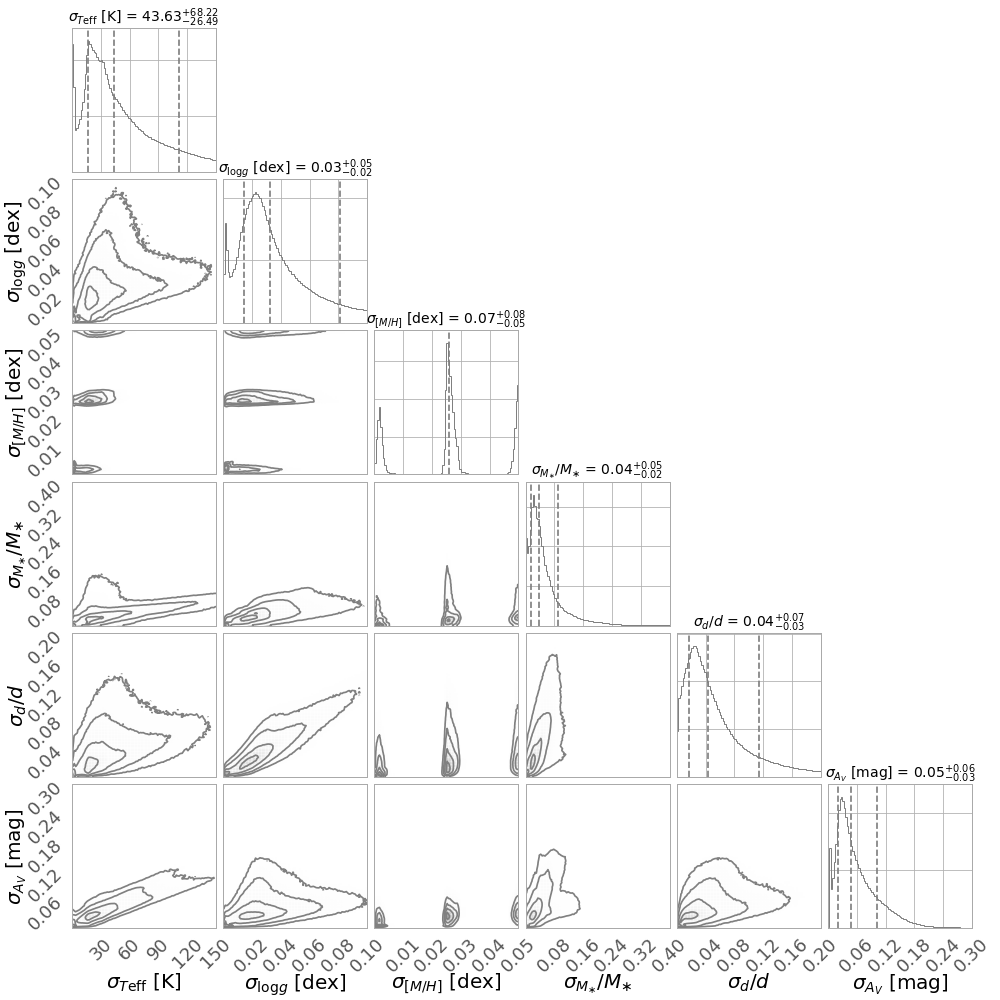}}  
  \end{overpic}
\caption{1D distributions and correlations between {\tt StarHorse} output parameters (bottom left corner plot) and their corresponding uncertainties (top right corner plot) for the LAMOST DR5 VAC sample.}
\label{fig:output_summary_lamost}
\end{figure*}

\begin{figure*}
\vspace{7cm}
  \begin{overpic}[scale=0.36]{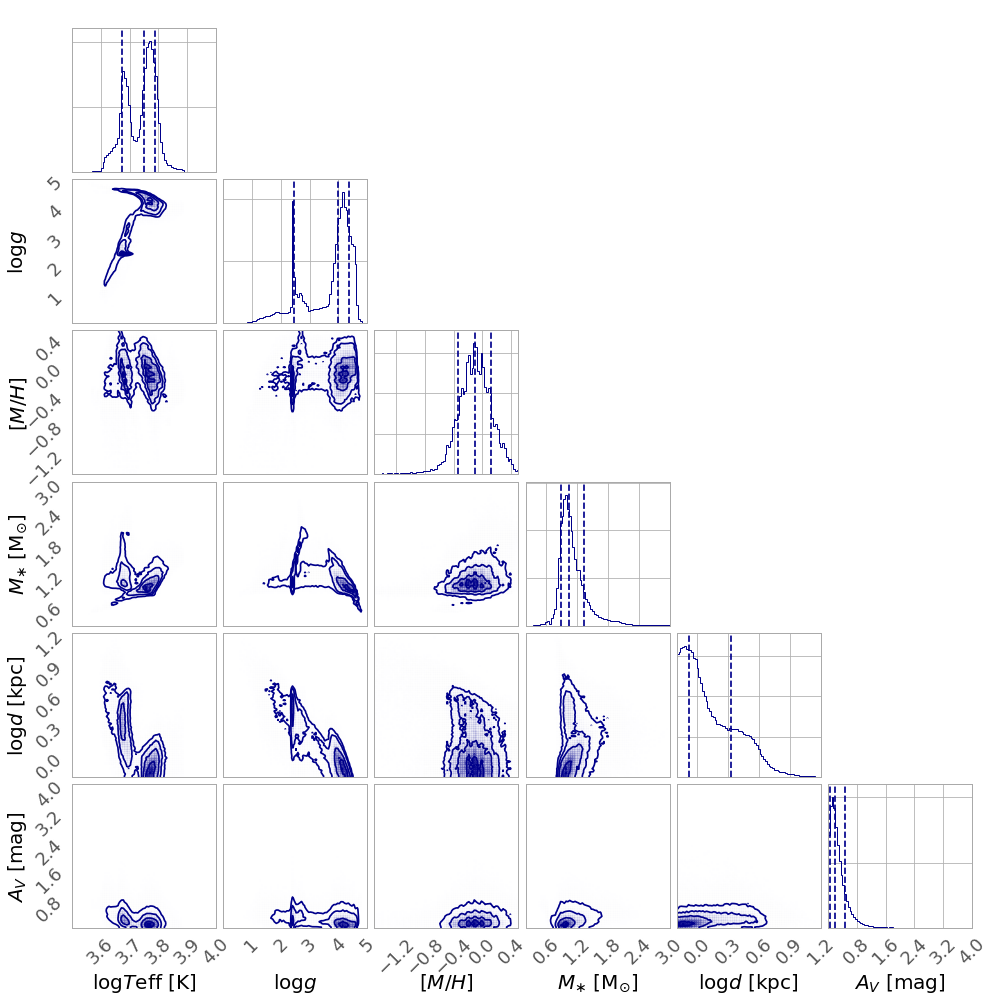}
     \put(52,52){\includegraphics[scale=0.36]{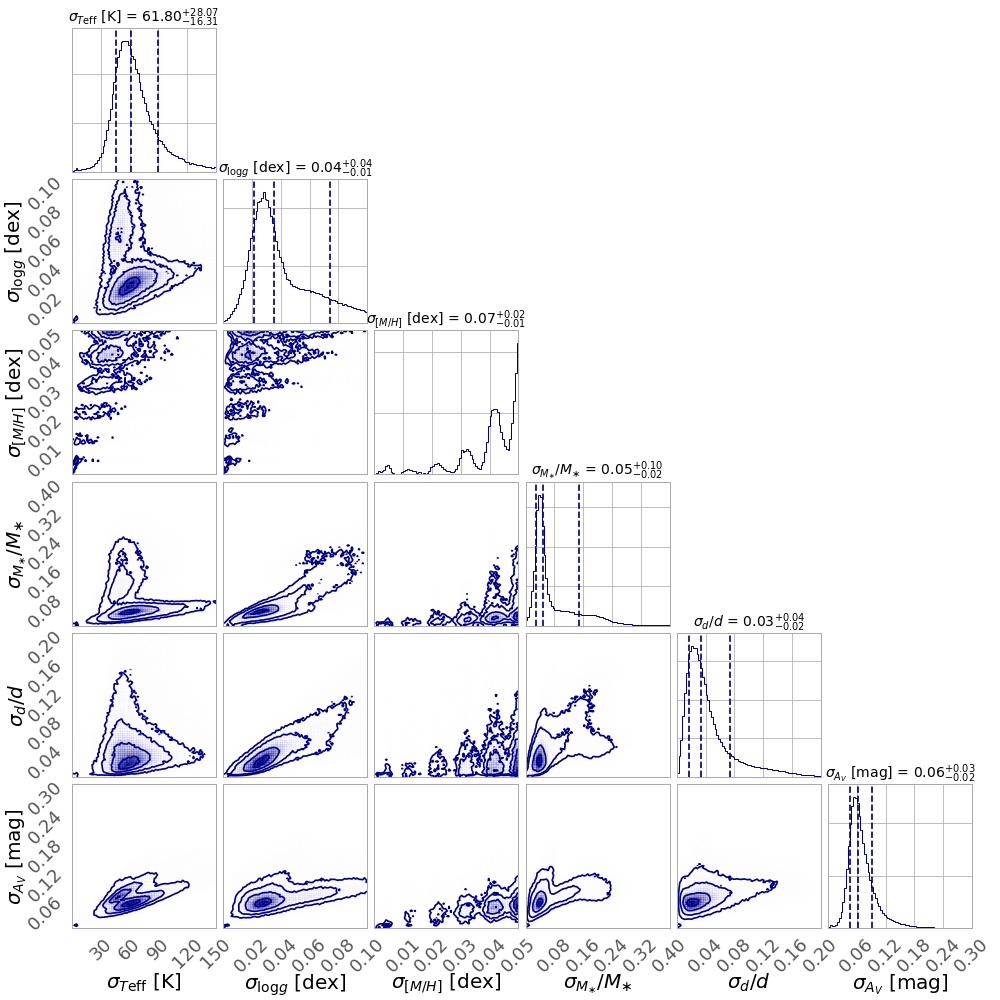}}
  \end{overpic}
\caption{1D distributions and correlations between {\tt StarHorse} output parameters (bottom left corner plot) and their corresponding uncertainties (top right corner plot) for the GALAH DR2 sample.}
\label{fig:output_summary_galah}
\end{figure*}

\begin{figure*}
\vspace{7cm}
  \begin{overpic}[scale=0.36]{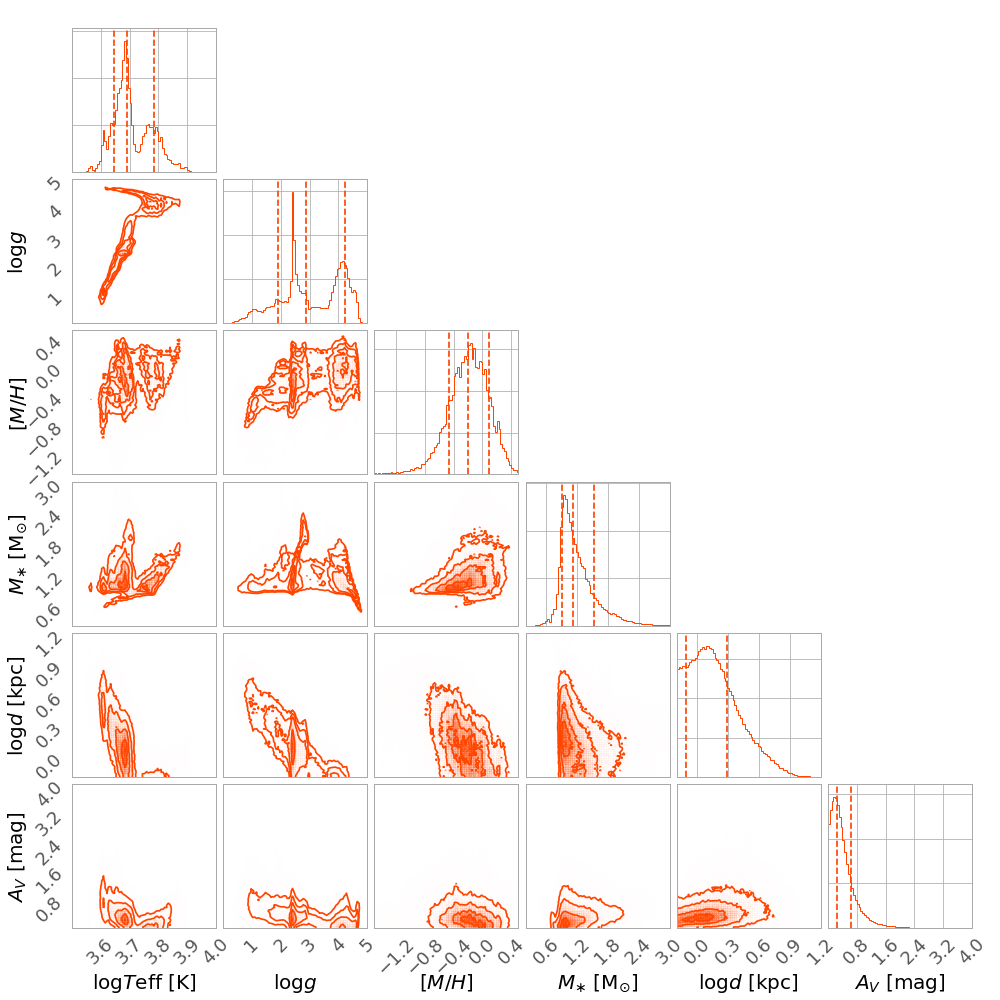}
     \put(52,52){\includegraphics[scale=0.36]{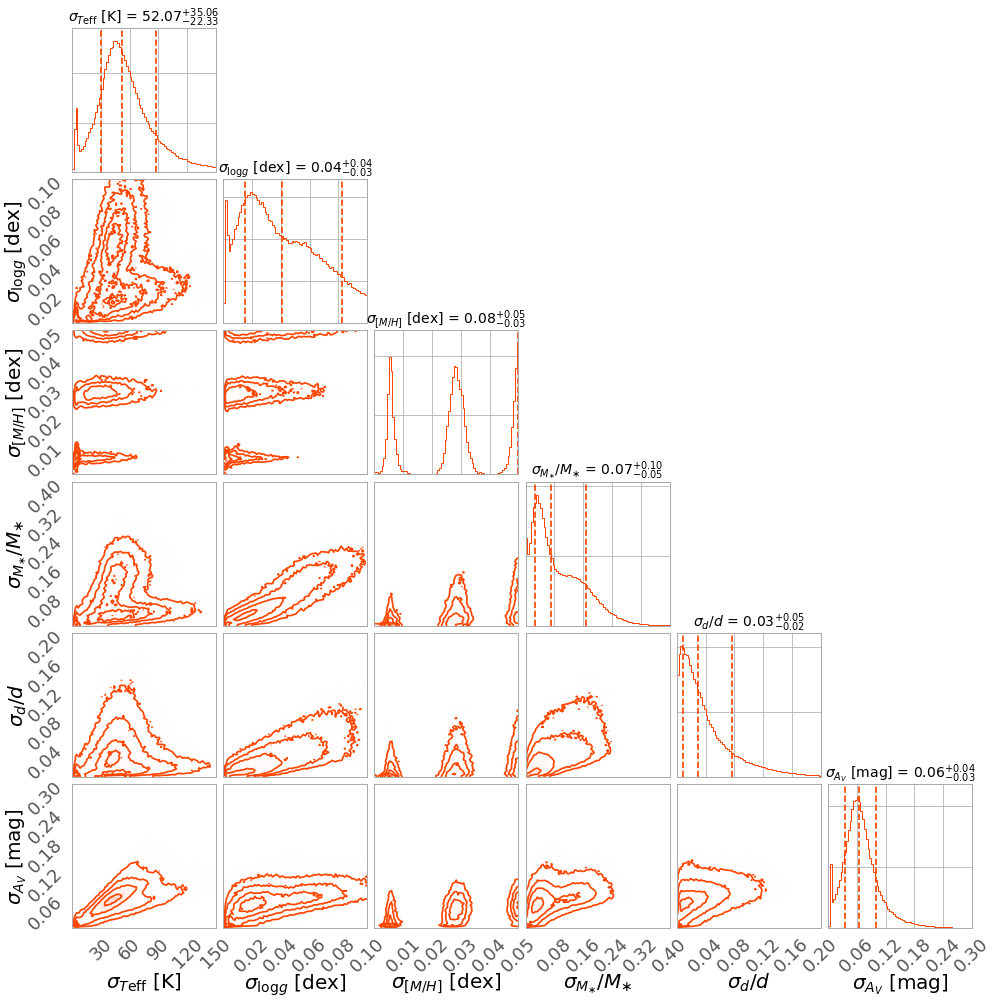}}  
  \end{overpic}
\caption{1D distributions and correlations between {\tt StarHorse} output parameters (bottom left corner plot) and their corresponding uncertainties (top right corner plot) for the RAVE DR6 sample.}
\label{fig:output_summary_rave}
\end{figure*}

\begin{figure*}
\vspace{7cm}
  \begin{overpic}[scale=0.36]{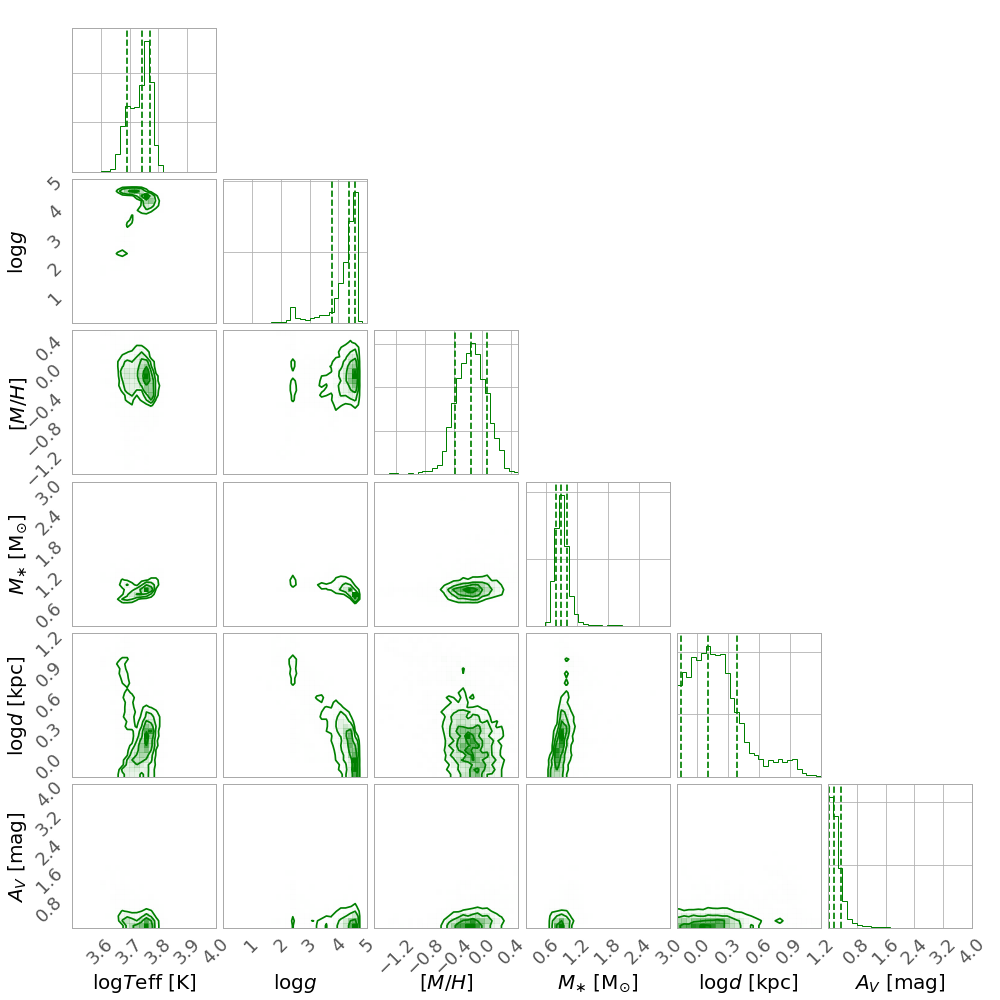}
     \put(52,52){\includegraphics[scale=0.36]{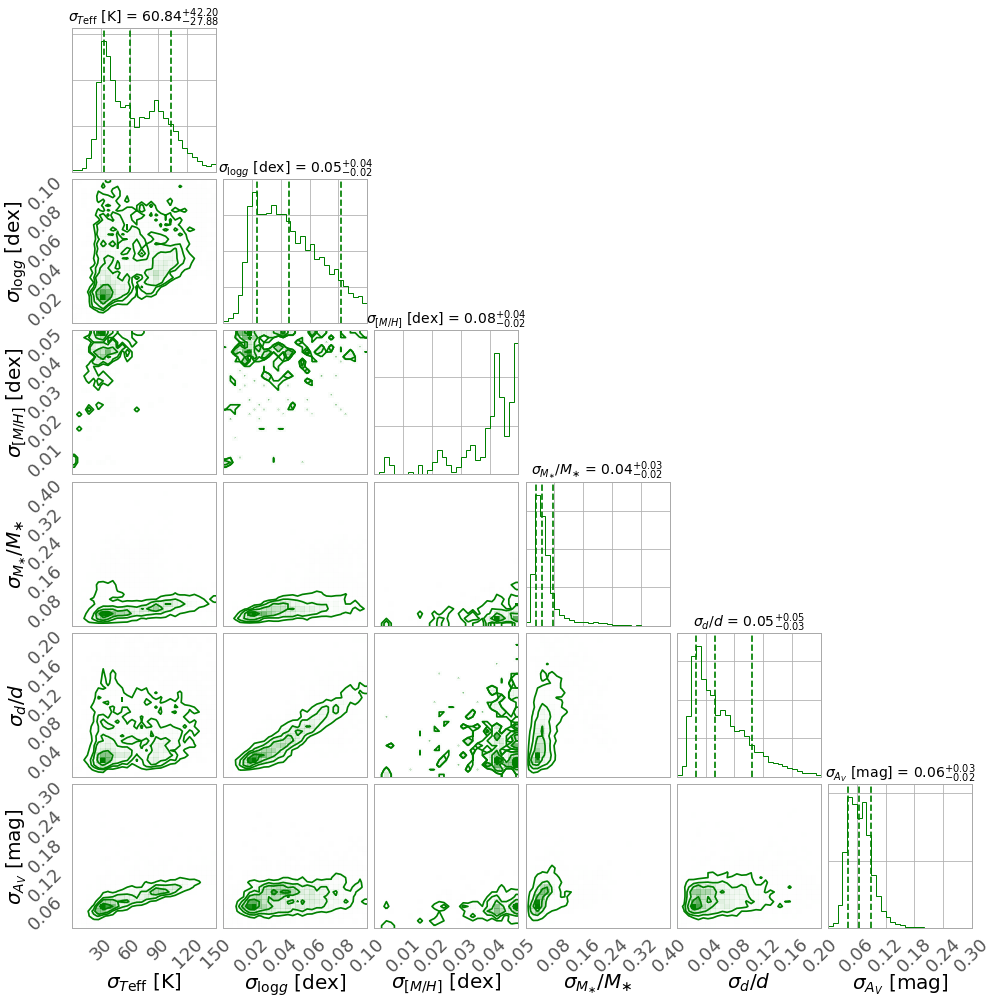}}  
  \end{overpic}
\caption{1D distributions and correlations between {\tt StarHorse} output parameters (bottom left corner plot) and their corresponding uncertainties (top right corner plot) for the GES DR3 sample.}
\label{fig:output_summary_ges}
\end{figure*}

In this section, we show some summary figures illustrating the quality of our new {\tt StarHorse} results for the surveys considered in this paper in addition to the APOGEE DR16 results. In particular, in Figs. \ref{galahmaps} through \ref{GESmaps} we provide sky plots and {\it Kiel} diagrams and CMDs for LAMOST DR5, GALAH DR2, RAVE DR6, and GES DR3, similar to Figs. \ref{fig:maps} and \ref{fig:kielcmd}. Figures \ref{fig:output_summary_lamost} through \ref{fig:output_summary_ges} display summary {\tt corner} plots of the {\tt StarHorse} output parameters for each survey, as shown for APOGEE DR16 in Fig. \ref{fig:output_summary}. The colour in each of those plots coincides with the colours used in Figs. \ref{survey_maps} and \ref{survey_dists}.

\end{document}